\documentclass[%
 12pt,
 aip,
 pof,
 amsmath,amssymb,
 preprint,%
 eqsecnum,
]{revtex4-1}

\usepackage{graphicx}
\usepackage{natbib}
\usepackage{combelow} 
\usepackage{bm} 
\usepackage{braket} 
\usepackage{amsmath} 
\usepackage{amsfonts}
\usepackage{amssymb}
\usepackage{epstopdf}
\usepackage[dvipsnames]{xcolor}

\def\feq{\ensuremath{f^{(\mathrm{eq})}}}

\def\erf{\ensuremath{\mathrm{erf}}}
\def\erfc{\ensuremath{\mathrm{erfc}}}

\def\e{\ensuremath{\mathrm{e}}}
\def\hh{\mathfrak{h}}
\def\abs#1{{\left|#1\right|}}
\def\wp{{\overline{p}}}
\def\vp{{\bm{p}}}

\def\vp{\bm{p}}
\def\vu{\bm{u}}

\def\arctanh{\rm arctanh}

\begin{document}


\title{Implementation of the force term in half-range lattice Boltzmann models} 



\author{Victor E.~Ambru\cb{s}}
\email[Corresponding author; ]{victor.ambrus@e-uvt.ro}
\affiliation{Center for Fundamental and Advanced Technical Research,
Romanian Academy, Bd.~Mihai Viteazul  24, 300223 Timi\cb{s}oara, Romania}
\affiliation{Department of Physics, West University of Timi\cb{s}oara, Bd.~Vasile P\^arvan 
4, 300223 Timi\cb{s}oara, Romania}

\author{Victor Sofonea}
\email[]{sofonea@acad-tim.tm.edu.ro}
\affiliation{Center for Fundamental and Advanced Technical Research,
Romanian Academy, Bd.~Mihai Viteazul  24, 300223 Timi\cb{s}oara, Romania}

\author{Richard Fournier}
\email[]{richard.fournier@laplace.univ-tlse.fr}
\affiliation{LAPLACE, Universit\'{e} de Toulouse, CNRS, INPT, UPS, France.}

\author{St\'{e}phane Blanco}
\email[]{stephane.blanco@laplace.univ-tlse.fr}
\affiliation{LAPLACE, Universit\'{e} de Toulouse, CNRS, INPT, UPS, France.}


\date{\today}

\begin{abstract}
In the frame of the Boltzmann equation, wall-bounded flows of rarefied gases require 
the implementation of boundary conditions at the kinetic level.
Such boundary conditions induce
a discontinuity in the distribution function with respect to the 
component of the momentum which is normal to the boundary. 
Expanding the distribution function with respect to 
half-range polynomials allows this discontinuity to be captured.
The implementation of this concept has been reported 
in the literature only for force-free flows.
In the case of general forces which can have non-zero components 
in the direction perpendicular to the walls, the implementation of 
the force term requires taking the momentum space 
gradient of a discontinuous function.
Our proposed method deals with this difficulty by 
employing the theory of distributions.
We validate our procedure by considering
the simple one-dimensional flow between diffuse-reflective walls of 
equal or different temperatures driven by the constant gravitational force. 
For this flow, a comparison between the results
obtained with the full-range and the half-range Gauss-Hermite LB models
is also presented.
\end{abstract}

\pacs{}

\maketitle 



%
%

%



\section{Introduction}\label{sec:intro}

Force-driven flows at microscale exhibit a plethora of applications \cite{gadelhaq06app,tabeling11},
such as 
mi\-cro\-e\-lec\-tro\-ki\-ne\-tics \cite{stone01,delgado01,voldman06}, microscale cell manipulation \cite{voldman06},
deposition in particle laden turbulent flows \cite{fede15}, as well as plasma flows \cite{balescu05}.
While at the Navier-Stokes
level, there are already established solvers for force-driven flows, the realm of 
microfluidics, where the Knudsen number Kn (defined as the ration between the mean free path 
of the constituents of the flow and the typical size of the flow domain) is non-negligible,
is still not sufficiently well explored. 
Such systems can be accurately described at a mesoscopic level, using the 
one-particle distribution function $f \equiv f(\bm{x}, \bm{p}, t)$ for 
particles with momentum $\bm{p}$ at position $\bm{x}$ and time $t$, which 
evolves according to the Boltzmann equation. 

At small values of Kn, the flow is close to equilibrium, such that the
lattice Boltzmann (LB) models constructed using the full-range Gauss-Hermite quadratures 
yield accurate results \cite{bangalore,shan06,deville}. As Kn is increased, the kinetic 
nature of the gas, as well as of the boundary conditions (which affect only the distribution 
of particles 
travelling from the wall back into the fluid), become important, giving rise to 
microfluidics-specific effects, such as velocity slip and temperature jump at 
the boundary \cite{grad,sone02,karniadakis05,struchtrup05,shen05,sone06}. 
Such effects are due to the development of a discontinuity
in the distribution function, which is best understood for the case of a collisionless gas:
the particles travelling towards the wall are distributed according to some essentially 
arbitrary incident flux, while the distribution of reflected particles (which travel 
away from the wall and back into the fluid domain) is determined by the mechanism of
interaction with the wall. In the case of diffuse reflection \cite{maxwell1879}, the 
particle-wall interaction causes
the incident flux of particle to relax towards the equilibrium distribution of the wall 
before being reemited back into the fluid domain. While other models are also possible,
in general similar discontinuities in the distribution function will be induced 
between the incident and emergent hemispheres of the momentum space, essentially because 
the particles emerging from the boundary carry some information about the wall 
which is unavailable to incident particles prior to their interaction with the wall.

The correct treatment of diffuse reflection boundary conditions and the subsequent
induced discontinuous nature of the distribution function requires 
the evaluation of half-space integrals, defined as integrals restricted to 
half of the momentum space \cite{grad,meng11jcp,adif,meng11pre,meng11pre2,meng13jfm,ambrus12}.

Such half-space integrals can be accurately recovered using LB models based on half-range quadratures
\cite{yang95,li03,li04,lorenzani07,li09,frezzotti09,frezzotti11,gibelli12,guo13pre,ghiroldi14,guo15pre,
gibelli15,sader15,ambrus14ipht,ambrus14pre,ambrus14ijmpc,ambrus16jcp,ambrus16jocs}.
Using one quadrature per hemisphere, wall-induced hemispheric discontinuities are appropriately
handled and we have shown, via a systematic set of simulation examples, that using 
half-range quadratures leads to accuracy gains in the rarefied regimes of the Couette
\cite{ambrus14pre,ambrus16jcp,ambrus17couette} and Poiseuille \cite{ambrus14ipht,ambrus16jocs} flows.
This theoretical framework cannot be extended in a straightforward manner 
to force-driven flows where the force has a non-vanishing component in the direction 
perpendicular to the wall. Instead, the extension must be made with care,
in order to avoid zeroth order errors (i.e. mass non-conservation) 
when the force is not strictly parallel to the boundary (we will address this point 
using a particular example starting from Eq.~\eqref{eq:boltz_hom}). 
Such problems were not present in the simulations of 
Ref.~\onlinecite{ambrus14ipht} because the forces were parallel to the boundaries. 
In order to allow the half-range Gauss quadrature concept to be applied for 
flows in more complex force fields, we therefore 
reconsidered our theoretical framework with a specific attention to the component 
of the external forces which is perpendicular to the boundary.

The implementation of the force term in LB simulations requires a model to represent 
the momentum gradient $\nabla_{\vp} f$ of the distribution function $f$
\cite{shan06}. 
For the case of LB models based on the full-range Hermite polynomials, 
$f$ and $\partial_p f$ are both projected onto 
the space of full-range Hermite polynomials, as described in Ref.~\onlinecite{martys}. 
Such a projection is possible since the full-range Hermite polynomials are 
continuous and differentiable everywhere in the momentum space.
We wish to apply a similar procedure to the case when half-range quadratures are employed,
with the aid of which physical configurations where $f$ is not 
differentiable with respect to $p$ at the interface between the two hemispheres 
(defined by the presence of the wall) can also be addressed.

To gain some insight on the difficulties associated with this discontinuity, 
it is instructive to consider the simple example of a one-dimensional uniform flow 
of a collisionless gas, subjected to a force $F$ along the $x$ axis. The Boltzmann equation 
in this case reads:
\begin{equation}
 \partial_t f + F\partial_p f = 0.\label{eq:boltz_hom}
\end{equation}
Taking the full moment with respect to $p$ of the above equation shows that the density remains 
constant through time:
\begin{equation}
 \partial_t n = 0.\label{eq:dtn_0}
\end{equation}
We now wish to implement Eq.~\eqref{eq:boltz_hom} using half-range quadratures. Taking the half-range 
moments of Eq.~\eqref{eq:boltz_hom} over the positive and negative semiaxes yields:
\begin{equation}
 \partial_t n_+ + F[f(\infty) - f(0_+)] = 0, \qquad 
 \partial_t n_- + F[f(0_-) - f(-\infty)] = 0,
 \label{eq:dtn_hemi}
\end{equation}
where $f(0_+) \equiv {\displaystyle \lim_{\substack{p \rightarrow 0\\p > 0}} f(p)}$ 
and $f(0_-) \equiv {\displaystyle \lim_{\substack{p \rightarrow 0\\p < 0}} f(p)}$, 
while $n_\pm$ are defined as:
\begin{equation}
 n_+ = \int_0^\infty dp\, f, \qquad 
 n_- = \int_{-\infty}^0 dp\, f.
\end{equation}
Assuming that $f(p)$ vanishes at $p \rightarrow \pm \infty$, 
the sum of the expressions in Eq.~\eqref{eq:dtn_hemi} yields:
\begin{equation}
 \partial_t n = f(0_+) - f(0_-),
\end{equation}
which reduces to Eq.~\eqref{eq:dtn_0} only when $f(p)$ is continuous at $p = 0$.
It is important to note that $f(0_+) -f(0_-)$ is fully independent of the quadrature 
refinement used to evaluate Eq.~\eqref{eq:dtn_hemi}. The difficulty identified here 
is not of numerical origin: it is a zeroth order bias that will never vanish by refining the 
numerical method (e.g.~by increasing the spatial or temporal resolution, or by increasing the 
quadrature order).
Thus, a strict application of the standard methodology presented in Ref.~\onlinecite{martys} to the 
case of half-range quadratures, when $f$ is allowed to develop a discontinuity at $p = 0$,
leads to spurious mass non-conservation as soon as the distribution function becomes 
discontinuous at $p = 0$. 

In this paper, we present a procedure for the construction of the momentum gradient $\partial_p f$ 
of the distribution function $f$ for force-driven flows when $f$ is projected on the space of 
half-range polynomials. Fundamental to this construction is the theory of distributions, which 
allows $f$ to be written using Heaviside step functions, the derivative of which are modelled as 
delta Dirac functions. The full details of this construction will be presented in section~\ref{sec:hh_force}.
Based on this analysis, we give the recipe for the numerical implementation of $\partial_p f$ in 
LB models based on half-range
Gauss quadratures, such as those introduced in 
Refs.~\onlinecite{ambrus14pre,ambrus16jcp}. 

In this paper, we restrict our analysis with no loss of generality to one-dimensional flows. 
The extension to $2$- or $3$-dimensional
flows is straightforward through a direct-product procedure, as described in Refs.~\cite{ambrus16jcp,ambrus16jocs}.
The details regarding this extension are summarised in Appendix~\ref{app:multi}. 

The outline of this paper is as follows. In Sec.~\ref{sec:force_full}, the expansion of 
$\partial_p f$ with respect to the full-range Hermite polynomials is discussed. 
The expansion of $\partial_p f$ with respect to general half-range polynomials is presented in 
Sec.~\ref{sec:hh_force}. This is the main result of this paper. 
The construction of lattice Boltzmann models for force-driven flows using either full-range or 
the half-range Gauss quadratures is presented in Sec.~\ref{sec:LB}.
The numerical results are presented in Sec.~\ref{sec:num} and Sec.~\ref{sec:conc} concludes this paper.

Throughout this paper, we employ only non-dimensional quantities, following the 
convention of Refs.~\onlinecite{meng11jcp,meng11pre,meng11pre2,meng13jfm,ambrus17couette}.

\section{Expansion of $\partial_p f$ with respect to the full-range Hermite polynomials}
\label{sec:force_full}

The expansion of the momentum derivative of the distribution function with respect to the 
full-range Hermite polynomials has already been discussed in, e.g., Refs.~\onlinecite{shan06,ambrus16jocs}.
In this section, we briefly review the main results in order to prepare the terrain for the 
equivalent construction when half-range polynomials are considered. The starting point for 
our analysis is the one-dimensional Boltzmann-BGK equation in the presence of an external force $F$:
\begin{equation}
 \partial_t f + \frac{p}{m} \partial_x f + F (\partial_p f) = -\frac{1}{\tau} (f - \feq),
 \label{eq:boltz}
\end{equation}
where $f \equiv f(p, x, t)$ is the distribution function for particles of momentum $p$, 
$\feq$ is the Maxwell-Boltzmann equilibrium distribution function:
\begin{equation}
 \feq = n g, \qquad 
 g \equiv g(u, T; p) = \frac{1}{\sqrt{2\pi m T}} \exp\left[-\frac{(p - mu)^2}{2mT}\right],
 \label{eq:feq}
\end{equation}
while $\tau$ is the relaxation time, which we implement as \cite{sone06}:
\begin{equation}
 \tau = \frac{\rm Kn}{n}.\label{eq:tau}
\end{equation}
In this paper, we restrict our analysis without loss of generality to one-dimensional flows.
The procedure presented herein is easily extendible to two- or three-dimensional flows,
following the methodology presented in Appendix~\ref{app:multi}.

We first start with the expansion of $f$ with respect to the set 
$\{H_\ell(\wp), \ell = 0, 1, \dots\}$ of full-range Hermite polynomials:
\begin{equation}
 f = \frac{\omega(\wp)}{p_0} \sum_{\ell = 0}^\infty \frac{1}{\ell!} \mathcal{F}_\ell H_\ell(\wp),
 \label{eq:H_f}
\end{equation}
where $\wp = p/p_0$ is the ratio of the momentum $p$ and some arbitrary momentum scale $p_0$. 
We employ the following weight function:
\begin{equation}
 \omega(\wp) = \frac{1}{\sqrt{2\pi}} e^{-\wp^2 / 2},\label{eq:H_omega}
\end{equation}
while the normalisation condition for the Hermite polynomials reads:
\begin{equation}
 \braket{H_\ell, H_{\ell'}} \equiv \int_{-\infty}^\infty dx\, H_\ell(x) H_{\ell'}(x) = 
 \ell!\,\delta_{\ell,\ell'}.\label{eq:H_ortho}
\end{equation}
The orthogonality relation \eqref{eq:H_ortho} allows the expansion coefficients 
$\mathcal{F}_\ell$ to be obtained as:
\begin{equation}
 \mathcal{F}_\ell = \int_{-\infty}^\infty dp\, f\, H_\ell(\wp).\label{eq:H_F}
\end{equation}
The expansion \eqref{eq:H_f} is exact since the Hermite polynomials satisfy the following 
completeness relation:
\begin{equation}
 \sum_{\ell = 0}^\infty \frac{1}{\ell!} H_\ell(x) H_\ell(x') = \frac{\delta(x - x')}{\sqrt{\omega(x) \omega(x')}}.
\end{equation}

The momentum derivative $\partial_p f$ can be obtained by differentiating Eq.~\eqref{eq:H_f} 
with respect to $p$:
\begin{equation}
 \partial_p f = -\frac{\omega(\wp)}{p_0^2} \sum_{\ell = 0}^\infty \frac{1}{\ell!} 
 \mathcal{F}_\ell H_{\ell + 1}(\wp),
 \label{eq:H_df_F}
\end{equation}
where the following relation was used \cite{ambrus16jocs,nist}:
\begin{equation}
 \partial_x [\omega(x) H_{\ell}(x)] = - \omega(x) H_{\ell + 1}(x).
\end{equation}
Using the result \eqref{eq:H_F}, Eq.~\eqref{eq:H_df_F} can be written as:
\begin{equation}
 [\partial_p f](p) = \int_{-\infty}^\infty dp'\, \mathcal{K}^H(p, p') f(p'),
 \label{eq:H_K_def}
\end{equation}
where the kernel function $\mathcal{K}^H(p, p')$ is given by:
\begin{equation}
 \mathcal{K}^H(p,p') = -\frac{\omega(\wp)}{p_0^2} \sum_{\ell = 0}^\infty \frac{1}{\ell!}
 H_{\ell + 1}(\wp) H_\ell(\wp').\label{eq:H_K}
\end{equation}

\section{Force term in LB models based on half-range quadratures} \label{sec:hh_force}

In this section, we derive an expression for the momentum gradient of the distribution function $f$ 
in the case when $f$ is expanded with respect to a family of half-range
orthogonal polynomials. 
For convenience, we restrict our derivation to the
one-dimensional case, as done in the preceding section, and refer the interested reader to Appendix~\ref{app:multi}
for the extension to multiple dimensions. Before beginning,
we would like to mention that the results derived in this section are
very general and hold for any family of orthogonal polynomials
defined on the Cartesian semiaxes and used in half-range LB models
(e.g., the Laguerre polynomials\cite{ambrus14pre} or the
half-range Hermite polynomials\cite{ambrus16jcp,ambrus16jocs}).

\subsection{Piece-wise decomposition of $f$}\label{sec:hh_force:piece}

As suggested by Gross et al.~\cite{gross}, in wall-bounded flows, the distribution function $f(p)$ 
can be split with respect to the sign of $p$:
\begin{equation}
 f(p) = \theta(p) f^+(p) + \theta(-p) f^-(p),\label{eq:ftheta}
\end{equation}
where $\theta(p)$ is the Heaviside step function, defined as:
\begin{equation}
 \theta(p) = \begin{cases}
  1 & p > 0, \\
  0 & \text{otherwise},
 \end{cases} \qquad
 \theta(p) + \theta(-p) = 1.
\end{equation}
Such a piece-wise split is useful because it allows $f$ to become discontinuous 
at $p = 0$, as is the case in, e.g., the ballistic regime of the Couette flow 
\cite{ambrus14pre,ambrus16jcp,graur09}.
Since the functions $f^+(p)$ and $f^-(p)$ are only defined for $p > 0$ and $p < 0$, 
respectively, the values $f^-(p = 0)$ and $f^+(p = 0)$ can be defined using the following limits:
\begin{equation}
 f^-(0) = \lim_{\substack{p \rightarrow 0\\ p < 0}} f^-(p), \qquad
 f^+(0) = \lim_{\substack{p \rightarrow 0\\ p > 0}} f^+(p).
 \label{eq:fpm0}
\end{equation}
In wall-bounded flows, $f^-(0)$ and $f^+(0)$ are in general not equal.

The two halves $f^{+}$ and $f^-$ of the distribution function $f$ can be expanded 
on $[0, \infty)$ and $(-\infty, 0]$, respectively, with respect to the set 
$\{\phi_\ell(\abs{\wp}), \ell = 0, 1, \dots\}$ of half-range orthogonal polynomials 
\cite{ambrus16jcp,ambrus16jocs}
following the procedure used for the full-range case \eqref{eq:H_f}:
\begin{equation}
 f^\sigma(p) = \frac{\omega(|\wp|)}{p_0} \sum_{\ell = 0}^\infty \frac{1}{\gamma_\ell} 
 \mathcal{F}_\ell^\sigma \phi_\ell(|\wp|),\label{eq:hh_f}
\end{equation} 
where $\sigma = \pm$ refers to the sign of the argument $p$ of $f$, as shown in Eq.~\eqref{eq:ftheta}, while $\gamma_\ell$ denotes the squared norm of
$\phi_\ell\, \equiv \phi_\ell(z), \, z \geq 0$:
\begin{equation}
 \braket{\phi_\ell, \phi_{\ell'}} = \int_0^\infty dz\, \omega(z) \phi_\ell(z) \phi_{\ell'}(z) 
 = \gamma_\ell\, \delta_{\ell,\ell'}.
 \label{eq:phi_ortho}
\end{equation}
The modulus in Eq.~\eqref{eq:hh_f} ensures that the arguments of $\omega$ and 
$\phi_\ell$ are always positive. 
The coefficients $\mathcal{F}_\ell^\sigma \equiv \mathcal{F}_\ell^\sigma(x, t)$ are independent of $p$
and their exact expression can be obtained using the orthogonality relation \eqref{eq:phi_ortho}:
\begin{align}
 \mathcal{F}_\ell^+ =& \int_0^\infty dp\, f(p)\, \phi_\ell(\wp),\nonumber\\
 \mathcal{F}_\ell^- =& \int_{-\infty}^0 dp\, f(p)\, \phi_\ell(-\wp)\nonumber\\
 =& \int_0^\infty dp\, f(-p)\, \phi_\ell(\wp).\label{eq:hh_F}
\end{align}
The equality in Eq.~\eqref{eq:hh_f} is exact if the polynomials $\phi_\ell(z)$ obey 
the completeness relation:
\begin{equation}
 \sum_{\ell = 0}^\infty \frac{1}{\gamma_\ell} \phi_\ell(z) \phi_\ell(z') = 
 \frac{\delta(z - z')}{\sqrt{\omega(z) \omega(z')}},
 \label{eq:phi_compl}
\end{equation}

\subsection{Derivative of $f$ with respect to the momentum}\label{sec:hh_force:df}

Applying the derivative operator with respect to $p$ on Eq.~\eqref{eq:ftheta} yields:
\begin{equation}
 \frac{\partial f}{\partial p} = \theta(p) \frac{\partial f^+}{\partial p} + 
 \theta(-p) \frac{\partial f^-}{\partial p} +
 \delta(p) [f^+(p) - f^-(p)],
 \label{eq:hh_dftheta}
\end{equation}
where the Dirac delta function $\delta(p)$ is linked to the derivative of the step functions 
$\theta(\pm p)$ through \cite{nist}:
\begin{equation}
 \partial_p \theta(p) = \delta(p), \qquad \partial_p \theta(-p) = -\delta(p).
\end{equation}
It should be noted that the $\delta$ function is essential in order to preserve the 
standard requirement that the zeroth order moment of $\partial_p f$ vanishes:
\begin{align}
 \int_{-\infty}^\infty dp\, \frac{\partial f}{\partial p} =& \int_{0}^\infty dp\, \frac{\partial f^+}{\partial p} +
 \int_{-\infty}^0 dp\, \frac{\partial f^-}{\partial p} + [f^+(0) - f^-(0)]\nonumber\\
 =& 0.
\end{align}
We now wish to project Eq.~\eqref{eq:hh_dftheta} on the semiaxes $[0,\infty)$ and $(-\infty, 0]$. 
Since $\theta(p) + \theta(-p) = 1$, the delta term can be included in both semiaxes:
\begin{equation}
 \frac{\partial f}{\partial p} = 
 \theta(p) \left(\frac{\partial f}{\partial p}\right)^+ + 
 \theta(-p) \left(\frac{\partial f}{\partial p}\right)^-, \qquad 
 \left(\frac{\partial f}{\partial p}\right)^\sigma = 
 \frac{\partial f^\sigma}{\partial p} + [\delta(p)]^\sigma [f^+(0) - f^-(0)],
 \label{eq:dftheta_split}
\end{equation}
where the factor multiplying the $\delta$ function is evaluated at $p = 0$ since 
at $p \neq 0$, the $\delta$ term vanishes. The distribution functions $f^\pm(0)$ 
are considered in the sense introduced in Eq.~\eqref{eq:fpm0}. 

We next discuss the expansion of the $\delta$ term (Subsec.~\ref{sec:hh_force:delta}) and 
of $\partial_p f^\pm$ (Subsec.~\ref{sec:hh_force:fpm}). The final result is 
summarised in Subsec.~\ref{sec:hh_force:result}.

\subsection{Expansion of the delta term} \label{sec:hh_force:delta}

To find the expansion of $\delta(p)$ with respect to 
half-range polynomials, it is convenient to work with the following alternative definition of $\delta$ 
\cite{nist}:
\begin{equation}
 \delta(p) = \lim_{\substack{\varepsilon \rightarrow 0\\ \varepsilon > 0}} \delta_\varepsilon(p), \qquad
 \delta_\varepsilon(p) = \frac{1}{\varepsilon \sqrt{\pi}} \exp\left(-\frac{p^2}{\varepsilon^2}\right).
\end{equation}
Clearly, $\delta_\varepsilon(p)$ is a well-defined, even function everywhere, including at $p = 0$, 
as long as $\varepsilon> 0$. Its expansion with respect to half-range polynomials can be written as:
\begin{equation}
 \delta_\varepsilon(p) = \theta(p) \frac{\omega(\wp)}{p_0} \sum_{\ell = 0}^\infty 
 \frac{1}{\gamma_\ell} \mathcal{F}_{\varepsilon, \ell}^{\delta, +} \phi_\ell(\wp) +
 \theta(-p) \frac{\omega(-\wp)}{p_0} \sum_{\ell = 0}^\infty 
 \frac{1}{\gamma_\ell} \mathcal{F}_{\varepsilon, \ell}^{\delta, -} \phi_\ell(-\wp).
\end{equation}
Since $\delta_\varepsilon(-p) = \delta_\varepsilon(p)$, the expansion coefficients satisfy
$\mathcal{F}_{\varepsilon, \ell}^{\delta, +} = \mathcal{F}_{\varepsilon, \ell}^{\delta, -} =
\mathcal{F}_{\varepsilon, \ell}^\delta$. Their value can be calculated as in Eq.~\eqref{eq:hh_F}:
\begin{align}
 \mathcal{F}^{\delta}_{\varepsilon,\ell} =& \int_0^\infty dp\, \delta_{\varepsilon}(\pm p) \phi_\ell(\wp)\nonumber\\
 =& \int_0^\infty \frac{dz}{\sqrt{\pi}} e^{-z^2} \phi_\ell\left(\frac{z\varepsilon}{p_0}\right).
\end{align}
Taking the limit $\varepsilon \rightarrow 0$ in the above equation, the argument of 
$\phi_\ell$ goes to $0$, reducing $\mathcal{F}^{\delta}_{\varepsilon,\ell}$ to:
\begin{equation}
 \lim_{\varepsilon \rightarrow 0} \mathcal{F}^{\delta}_{\varepsilon,\ell} = \frac{1}{2} \phi_{\ell, 0},
\end{equation}
where the notation $\phi_{\ell,s}$ is introduced in Eq.~\eqref{eq:phi_expl}.

Thus, the expansion of $\delta(p)$ with respect to half-range polynomials is:
\begin{equation}
 [\delta(p)]^\sigma = \frac{\omega(\abs{\wp})}{2p_0} \Phi^{\infty}_0(\abs{\wp}),
 \label{eq:deltaf}
\end{equation}
where the notation $\Phi^\infty_s(z) = \lim_{n \rightarrow \infty} \Phi^n_s(z)$ is defined according to \cite{ambrus16jcp}:
\begin{equation}
 \Phi^n_s(z) = \sum_{\ell = s}^n \frac{1}{\gamma_\ell} \phi_{\ell, s} \phi_\ell(z).
 \label{eq:Phi_def}
\end{equation}
The coefficient of $\delta(p)$ in Eq.~\eqref{eq:dftheta_split} can also be projected with 
respect to the half-range polynomials by setting $p = 0$ in Eq.~\eqref{eq:hh_f}:
\begin{equation}
 f^+(0) - f^-(0) = 
 \frac{\omega(0)}{p_0} \sum_{\ell = 0}^\infty \frac{1}{\gamma_\ell} 
 (\mathcal{F}_\ell^+ - \mathcal{F}_\ell^-) \phi_{\ell,0},\label{eq:fpmfm0}
\end{equation}
such that the projection of the full term $\delta(p) [f^+(0) - f^-(0)]$ 
appearing in Eq.~\eqref{eq:dftheta_split} with respect to 
half-range polynomials becomes:
\begin{equation}
 [\delta(p)]^\sigma [f^+(p) - f^-(p)] = 
 \frac{\omega(\abs{\wp})}{2p_0} \Phi^{\infty}_0(\abs{\wp}) \times 
 \left[\frac{\omega(0)}{p_0} \sum_{\ell = 0}^\infty (\mathcal{F}^+_\ell - 
 \mathcal{F}^-_\ell) \phi_{\ell,0}\right].
 \label{eq:deltaf_full}
\end{equation}

\subsection{The derivative of $f^\sigma$}\label{sec:hh_force:fpm}

Taking the derivative with respect to $p$ of Eq.~\eqref{eq:hh_f} gives:
\begin{equation}
 \frac{\partial f^\sigma}{\partial p} = \frac{\sigma}{p_0^2} \sum_{\ell = 0}^\infty 
 \frac{1}{\gamma_\ell} \mathcal{F}_\ell^\sigma 
 \left(\frac{\partial [\omega(z) \phi_\ell(z)]}{\partial z}\right)_{z=\abs{\wp}}.
\end{equation}
Equation~\eqref{eq:ophip} can be used to eliminate the above derivative with respect to $z$, giving:
\begin{equation}
 \frac{\partial f^\sigma}{\partial p} = \frac{\sigma \omega(\abs{\wp})}{p_0^2} \sum_{\ell = 0}^\infty \mathcal{F}_\ell^\sigma
 \sum_{s = \ell + 1}^\infty \frac{1}{\gamma_s} \varphi_{s, \ell} \phi_s(\abs{\wp})
 - \frac{\sigma \omega(\abs{\wp})}{p_0} \Phi_0^\infty(\abs{\wp})
 \left[\frac{\omega(0)}{p_0} \sum_{\ell = 0}^\infty \frac{1}{\gamma_\ell} 
 \mathcal{F}_\ell^\sigma \phi_{\ell,0}\right],
 \label{eq:dfpm}
\end{equation}
where the term in the square braket is equal to $f^\sigma(0)$, as can be seen by setting 
$p = 0$ in Eq.~\eqref{eq:hh_f}.

\subsection{Final result}\label{sec:hh_force:result}
Adding Eqs.~\eqref{eq:deltaf} and \eqref{eq:dfpm} gives:
\begin{multline}
 \frac{\partial f}{\partial p} = 
 \frac{\omega(\abs{\wp})}{p_0^2} [\theta(p) - \theta(-p)] \Bigg[
 \sum_{\ell = 0}^\infty \mathcal{F}_\ell^\sigma
 \sum_{s = \ell + 1}^\infty \frac{1}{\gamma_s} \varphi_{s, \ell} \phi_s(\abs{\wp})\\
 - \frac{\omega(0)}{2} \Phi_0^\infty(\abs{\wp}) 
 \sum_{\ell = 0}^\infty \frac{1}{\gamma_\ell}
 (\mathcal{F}^+_\ell + \mathcal{F}^-_\ell)\phi_{\ell,0}\Bigg].
 \label{eq:dfpm2} 
\end{multline}
The above result can be put into the form \eqref{eq:H_K_def}:
\begin{equation}
 \frac{\partial f}{\partial p} = \int_{-\infty}^\infty dp'\, \mathcal{K}^{\rm half}(p, p') f(p'),
 \label{eq:hh_K_def}
\end{equation}
where the kernel function $\mathcal{K}^{\rm half}(p, p')$ is given by:
\begin{multline}
 \mathcal{K}^{\rm half}(p,p') = \frac{\omega(\abs{\wp})}{p_0^2} 
 [\theta(p) - \theta(-p)] \\
 \times \left[
 \theta(pp') \sum_{\ell = 0}^\infty \phi_\ell(\abs{\wp'}) 
 \sum_{s = \ell + 1}^\infty \frac{1}{\gamma_s} \phi_s(\abs{\wp}) \varphi_{s,\ell} - 
 \frac{\omega(0)}{2} \Phi_0^\infty(\abs{\wp}) \Phi_0^\infty(\abs{\wp'})\right].
 \label{eq:hh_K}
\end{multline}
Equation~\eqref{eq:hh_K} represents the main result of this paper. 
It can be seen that the mixing between the regions with $p > 0$ and $p < 0$
occurs only due to the second term inside the square braket.
The implementation of Eq.~\eqref{eq:hh_K} will be further
discussed in Sec.~\ref{sec:LB:hh}.

\section{Lattice Boltzmann models for force-driven flows}\label{sec:LB}

The basic ingredients for the construction of a lattice Boltzmann model to solve Eq.~\eqref{eq:boltz} are: 
1) discretising the momentum space; 2) replacing the equilibrium distribution function $\feq$ 
in the collision term of Eq.~\eqref{eq:boltz}
by a truncated polynomial; 3) replacing the momentum 
derivative of the distribution function in Eq.~\eqref{eq:boltz}
using a suitable expression; 
4) choosing a numerical method for the time evolution and spatial advection;
and 5) implementation of the boundary conditions.
A common feature of all lattice Boltzmann models is that the conservation 
equations for the particle number density $n$, macroscopic momentum densiy 
$n m \vu$ and temperature $T$ (for thermal models) are exactly recovered.

Regardless of the chosen discretisation of the momentum space, 
the Boltzmann equation \eqref{eq:boltz} becomes:
\begin{equation}
 \partial_t f_k + \frac{p_k}{m} \partial_x f_k + 
 F (\partial_p f)_k = -\frac{1}{\tau}(f_k - \feq_k),\label{eq:boltzk}
\end{equation}
where $f_k$ ($k =1, 2, \dots \mathcal{Q}$) represents the set of distribution functions corresponding to 
the discrete momenta $p_k$. The total number $\mathcal{Q}$ of discrete momenta depends
on the particular quadrature. We will discuss in detail the implementation of
steps 1) -- 3) outlined above in LB models based on the full-range Gauss-Hermite quadratures
(Subsec.~\ref{sec:LB:H}) and on general half-range quadratures, specilising the latter for 
the case of the Gauss-Laguerre and half-range Gauss-Hermite quadratures.
(Subsec.~\ref{sec:LB:hh}). In Subsec.~\ref{sec:LB:sch} 
we present the numerical scheme employed in this paper for the time stepping and 
advection (step 4), as well as our implementation of the diffuse reflection boundary 
conditions (step 5). Finally, in Subsec.~\ref{sec:LB:notation} we introduce the 
notation which can be used to distinguish between the 1D models
employed in our simulations.

\subsection{Full-range Gauss-Hermite quadrature} \label{sec:LB:H}

A full-range lattice Boltzmann model is constructed such that the full-range moments 
$M_s$ of finite orders $s$ of the distribution function $f$ are exactly recovered:
\begin{equation}
 M_s \equiv \int_{-\infty}^\infty dp\, f\, p^s = 
 \sum_{k = 1}^Q f_k p^s_k.\label{eq:moms}
\end{equation}

\subsubsection{Discretisation of the momentum space}\label{sec:LB:H:discrete}

Considering the expansion \eqref{eq:H_f} with respect to the Gauss-Hermite polynomials, 
Eq.~\eqref{eq:moms} becomes:
\begin{equation}
 M_s = \sum_{\ell = 0}^\infty \frac{1}{\ell!} \mathcal{F}_\ell \int_{-\infty}^\infty 
 d\wp\,\omega(\wp)\, H_\ell(\wp) p^s,\label{eq:moms_aux}
\end{equation}
where the weight function $\omega(z)$ is defined in Eq.~\eqref{eq:H_omega}.
The integral in Eq.~\eqref{eq:moms_aux} can be recovered using the Gauss-Hermite quadrature:
\begin{equation}
 \int_{-\infty}^\infty dz\, \omega(z) P_s(z) \simeq \sum_{k = 1}^Q w_k P_s(z_k),
 \label{eq:H_quad}
\end{equation}
where $P_s(z)$ is a polynomial of order $s$ in $z$. 
The equality in Eq.~\eqref{eq:H_quad} is exact if the number of quadrature points satisfies $2Q > s$.
The quadrature points $z_k$ are the roots of the Hermite polynomial $H_Q(z)$ of order $Q$, 
while their associated weights $w_k$ are given by:
\begin{equation}
 w_k = \frac{Q!}{[H_{Q + 1}(\wp_{k})]^2}. \label{eq:H_wk}
\end{equation}
Thus, Eq.~\eqref{eq:moms} is exact for $0 \le s < Q$ if the expansion \eqref{eq:H_f} 
is truncated at $\ell = Q$ and $f_k$ is defined as:
\begin{equation}
 f_k = \frac{w_k p_0}{\omega(p_k)} f^Q(p_k) 
 = w_k \sum_{\ell = 0}^{Q - 1} \frac{1}{\ell!} \mathcal{F}_\ell H_\ell(\wp_k).
 \label{eq:H_fk}
\end{equation}
It is worth mentioning that $\wp_k = 0$, corresponding to stationary particles, is 
allowed in the set of discrete velocities only for odd quadrature orders $Q$. 
The implementation of diffuse reflection for such particles is problematic 
and the results obtained using models employing similar number of velocities are worse 
if the velocity set contains velocities which are parallel to the walls than in the case 
when such velocities are absent\cite{meng11pre,
ambrus12,kim08,watari07,watari09,watari10,shi11,izarra11,ambrus16jcp}.
We will therefore avoid the use of the ${\rm HLB}(N;Q)$ models with odd $Q$ in this work.

\subsubsection{Hyperbolicity of the resulting scheme}\label{sec:LB:H:hyp}

\begin{figure}
\begin{center}
\begin{tabular}{cc}
 \includegraphics[angle=270,width=0.45\columnwidth]{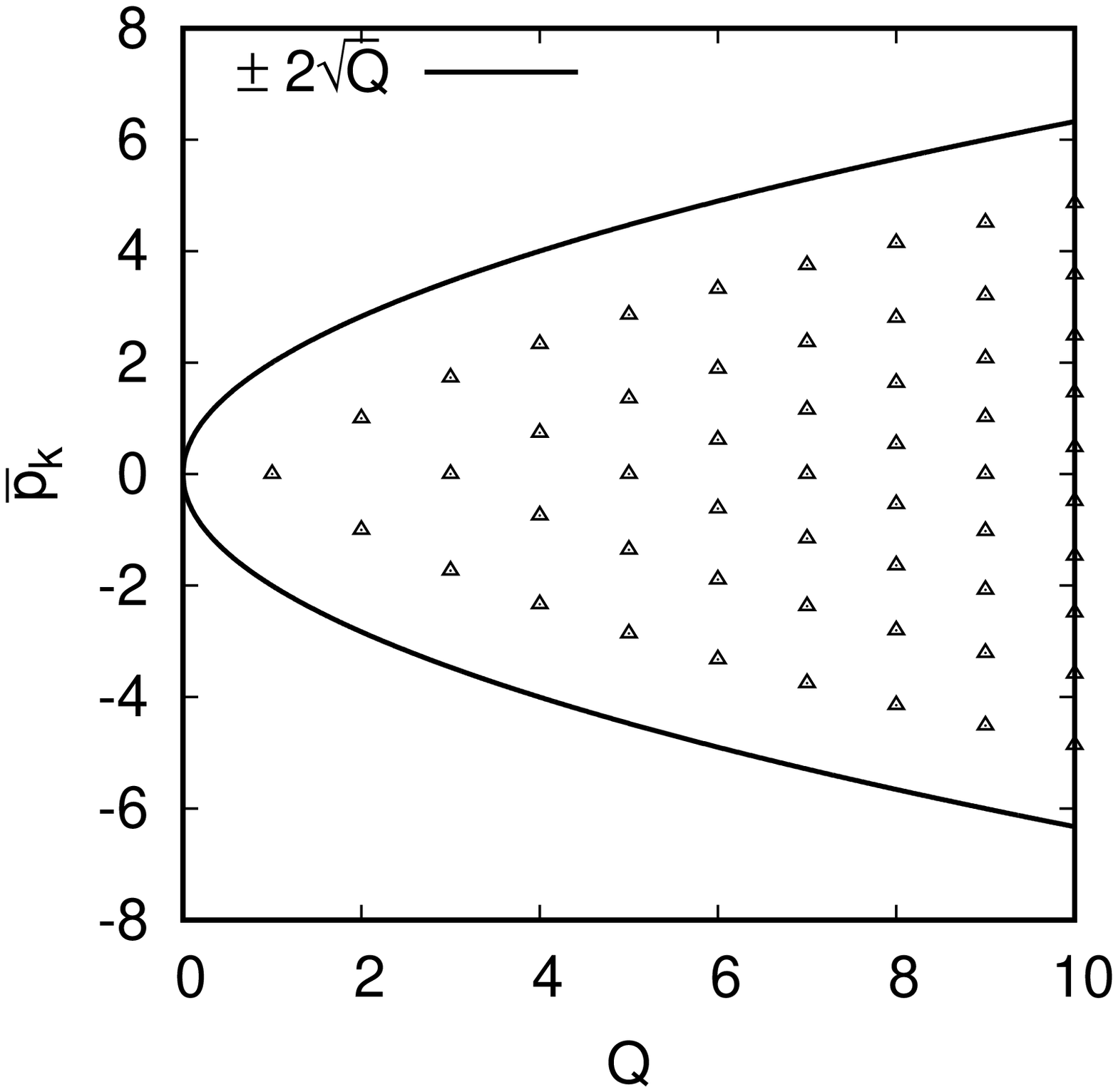} &
 \includegraphics[angle=270,width=0.45\columnwidth]{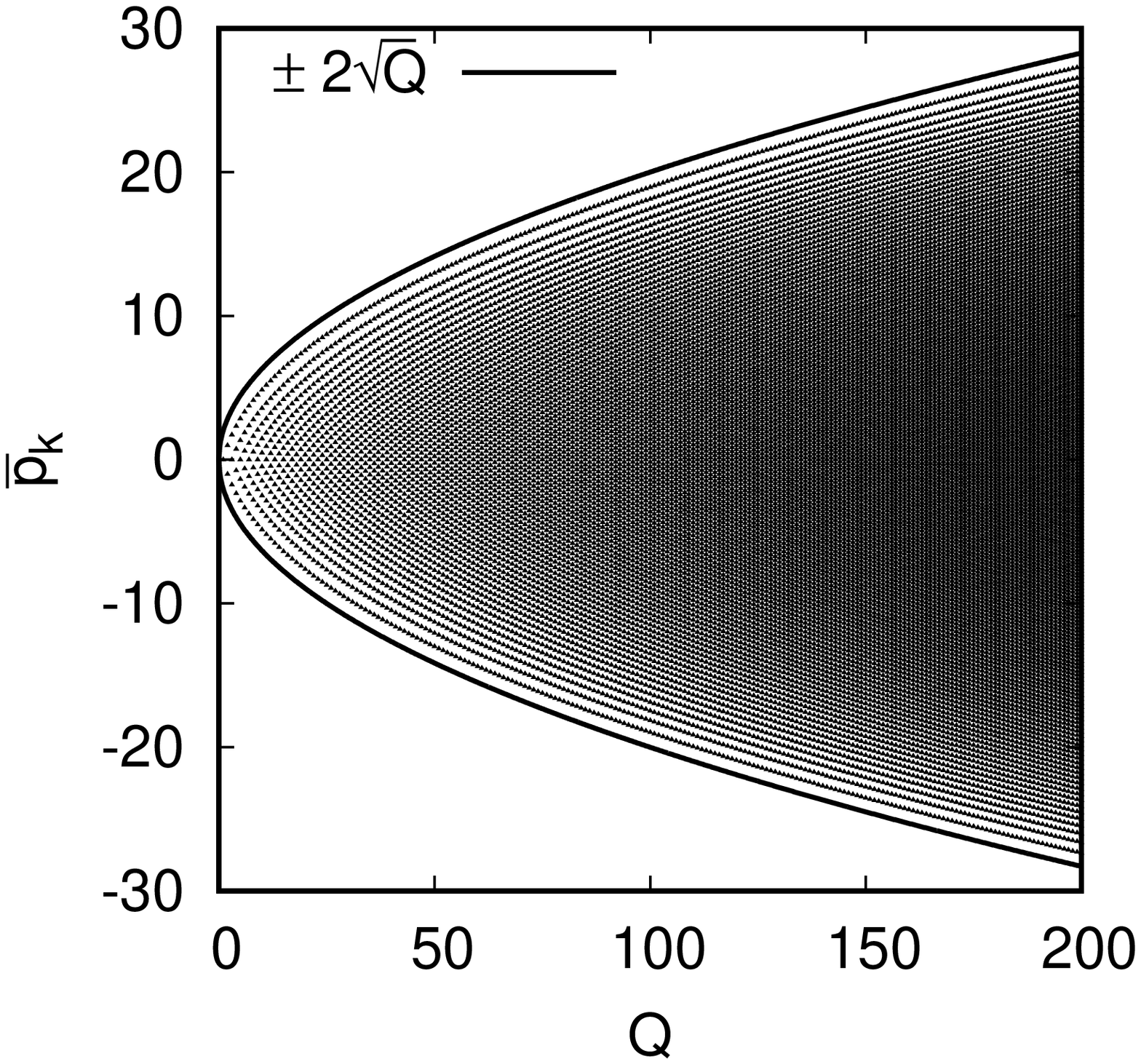} \\
 (a) & (b)
\end{tabular}
\end{center}
\caption{Scatter plot of the quadrature points $\wp_k$ ($1 \le k \le Q$) 
occuring in the Gauss-Hermite quadrature of order $Q$ for 
(a) $1 \le Q \le 10$ and (b) $1 \le Q \le 200$. At each value of 
$Q$ represented on the horizontal axis, 
there are $Q$ points on the corresponding vertical line, representing the 
values of the quadrature points $\wp_k$. All quadrature points are
contained inside the outer envelope $\pm 2\sqrt{Q}$.
}
\label{fig:H_hyp}
\end{figure}

As remarked in Ref.~\onlinecite{fede15}, choosing the discrete momenta to be 
linked to the roots of $H_Q(z)$ through $p_k = p_0 z_k$ automatically 
ensures that the truncation \eqref{eq:H_fk} of $f_k$ with respect to the Hermite 
polynomials is preserved by the Boltzmann equation \eqref{eq:boltzk}, 
provided $f_k$ is initialised in the form in Eq.~\eqref{eq:H_fk}. This can be seen 
by casting the Boltzmann equation \eqref{eq:boltz} in the following form:
\begin{equation}
 \partial_t f + \frac{p}{m} \partial_x f = S,\label{eq:boltz_S}
\end{equation}
where $S = -F \partial_p f - \frac{1}{\tau}(f - \feq)$ is completely local (i.e.,
it does not involve spatial derivatives of $f$). Multiplying Eq.~\eqref{eq:boltz_S} 
by $H_\ell(\wp)$ and integrating with respect to the momentum space yields:
\begin{equation}
 \partial_t \mathcal{F}_\ell + \partial_x A_{\ell,s} \mathcal{F}_s = \mathcal{S}_\ell,
 \label{eq:boltz_Sl}
\end{equation}
where $\mathcal{F}_\ell$ is the coefficient corresponding to $H_\ell$ in the expansion \eqref{eq:H_f}
and $\mathcal{S}_\ell$ is defined in a similar manner:
\begin{equation}
 S = \frac{\omega(\wp)}{p_0} \sum_{\ell = 0}^\infty \frac{1}{\ell!} \mathcal{S}_\ell H_\ell(\wp),
 \qquad
 \mathcal{S}_\ell = \int_{-\infty}^\infty dp\, S\, H_\ell(\wp).
\end{equation}
The spatial derivative acts on a linear combination $A_{\ell,s} \mathcal{F}_s$  of $\mathcal{F}_s$, 
where the non-zero coefficients $A_{\ell, s}$ are:
\begin{equation}
 A_{\ell, \ell + 1} = \frac{p_0}{m}, \qquad 
 A_{\ell, \ell - 1} = \frac{\ell p_0}{m}.\label{eq:H_A_el}
\end{equation}
The above result is obtained by noting that $p H_\ell(\wp) = p_0[H_{\ell + 1}(\wp) + \ell H_{\ell - 1}(\wp)]$.

Employing a quadrature of order $Q$ is equivalent to setting $\mathcal{F}_\ell = 0$ for all $\ell \ge Q$.
This allows $A_{\ell, s}$ to be represented as a $Q \times Q$ matrix of the form:
\begin{equation}
 \mathbb{A} = \frac{p_0}{m}
 \begin{pmatrix}
  0 & 1 & 0 & 0 & 0 & \cdots \\
  1 & 0 & 2 & 0 & 0 & \cdots \\
  0 & 1 & 0 & 3 & 0 & \cdots \\
  0 & 0 & 1 & 0 & 4 & \cdots \\
  0 & 0 & 0 & 1 & 0 & \cdots \\
  \dots & \dots & \dots & \dots & \dots & \dots
 \end{pmatrix}\label{eq:H_A_matrix}
\end{equation}
The system \eqref{eq:boltz_Sl} is said to be hyperbolic if the matrix 
$\mathbb{A}$ has only real eigenvalues $\lambda_s \in \mathbb{R}$ ($s = 1, 2, \dots Q$). 
The system is strictly hyperbolic if the eigenvalues and their corresponding right eigenvectors 
$\mathbb{K}_s$ are also distinct \cite{toro09,rezzolla13}.
At a fixed value of $Q$, the matrix $\mathbb{A}$ \eqref{eq:H_A_matrix} is truncated 
at $Q \times Q$ elements. In the following, we will consider the first few values of $Q$.

At $Q = 1$, $\mathbb{A} = (0)$ and $\lambda = 0$ with $\mathbb{K} = (1)$.

When $Q= 2$, we have:
\begin{equation}
 \mathbb{A} = \frac{p_0}{m}
 \begin{pmatrix}
  0 & 1 \\
  1 & 0
 \end{pmatrix},
 \qquad \lambda_\pm = \pm \frac{p_0}{m}, \qquad 
 \mathbb{K}_\pm = 
 \begin{pmatrix}
  1\\
  \pm 1
 \end{pmatrix}.
\end{equation}

When $Q = 3$, the following values are obtained:
\begin{equation}
 \mathbb{A} = \frac{p_0}{m}
 \begin{pmatrix}
  0 & 1 & 0\\
  1 & 0 & 2\\
  0 & 1 & 0
 \end{pmatrix},
 \qquad \lambda_0 = 0, \quad \lambda_\pm = \pm \frac{p_0 \sqrt{3}}{m}, \qquad 
 \mathbb{K}_0 = 
 \begin{pmatrix}
  1\\
  0 \\ 
  -1/2
 \end{pmatrix}, \quad 
 \mathbb{K}_\pm = 
 \begin{pmatrix}
  1 \\
  \pm\sqrt{3}\\
  1
 \end{pmatrix}. 
\end{equation}

It can be seen that, at fixed $Q$, the eigenvalues $\lambda_s$ ($1 \le \lambda_s \le Q$) 
are linked to the roots $\wp_s$ of the Hermite polynomial $H_Q(\wp)$ of order $Q$ through:
\begin{equation}
 \lambda_s = \frac{p_0 \wp_s}{m} = \frac{p_s}{m}.
\end{equation}
This is also obvious since the projection of the Boltzmann equation \eqref{eq:boltzk} 
after discretisation onto the space of Hermite polynomials is equivalent to performing a linear 
transformation using the matrix:
\begin{equation}
 \mathcal{H}_{\ell,k} = H_\ell(\wp_k).
\end{equation}
Applying the above transformation to the state vector $f_k$ yields the state vector of the system 
of moment equations \eqref{eq:boltz_Sl}:
\begin{equation}
 \sum_{k = 1}^Q \mathcal{H}_{\ell,k} f_k = \mathcal{F}_\ell.
\end{equation}
It is now clear that calculating the eigenvalues of the matrix $\mathcal{A}_{\ell,s}$ \eqref{eq:H_A_el} is 
equivalent to calculating the eigenvalues of the matrix $\mathcal{A}_{k,k'}$, defined by writing 
Eq.~\eqref{eq:boltzk} as:
\begin{equation}
 \partial_t f_k + \partial_x \left(\mathcal{A}_{k,k'} f_{k'}\right) = S_k,
 \label{eq:boltzk_Ak}
\end{equation}
where it can be seen that $\mathcal{A}_{k,k'} = \frac{p_k}{m} \delta_{k,k'}$ is in diagonal form.
Thus, the eigenvalues of $\mathbb{A}$ \eqref{eq:H_A_matrix} are nothing but the discrete velocities 
$p_k / m$ corresponding to the Gauss-Hermite quadrature of order $Q$. These discrete velocities 
are necessarily distinct since they are linked to the roots of the Hermite polynomial of order $Q$. 
Thus, it can be seen that the system of lattice Boltzmann equations \eqref{eq:boltzk} 
is strictly hyperbolic for any choice of the quadrature order $Q$.

We end this section by illustrating the quadrature points corresponding to the Gauss-Hermite 
quadrature of various orders $Q$ through a scatter plot in Fig.~\ref{fig:H_hyp}.
It can be seen that for a fixed quadrature order $1\le Q \le 200$, all quadrature points lie between 
$\pm 2 \sqrt{Q}$. For completeness, we include in the supplementary material the roots and weights 
of the Gauss-Hermite quadrature up to $Q = 200$, as described in Appendix~\ref{app:supp}.

\subsubsection{Expansion of the equilibrium distribution function} \label{sec:LB:H:feq}

Let us consider the expansion of $g$ \eqref{eq:feq} in the form given in Eq.~\eqref{eq:H_f}:
\begin{equation}
 g = \frac{\omega(\wp)}{p_0} \sum_{\ell = 0}^\infty \frac{1}{\ell!} 
 \mathcal{G}_\ell H_\ell(\wp).
\end{equation}
Following the discretisation of the momentum space, $g(p)$ is replaced by 
$g_k$ following the procedure in Eq.~\eqref{eq:H_fk}:
\begin{equation}
 g_k = w_k \sum_{\ell = 0}^{N} \frac{1}{\ell!} 
 \mathcal{G}_\ell H_\ell(\wp_k),\label{eq:H_gk}
\end{equation}
where the truncation order $N \le Q - 1$ is a parameter of the model, representing the 
order up to which the moments of $\feq$ are exactly recovered using the Gauss-Hermite quadrature 
rule. The coefficients $\mathcal{G}$ can be found using the orthogonality relation 
\eqref{eq:H_ortho} \cite{ambrus16jcp,ambrus16jocs}:
\begin{align}
 \mathcal{G}_\ell =& \int_{-\infty}^\infty dp\, g\, H_{\ell}(\wp)\nonumber\\
 =& \sum_{s = 0}^{\lfloor \ell / 2\rfloor} \frac{\ell!}{2^s s! (\ell - 2s)!} 
 \left(\frac{mT}{p_0^2} - 1\right)^2 \left(\frac{mu}{p_0}\right)^{\ell - 2s}.
\end{align}

\subsubsection{Force term}\label{sec:LB:H:force}

We now turn to the implementation of the momentum derivative. 
This can be achieved by finding the equivalent $\mathcal{K}_{k,k'}$
after the discretisation of 
the momentum space of the kernel function $\mathcal{K}^H(p,p')$ 
\eqref{eq:H_K}. The definition of $\mathcal{K}^H_{k,k'}$ is the discrete 
analogue of Eq.~\eqref{eq:H_K_def}:
\begin{equation}
 (\partial_p f)_k = \sum_{k' = 1}^Q \mathcal{K}^H_{k,k'} f_{k'}.
\end{equation}
The sum over $k'$ is obtained by converting the integral with respect to $p'$ 
appearing in Eq.~\eqref{eq:H_K_def} into a quadrature 
sum following the prescription \eqref{eq:H_quad}:
\begin{equation}
 \int_{-\infty}^\infty dp'\, \mathcal{K}^H(p, p') f(p') \simeq \sum_{k' = 1}^Q 
 \mathcal{K}^H(p, p'_k) f_{k'}, \label{eq:H_K_aux}
\end{equation}
where the relation \eqref{eq:H_fk} between $f(p)$ and $f_k$ was used.
Since, by virtue of Eq.~\eqref{eq:H_fk}, $f_{k'}$ is a polynomial of order 
$Q - 1$ in $p'$, Eq.~\eqref{eq:H_K_aux} is exact if $\mathcal{K}^H(p, p')$ is 
truncated at $\ell = Q - 1$. The terms corresponding to larger values of $\ell$ 
do not contribute to the integral \eqref{eq:H_K_aux} when $f$ is a polynomial
of order $Q - 1$, since the Hermite polynomials of orders $\ell > Q - 1$ are 
orthogonal to all polynomials of order $Q - 1$ or less by virtue of the orthogonality 
relation \eqref{eq:H_ortho}.
Applying the relation \eqref{eq:H_fk} between $f(p)$ and $f_k$ for the 
case of $\partial_p f$ and $(\partial_p f)_k$ allows $\mathcal{K}^H_{k,k'}$ 
to be written as:
\begin{equation}
 \mathcal{K}^H_{k,k'} = -\frac{w_k}{p_0} \sum_{\ell = 0}^{Q-2} \frac{1}{\ell!}
 H_{\ell + 1}(\wp_k) H_\ell(\wp_{k'}),\label{eq:H_Kk}
\end{equation}
where the sum over $\ell$ is truncated at $Q - 2$ since $H_{\ell + 1}(p_k)$ vanishes 
when $\ell = Q - 1$.
The resulting matrix $\mathcal{K}^H_{k,k'}$ has 
$Q \times Q$ elements, since $1 \le k, k' \le Q$. These elements can be 
computed at runtime, or they can be read from a data file. For the reader's convenience, 
we supply the matrix elements $\mathcal{K}^H_{k,k'}$ for $Q = 1, 2, \dots 200$ 
in the supplementary material attached to this manuscript
(more details are given in Appendix~\ref{app:supp}).

\subsection{Half-range quadratures} \label{sec:LB:hh}

Half-range quadratures are employed to ensure the recovery of the half-range 
moments $M^\pm_s$ of finite orders $s$ of the distribution function $f$:
\begin{equation}
 M_s^+ = \int_0^\infty dp\, f^+(p) p^s, \qquad
 M_s^- = \int_{-\infty}^0 dp\, f^-(p) p^s.\label{eq:momspm}
\end{equation}

\subsubsection{Discretisation of the momentum space}\label{sec:LB:hh:discrete}

The recovery of the half-range integrals in Eq.~\eqref{eq:momspm} can be achieved 
using half-range Gauss quadratures:
\begin{equation}
 \int_0^\infty dz\, \omega(z) P_s(z) \simeq \sum_{k = 1}^Q w_k P_s(z_k),
 \label{eq:hh_quad}
\end{equation}
where the equality is exact if $2Q > s$. 
The quadrature points $z_k$ ($k = 1, 2, \dots Q$) are the $Q$ (positive) roots 
of the half-range polynomial $\phi_Q(z)$, while the quadrature weights $w_k$ 
are in general given by \cite{hildebrand87,shizgal15,ambrus16jcp}:
\begin{equation}
 w_k = -\frac{a_Q \gamma_Q}{\phi_{Q+1}(\abs{\wp_k}) \phi_Q'(\abs{\wp_k})}, 
 \label{eq:wk_half_gen}
\end{equation}
where $a_Q = \phi_{Q+1,Q+1} / \phi_{Q,Q}$ and $\phi_{\ell,s}$ represents the 
coefficient of $z^s$ in $\phi_\ell(z)$, as described in Eq.~\eqref{eq:phi_expl}.

Considering the expansion \eqref{eq:hh_f} with respect to the half-range polynomials 
$\phi_\ell$, Eq.~\eqref{eq:momspm} becomes:
\begin{equation}
 \begin{pmatrix}
  M_s^+\\ 
  M_s^-
 \end{pmatrix}
 = \sum_{\ell = 0}^\infty \frac{1}{\ell!} 
 \begin{pmatrix}
  \mathcal{F}_\ell^+\\
  \mathcal{F}_\ell^-
 \end{pmatrix}
 \int_{0}^\infty d\wp\,\omega(\wp)\, \phi_\ell(\wp) (\pm p)^s.\label{eq:momspm_aux}
\end{equation}
Truncating the expansion \eqref{eq:hh_f} at $\ell = Q - 1$ ensures that 
a quadrature of order $Q$ can recover the moments \eqref{eq:momspm_aux} for 
$0 \le s \le Q$. Since $Q$ quadrature points are required on each semiaxis of the momentum space, 
the total number of elements in the momentum set is $\mathcal{Q} = 2Q$, defined as:
\begin{equation}
 p_k = p_0 z_k, \qquad p_{k + Q} = -p_k \qquad 
 (1 \le k \le Q).\label{eq:hh_pk}
\end{equation}
Thus, the half-range moments \eqref{eq:momspm} are recovered as:
\begin{equation}
 M_s^+ = \sum_{k = 1}^Q f_k p_k^s, \qquad 
 M_s^- =\sum_{k = Q+1}^{2Q} f_k p_k^s,
\end{equation}
where 
\begin{equation}
 f_k = \frac{w_k p_0}{\omega(\wp_k)} f(p_k), \qquad 
 f_{k + Q} = \frac{w_k p_0}{\omega(\wp_k)} f(-p_k) \qquad 
 (1 \le k \le Q).
\end{equation}
The quadrature weights $w_k$ \eqref{eq:wk_half_gen} are given in the case of the Gauss-Laguerre ($w_k^L$) \cite{ambrus14pre,ambrus14ipht}
and half-range Gauss-Hermite ($w_k^\hh$) \cite{ambrus16jcp,ambrus16jocs} quadratures through:
\begin{equation}
 w_k^L = \frac{\wp_k}{(Q + 1)^2 [L_{Q + 1}(\wp_k)]^2}, \qquad 
 w_k^\hh = \frac{\wp_k a_Q^2}{\hh_{Q+1}^2(\wp_k) \left[\wp_k + \hh_{Q,0}^2/\sqrt{2\pi}\right]},
 \label{eq:wk_half}
\end{equation}
where $a_\ell$ is defined in Eq.~\eqref{eq:a_def}. The numerical values of the roots and quadrature 
weights for the half-range Gauss-Hermite quadratures with $1 \le Q \le 200$ 
can be found in the supplementary material, as described in Appendix~\ref{app:supp}.

\subsubsection{Hyperbolicity of the resulting scheme}\label{sec:LB:hh:hyp}

\begin{figure}
\begin{center}
\begin{tabular}{cc}
\includegraphics[angle=270,width=0.45\columnwidth]{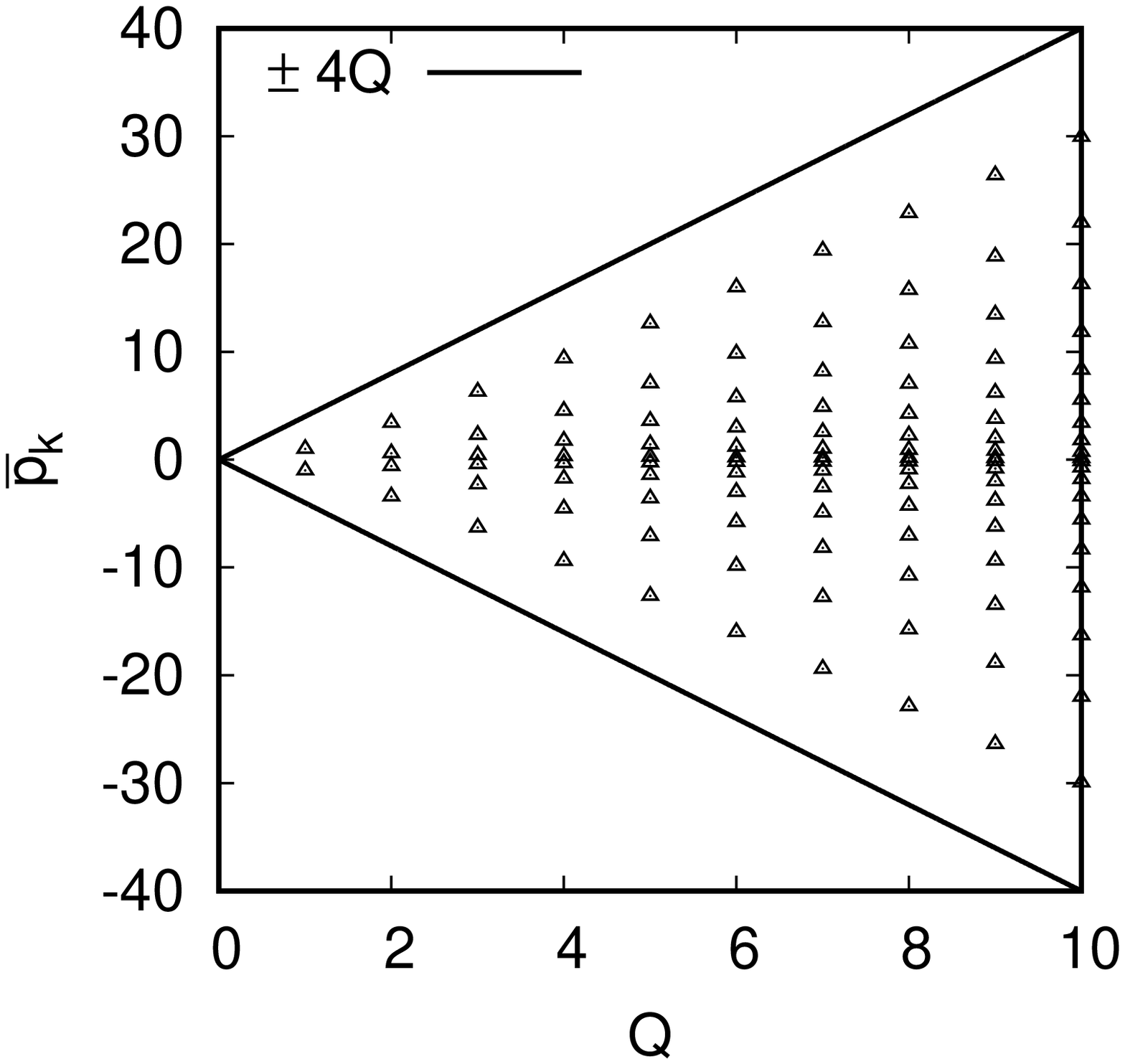} &
\includegraphics[angle=270,width=0.45\columnwidth]{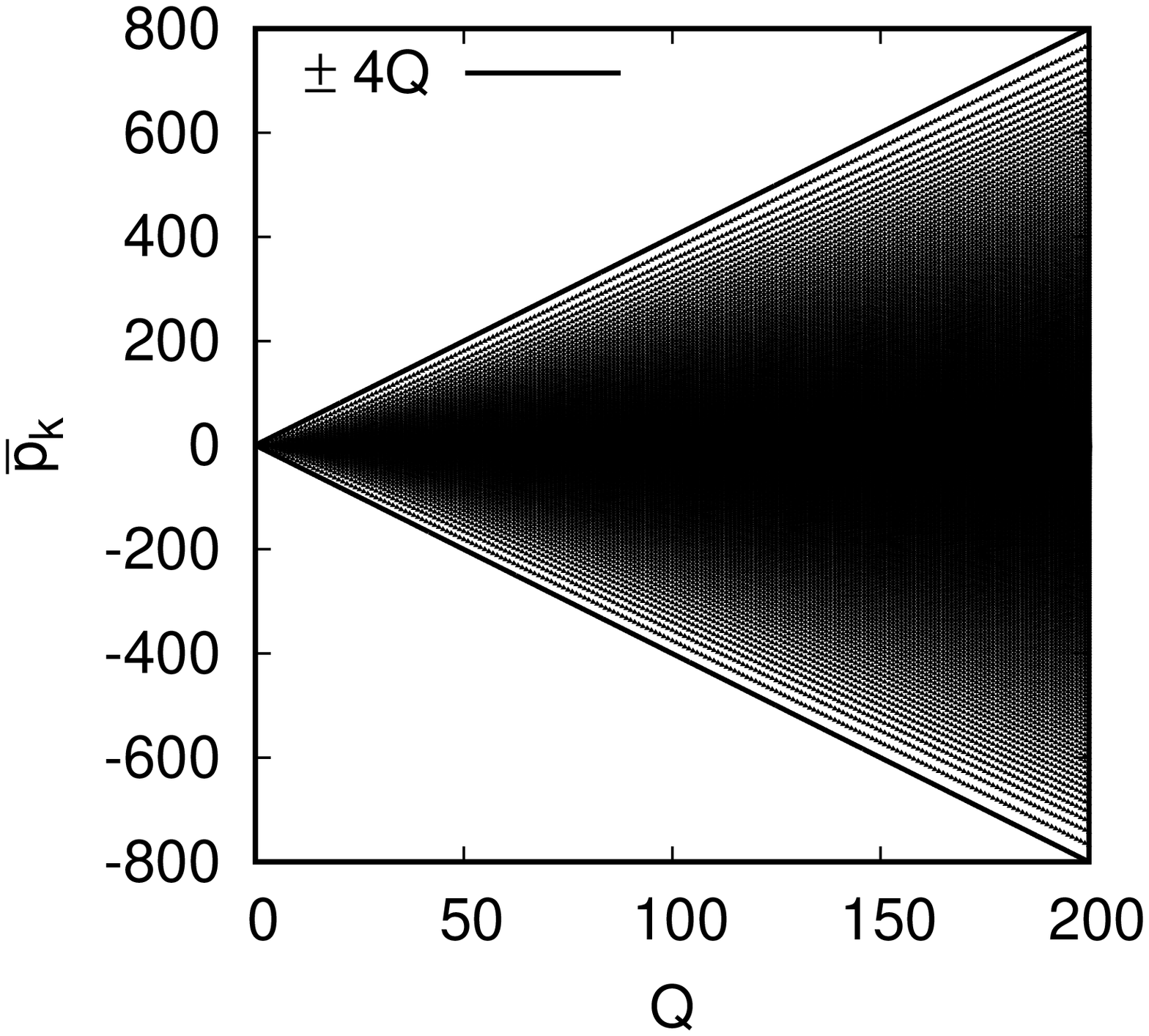} \\
(a) & (b) \\
\includegraphics[angle=270,width=0.45\columnwidth]{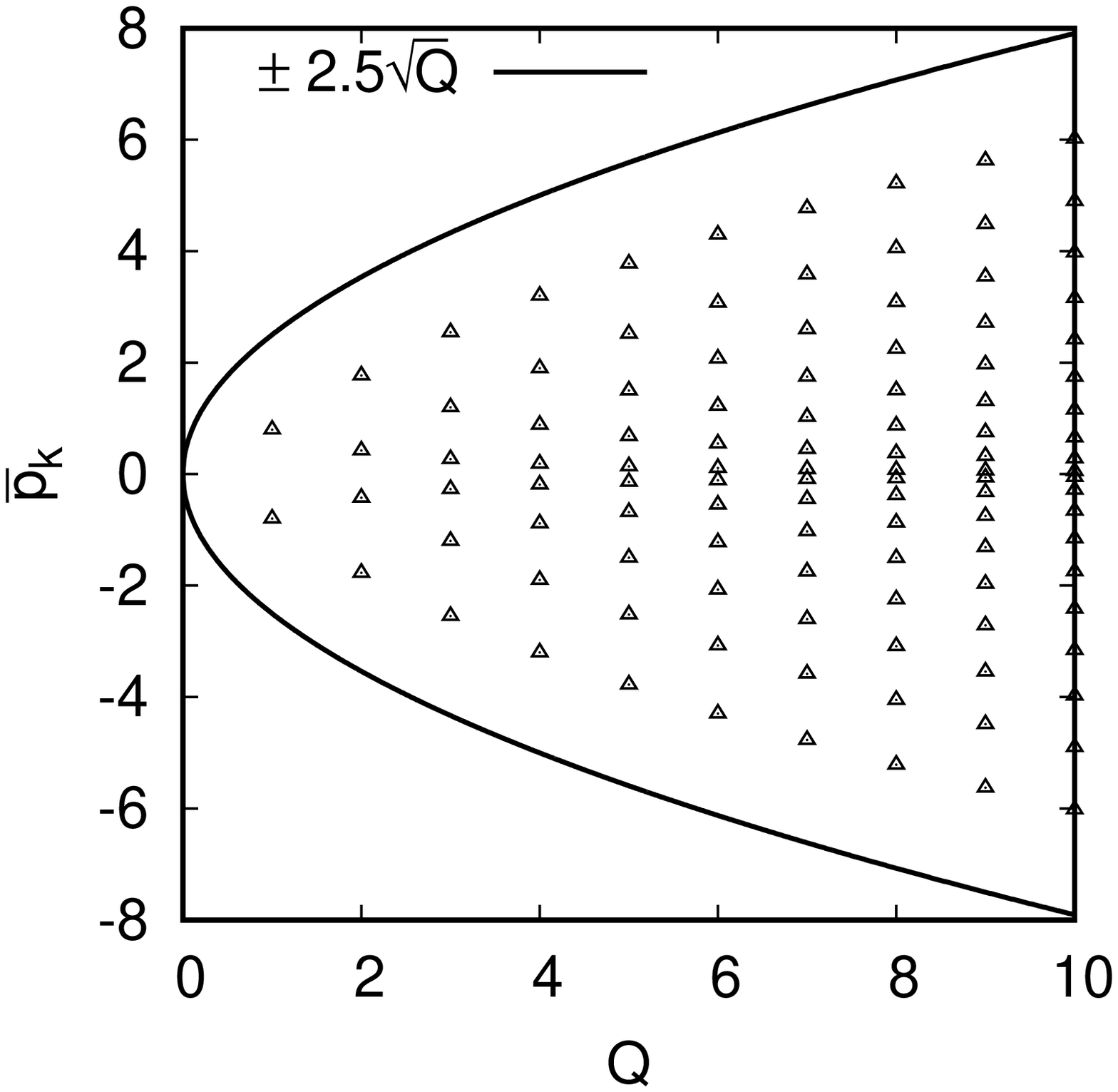} &
\includegraphics[angle=270,width=0.45\columnwidth]{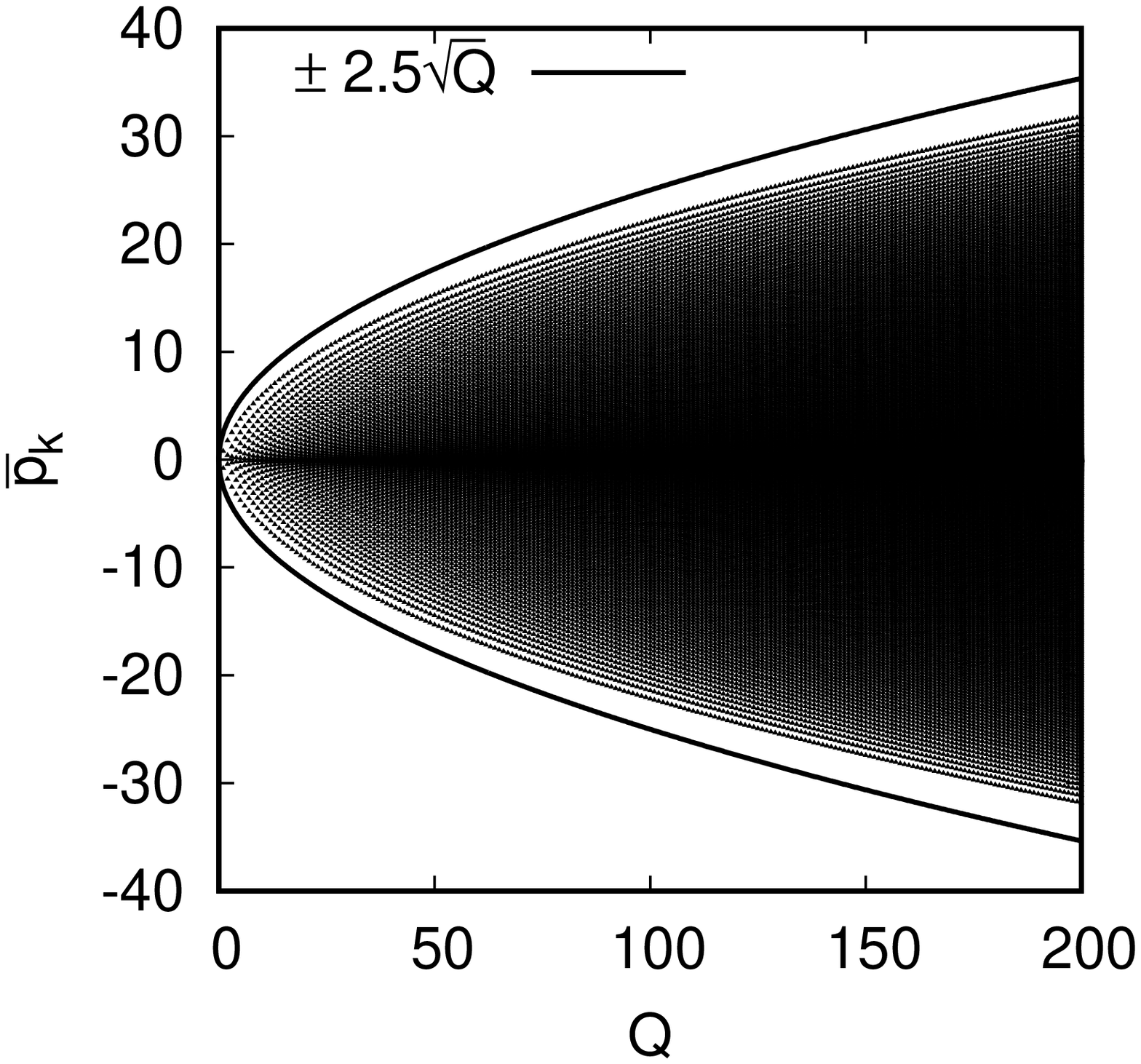} \\
(c) & (d)
\end{tabular}
\end{center}
\caption{Scatter plot of the quadrature points $\pm \wp_k$ ($1 \le k \le Q$) 
occuring in the Gauss-Laguerre (top) and half-range Gauss-Hermite (bottom) quadratures of order $Q$,
where $1 \le Q \le 10$ (left) and $1 \le Q \le 200$ (right). At each value of 
$Q$ represented on the horizontal axis, 
there are $2Q$ points on the vertical line, representing the 
values of the quadrature points $\wp_k$ corresponding to the positive ($1 \le k \le Q$)
and negative ($Q+ 1 \le k \le 2Q$) semiaxes. All quadrature points are 
contained inside the outer envelope (a) $\pm 4Q$ for the Gauss-Laguerre quadrature 
and (b) $\pm 2.5\sqrt{Q}$ for the half-range Gauss-Hermite quadrature.
}
\label{fig:hh_hyp}
\end{figure}

We now perform the analysis presented in Sec.~\ref{sec:LB:H:hyp} for the case of the LB models 
based on half-range quadratures. Multiplying Eq.~\eqref{eq:boltz_S} by $\phi_\ell(\abs{\wp})$ and 
integrating with respect to $\wp$ over $[0, \infty)$ ($+$) and $(-\infty,0]$ ($-$) yields:
\begin{align}
 \partial_t \mathcal{F}^+_\ell + \partial_x A_{\ell,s}^+ \mathcal{F}^+_s =& \mathcal{S}_\ell^+, \nonumber\\
 \partial_t \mathcal{F}^-_\ell + \partial_x A_{\ell,s}^- \mathcal{F}^-_s =& \mathcal{S}_\ell^-,
 \label{eq:boltz_Slpm}
\end{align}
where the matrices $A_{\ell, s}^\pm$ have the following non-zero elements:
\begin{equation}
 A_{\ell, \ell - 1}^\pm = \mp \frac{c_\ell p_0}{a_\ell m}, \qquad 
 A_{\ell, \ell}^\pm = \mp \frac{b_\ell p_0}{a_\ell m}, \qquad 
 A_{\ell, \ell + 1}^\pm = \pm \frac{p_0}{a_\ell m},
 \label{eq:hh_A}
\end{equation}
which are induced by the following recurrence relation:
\begin{equation}
 \phi_{\ell + 1}(z) = (a_\ell z + b_\ell) \phi_\ell(z) + \phi_{\ell - 1}(z).
 \label{eq:phi_rec}
\end{equation}
When the momentum space is discretized using a quadrature of order $Q$, the coefficients 
$\mathcal{F}^\pm_Q = 0$ and the matrices $A_{l,s}^\pm$ are of size $Q \times Q$.
The functions $\mathcal{F}^+_\ell$ and $\mathcal{F}^-_\ell$ corresponding to the 
projection of $f$ on the positive and negative momentum semiaxes mix only due to the source function 
$\mathcal{S}_\ell$, which can be expressed in completely local form with respect to $\mathcal{F}^\pm_s$.
Thus, the hyperbolicity of the system of equations \eqref{eq:boltz_Slpm} can be analysed on each 
semiaxis independently. Moreover, Eq.~\eqref{eq:hh_A} shows that $\mathbb{A}^+ = -\mathbb{A}^- = \mathbb{A}$, such that
the eigenvalues of $\mathbb{A}^-$ can be obtained from the eigenvalues of $\mathbb{A}^+$ by multiplying the 
latter by a factor of $-1$. Hence, we will only focus on $\mathbb{A}^+ \equiv \mathbb{A}$, which can be written as follows:
\begin{equation}
 \mathbb{A} = \frac{p_0}{m}
 \begin{pmatrix}
  -\frac{b_0}{a_0} & \frac{1}{a_0} & 0 & 0 & 0 & \cdots \\
  -\frac{c_1}{a_1} & -\frac{b_1}{a_1} & \frac{1}{a_1} & 0 & 0 & \cdots \\
  0 & -\frac{c_2}{a_2} & -\frac{b_2}{a_2} & \frac{1}{a_2} & 0 & \cdots \\
  0 & 0 & -\frac{c_3}{a_3} & -\frac{b_3}{a_3} & \frac{1}{a_3} & \cdots \\
  0 & 0 & 0 & -\frac{c_4}{a_4} & -\frac{b_4}{a_4} & \cdots \\
  \dots & \dots & \dots & \dots & \dots & \dots
 \end{pmatrix}\label{eq:hh_A_matrix}
\end{equation}

At $Q = 1$, the eigenvalue equation yields $\lambda = -\frac{p_0}{m} \frac{b_0}{a_0}$, 
which represents the root of $\phi_1(\wp)$ multiplied by $p_0 / m$.
Next, at $Q = 2$, the eigenvalue equation reads 
\begin{equation}
 a_0 a_1 \lambda^2 + (a_0 b_1 + a_1 b_0) \lambda + b_0 b_1 + c_1 = 
 \frac{\phi_2(\lambda)}{\phi_0(\lambda)} = 0,
 \label{eq:lambda_Q2}
\end{equation}
where the intermediate equality follows by applying the 
the recurrence relation \eqref{eq:phi_rec} twice to obtain $\phi_2(\lambda)$
in terms of $\phi_0(\lambda)$.
Eq.~\eqref{eq:lambda_Q2} shows that the eigenvalues of $\mathbb{A}^+$ when $Q = 2$ 
are the roots of $\phi_2(\lambda)$.

Thus we reach the same conclusion as for the case of the full-range Gauss-Hermite quadrature, which we 
summarise as follows.
The system of moment equations \eqref{eq:boltz_Slpm} is obtained by applying a linear transformation on 
the lattice Boltzmann equation \eqref{eq:boltzk} using the matrices
\begin{equation}
 \mathcal{H}_{\ell,k}^+ = 
 \begin{cases}
  \phi_\ell(\wp_k), & 1 \le k \le Q,\\
  0, & {\rm otherwise}
 \end{cases}\qquad
 \mathcal{H}_{\ell,k}^- = 
 \begin{cases}
  \phi_\ell(-\wp_k), & Q < k \le 2Q,\\
  0,& {\rm otherwise}.
 \end{cases}
\end{equation}
Putting the Boltzmann equation in the form \eqref{eq:boltzk_Ak}
shows that calculating the eigenvalues of the matrices $\mathcal{A}_{\ell, s}^+$
and $\mathcal{A}_{\ell, s}^-$ is equivalent to obtaining the eigenvalues of 
$\mathcal{A}_{k,k'} = \frac{p_k}{m} \delta_{k,k'}$, which are 
just the discrete velocities corresponding to the half-range Gauss quadrature of order
$Q$, together with their reflection with respect to the origin. Since $p_k = p_0 \wp_k$ 
($1 \le k \le Q$) is linked to the roots of $\phi_Q(\wp)$, it is clear that the 
resulting eigenvalues are real and distinct and thus, the system of lattice Boltzmann 
equations \eqref{eq:boltzk} is strictly hyperbolic. 

We end this section by showing a scatter plot of the quadrature points $\wp_k$ 
($1 \le k \le 2Q$) for the Gauss-Laguerre [Fig.~\ref{fig:hh_hyp}(a,b)]
and half-range Gauss-Hermite [Fig.~\ref{fig:hh_hyp}(c,d)] quadratures. 
The solid lines indicate that, at fixed $Q$, the quadrature points satisfy 
$-4Q < \wp_k < 4Q$ and $-2.5\sqrt{Q} < \wp_k < 2.5\sqrt{Q}$ in the case of the 
Gauss-Laguerre and half-range Gauss-Hermite quadratures, respectively. 

\subsubsection{Expansion of the equilibrium distribution function} \label{sec:LB:hh:feq}

In the case of half-range quadratures, the function $g$ defined in Eq.~\eqref{eq:feq} 
must be expanded separately on the positive and negative momentum semiaxes:
\begin{equation}
 g_\sigma = \frac{\omega(\abs{\wp})}{p_0} \sum_{\ell = 0}^\infty \mathcal{G}^\sigma_\ell
 \phi_\ell(\abs{\wp}),
\end{equation}
where $\sigma = 1$ when $p > 0$ and $\sigma = -1$ when $p < 0$. 
Following the convention of Eq.~\eqref{eq:hh_pk}, the momentum space is discretised using 
$\mathcal{Q} = 2Q$ elements with $p_k > 0$ and $p_{k + Q} = -p_k$ for $1 \le k \le Q$.
The corresponding equilibrium distributions $\feq_k = n g_k$ are constructed 
using
\begin{equation}
 g_k = w_k \sum_{\ell = 0}^N \mathcal{G}^+_\ell \phi_\ell(\wp_k),\qquad
 g_{k + Q} = w_k \sum_{\ell = 0}^N \mathcal{G}^-_\ell \phi_\ell(\wp_k),
\end{equation}
where the expansion order $0 \le N < Q$ is a free parameter of the model which 
represents the order up to which the half-range moments of $\feq$ are exactly recovered.
The coefficients $\mathcal{G}^\pm_\ell$ can be found using the orthogonality 
relation \eqref{eq:phi_ortho}:
\begin{equation}
 \mathcal{G}^+_\ell = \int_0^\infty dp\, g\, \phi_\ell(\wp), \qquad 
 \mathcal{G}^-_\ell = \int_{-\infty}^0 dp\, g\, \phi_\ell(-\wp).
\end{equation}
Irrespective of the choice of quadrature, $\mathcal{G}^\pm_\ell$ can be expressed analytically 
and $g_k$ becomes \cite{ambrus16jcp,ambrus16jocs}:
\begin{align}
 g_k =& \frac{w_k}{2} \sum_{s = 0}^{N} \left(\frac{mT}{2p_0^2}\right)^{s/2} \Phi_s^{N}(\wp_k) 
 \left[(1 + \erf \, \zeta) P_s^+(\zeta) + \frac{2}{\sqrt{\pi}} \e^{-\zeta^2} P_s^*(\zeta)\right],\nonumber\\
 g_{k+Q} =& \frac{w_k}{2} \sum_{s = 0}^{N} \left(\frac{mT}{2p_0^2}\right)^{s/2} \Phi_s^{N}(\wp_k)
 \left[(1 - \erf \, \zeta) P_s^+(-\zeta) + \frac{2}{\sqrt{\pi}} \e^{-\zeta^2} P_s^*(-\zeta)\right],
 \label{eq:hh_gk}
\end{align}
where $\zeta = u \sqrt{m / 2T}$, $\Phi_s^N(\wp_k)$ is defined in Eq.~\eqref{eq:Phi_def}, while
$P_s^+(\zeta)$ and $P_s^*(\zeta)$ represent polynomials of orders $s$ and $s - 1$, respectively, 
defined through:
\begin{equation}
 P_s^\pm(\zeta) = e^{\mp \zeta^2} \frac{d^s}{d\zeta^s} e^{\pm \zeta^2}, \qquad 
 P_s^*(\zeta) = \sum_{j = 0}^{s-1} \binom{s}{j} P_j^+(\zeta) P_{s-j-1}^-(\zeta)..
\end{equation}

\subsubsection{Force term}\label{sec:LB:hh:force}

We have seen in Sec.~\ref{sec:hh_force:result} that the kernel $\mathcal{K}^{\rm half}(p, p')$ 
for the momentum derivative $\partial_p f$ mixes the distributions of 
particles on the two semiaxes. The full-range integral with respect 
to $p'$ in Eq.~\eqref{eq:hh_K_def} can be written as a sum of two half-range 
integrals over the domains $[0,\infty)$ and $(-\infty,0]$, which can be recovered as 
follows:
\begin{equation}
 (\partial_p f)_k = \sum_{k' = 1}^{2Q} \mathcal{K}^{\rm half}_{k,k'} f_{k'}.
 \label{eq:hh_LB_force_aux}
\end{equation}
In the above, $f_{k'}$ is a polynomial of order $Q - 1$ in $p_{k'}$. The 
equality in Eq.~\eqref{eq:hh_LB_force_aux} can be achieved only if $\mathcal{K}^{\rm half}(p,p')$ 
is truncated at order $Q - 1$ with respect to $\phi_\ell(p')$. This truncation 
still allows the exact recovery of the integral with respect to $p'$ in Eq.~\eqref{eq:hh_K_def} 
since the polynomials $\phi_\ell(\abs{\wp'})$ with $\ell \ge Q$ are orthogonal to all polynomials 
of order $Q - 1$ or less by virtue of the orthogonality relation \eqref{eq:phi_ortho}.

We furthermore require the truncation of $\mathcal{K}^{\rm half}(p,p')$ at $\ell = Q -1$ 
with respect to $\phi_\ell(\abs{\wp})$. This truncation allows the closure of the moment equation 
\eqref{eq:boltz_Slpm} with respect to $\mathcal{F}_\ell^\pm$ with $0 \le \ell \le Q - 1$. We note 
that the moments of order $0 \le s < Q$ of $\partial_p f$ are exactly recovered when using the 
above truncation of $\mathcal{K}^{\rm half}(p,p')$, since the polynomials $\phi_\ell(\abs{\wp})$ 
with $\ell \ge Q$ do not contribute to the moments of order $0 \le s < Q$ by virtue of the 
orthogonality relation \eqref{eq:phi_ortho}.

The resulting expression for the kernel $\mathcal{K}^{\rm half}_{k,k'}$ is:
\begin{equation}
 \mathcal{K}^{\rm half}_{k,k'} = 
 \frac{w_k \sigma_k}{p_0} \left[
 \frac{1 + \sigma_k \sigma_{k'}}{2} \sum_{\ell = 0}^{Q - 2} \phi_\ell(\abs{\wp_{k'}}) 
 \sum_{s = \ell + 1}^{Q - 1} \frac{1}{\gamma_s} \phi_s(\abs{\wp_k}) \varphi_{s,\ell} - 
 \frac{\omega(0)}{2} \Phi_0^Q(\abs{\wp_k}) \Phi_0^Q(\abs{\wp_{k'}})\right],
 \label{eq:hh_Kk}
\end{equation}
where $\sigma_k = 1$ when $1 \le k \le Q$ and $\sigma_{k} = -1$ when $Q < k \le 2Q$.
Let us now specialise Eq.~\eqref{eq:hh_Kk} to the case of the Gauss-Laguerre and 
half-range Gauss-Hermite quadratures.

When the Gauss-Laguerre quadrature is enmployed, $\omega(z)$, $\gamma_\ell$ and $\varphi_{\ell, s}$ 
take the following values:
\begin{equation}
 \omega(z) = e^{-z}, \qquad
 \gamma_\ell = 1, \qquad 
 \varphi_{s,\ell} = 
 \begin{cases}
  1 & 0 \le \ell < s, \\
  0 & {\rm otherwise.}
 \end{cases}
\end{equation}
In this case, Eq.~\eqref{eq:hh_Kk} becomes:
\begin{equation}
 \mathcal{K}^L_{k,k'} = 
 \frac{w_k \sigma_k}{p_0} \left[
 \frac{1 + \sigma_k \sigma_{k'}}{2} \sum_{\ell = 0}^{Q - 2} L_\ell(\abs{\wp_{k'}}) 
 \sum_{s = \ell + 1}^{Q - 1} L_s(\abs{\wp_k}) - 
 \frac{1}{2} \Phi_0^Q(\abs{\wp_k}) \Phi_0^Q(\abs{\wp_{k'}})\right],
 \label{eq:hh_Kk_L}
\end{equation}
where $\Phi_0^Q(\abs{\wp_k})$, defined in Eq.~\eqref{eq:Phi_def}, reduces to:
\begin{equation}
 \Phi_0^Q(\abs{\wp_k}) = \sum_{s = 0}^{Q - 1} L_s(\abs{\wp}_k),
\end{equation}
where the upper limit of the sum over $s$ is $Q - 1$ since $L_Q(\abs{\wp}_k) = 0$.

In the case of the half-range Hermite polynomials, $\gamma_\ell = 1$ and
$\varphi_{\ell,s}$ can be found in Eq.~\eqref{eq:hh_varphi}.
Thus, Eq.~\eqref{eq:hh_Kk} becomes:
\begin{multline}
 \mathcal{K}^\hh_{k,k'} = 
 \frac{w_k \sigma_k}{p_0} \Bigg\{
 \frac{1 + \sigma_k \sigma_{k'}}{2} \sum_{\ell = 0}^{Q - 2} 
 \hh_\ell(\abs{\wp_{k'}}) \left[
 \frac{\hh_{\ell,0}}{\sqrt{2\pi}} \sum_{s = \ell + 1}^{Q - 1} \hh_{s,0} \hh_s(\abs{\wp_k}) 
 - \frac{\hh_{\ell, \ell}}{\hh_{\ell+1,\ell+1}} \hh_{\ell + 1}(\abs{\wp_k})\right]\\
 - \frac{1}{2\sqrt{2\pi}} \Phi^Q_0(\abs{\wp_k}) \Phi^Q_0(\abs{\wp_{k'}})\Bigg\}.
 \label{eq:hh_Kk_hh}
\end{multline}
where $\Phi_0^Q(\abs{\wp_k})$ \eqref{eq:Phi_def} reduces to:
\begin{equation}
 \Phi_0^Q(\abs{\wp_k}) = \sum_{s = 0}^{Q-1} \hh_{s,0} \hh_s(\abs{\wp_k}).
 \label{eq:Phi_def_hh}
\end{equation}
For the reader's convenience, we supply the matrix elements $\mathcal{K}^\hh_{k,k'}$ 
for $1 \le Q \le 200$ as supplemental material (more details can be found in 
Appendix~\ref{app:supp}).

\subsection{Numerical method} \label{sec:LB:sch}

In order to numerically solve the lattice Boltzmann equation \eqref{eq:boltzk}, we employ an 
explicit finite difference scheme based on the total variation diminishing (TVD) third-order 
Runge-Kutta (RK-3) method \cite{shu88,gottlieb98,henrick05,trangenstein07,blaga16jcp} 
and the fifth order weighted essentially non-oscillatory (WENO-5) scheme
\cite{jiang96,shu99,gan11,rezzolla13,blaga16jcp,hejranfar17pre}.

\subsubsection{Time stepping}\label{sec:LB:sch:RK3}

\begin{table}
\begin{tabular}{r|rrr}
0 & & & \\
1 & 1 & & \\
1/2 & 1/4 & 1/4 & \\\hline
& 1/6 & 1/6 & 2/3
\end{tabular}
\caption{Butcher tableau for the third-order Runge-Kutta 
time-stepping procedure described in Eq.~\eqref{eq:rk3}.\label{tab:rk3}}
\end{table}

In order to apply the RK-3 method, Eq.~\eqref{eq:boltzk} can be written as:
\begin{equation}
 \partial_t f_k = L[f]_k, \qquad  
 L[f]_k = -\frac{p_k}{m} \partial_x f_k + m g (\partial_p f)_k - \frac{1}{\tau}(f_k - \feq_k),
 \label{eq:L_def}
\end{equation}
where the advective term $\frac{p_k}{m} \partial_x f_k$ is discussed in Subsec.~\ref{sec:LB:sch:WENO5}, while
the momentum derivative is computed as follows:
\begin{equation}
 (\partial_p f)_k = \sum_{k' = 1}^{\mathcal{Q}} \mathcal{K}_{k,k'} f_{k'}.
\end{equation}
When the full-range Gauss-Hermite quadrature is employed, 
$\mathcal{Q} = Q$ and the kernel $\mathcal{K}_{k,k'}$ is given in Eq.~\eqref{eq:H_Kk}.
In the cases when the Gauss-Laguerre and half-range Gauss-Hermite 
are employed, $\mathcal{Q} = 2Q$ and the corresponding kernels 
$\mathcal{K}_{k,k'}^L$ and $\mathcal{K}_{k,k'}^\hh$ are given in 
Eqs.~\eqref{eq:hh_Kk_L} and \eqref{eq:hh_Kk_hh}, respectively.

We discretise the time coordinate using the time step $\delta t$.
The total variation diminishing (TVD) preserving RK-3 method proposed in 
Refs.~\onlinecite{shu88,gottlieb98,henrick05,trangenstein07,rezzolla13,blaga16jcp,busuioc17} 
requires three steps of the form \eqref{eq:L_def}, which can be summarised as follows:
\begin{align}
 f^{(1)}_k =& f_k(t) + \delta t \, L[f]_k, \nonumber\\
 f^{(2)}_k =& \frac{3}{4} f_k(t) + \frac{1}{4} f^{(1)}_k + \frac{1}{4} \delta t\, L[f^{(1)}]_k,\nonumber\\
 f(t + \delta t)_k =& \frac{1}{3} f_k + \frac{2}{3} f^{(2)}_k + \frac{2}{3} \delta t\, L[f^{(2)}]_k. 
 \label{eq:rk3}
\end{align}
The Butcher tableau associated to the above scheme is given in Tab.~\ref{tab:rk3}.

\subsubsection{Advection}\label{sec:LB:sch:WENO5}

\begin{table}
\begin{tabular}{r|rrr}
 & $\overline{\omega}_1$ & $\overline{\omega}_2$ & $\overline{\omega}_3$ \\\hline
$\sigma_1 = \sigma_2 = \sigma_3$ & $0.1$ & $0.6$ & $0.3$ \\\hline
$\sigma_2 = \sigma_3 = 0$ & $0$ & $2/3$ & $1/3$ \\
$\sigma_3 = \sigma_1 = 0$ & $1/4$ & $0$ & $3/4$ \\
$\sigma_1 = \sigma_2 = 0$ & $1/7$ & $6/7$ & $0$ \\\hline
$\sigma_1 = 0$ & $1$ & $0$ & $0$\\
$\sigma_2 = 0$ & $0$ & $1$ & $0$\\
$\sigma_3 = 0$ & $0$ & $0$ & $1$
\end{tabular}
\caption{The weighting factors $\overline{\omega}_q$ \eqref{eq:weno5_omega} 
as one, two or all three values of $\sigma_s$ vanish.\label{tab:weno}}
\end{table}

The domain between the left and right walls is discretised using $N_x$ nodes. In order to better capture 
the macroscopic profiles in the vicinity of the boundaries, we follow Refs.~\onlinecite{mei98,guo03,busuioc17}
and employ the stretching of the coordinate system induced by the following transformation:
\begin{equation}
 x(\eta) = \frac{L}{2A} \tanh\eta,\label{eq:x_eta}
\end{equation}
where $A \in (0, 1)$ controls the degree of stretching. We construct the spatial grid by employing 
$N_x$ equidistant values of $\eta \in [-\arctanh \,A, \arctanh \,A]$, i.e.:
\begin{equation}
 \eta_i = \left(i - \frac{N_x + 1}{2}\right) \delta \eta, \qquad 
 \delta \eta = \frac{2}{N_x} \arctanh\,A,\label{eq:etai}
\end{equation}
where $i = 1, 2, \dots N_x$. 
The coarsest part of the resulting grid lies near $x = 0$,
while the densest part lies near the two boundries located at $x = \pm L/ 2$.

We now write the advection term using the fluxes $\mathcal{F}_{i \pm 1/2}$ as follows:
\begin{equation}
 \left(\frac{p_k}{m} \partial_x f_k\right)_i \simeq \frac{\mathcal{F}_{k;i + 1/2} - \mathcal{F}_{k;i - 1/2}}
 {x(\eta_{i + 1/2}) - x(\eta_{i - 1/2})},
\end{equation}
where $\eta_{i\pm 1/2} = \eta_i \pm \delta \eta / 2$, while $x(\eta_{i\pm 1/2})$ 
is defined through Eq.~\eqref{eq:x_eta} and $\delta \eta$ is given in Eq.~\eqref{eq:etai}.
In this paper, we employ the fifth-order weighted essentially non-oscillatory (WENO-5) scheme
\cite{jiang96,shu99,gan11,rezzolla13,blaga16jcp,hejranfar17pre,busuioc17}
for the computation of the fluxes $\mathcal{F}_{i \pm 1/2}$. For completeness, 
we summarise our implementation of the WENO-5 scheme below.

Without loss of generality, we consider the case when $p_k > 0$. 
The flux $\mathcal{F}_{k;i+1/2}$ can be written as:
\begin{equation}
\mathcal{F}_{k; i+1/2} = \overline{\omega}_1\mathcal{F}^1_{k; i+1/2} +
\overline{\omega}_2\mathcal{F}^2_{k; i+1/2} + \overline{\omega}_3\mathcal{F}^3_{k; i+1/2},
\label{eq:weno5_flux}
\end{equation}
where the interpolating functions $\mathcal{F}^q_{k; i + 1/2}$ are given by:
\begin{align}
\mathcal{F}^1_{k; i+1/2} =& \frac{p_k}{m} \left(\frac{1}{3}f_{k; i-2} - 
\frac{7}{6} f_{k; i-1} + \frac{11}{6} f_{k; i}\right), \nonumber \\
\mathcal{F}^2_{k; i+1/2} =& \frac{p_k}{m} \left(-\frac{1}{6}f_{k; i-1} + 
\frac{5}{6} f_{k; i} + \frac{1}{3} f_{k; i+1}\right), \nonumber \\
\mathcal{F}^3_{k; i+1/2} =& \frac{p_k}{m} \left(\frac{1}{3}f_{k; i} + 
\frac{5}{6} f_{k; i+1} - \frac{1}{6} f_{k; i+2}\right),
\end{align}
while the weighting factors $\overline{\omega}_q$ are defined as:
\begin{equation}
\overline{\omega}_q = \frac{\widetilde{\omega}_q}{\widetilde{\omega}_1+\widetilde{\omega}_2+\widetilde{\omega}_3}, \qquad 
\widetilde{\omega}_q = \frac{\delta_q}{\sigma^2_q}.\label{eq:weno5_omega}
\end{equation}
The ideal weights $\delta_q$ are:
\begin{equation}
 \delta_1 = 1/10, \qquad \delta_2 = 6/10,\qquad \delta_3 = 3/10, 
\end{equation}
while the indicator of smoothness functions $\sigma_q$ ($q = 1, 2, 3$) are given by:
\begin{align}
\sigma_1 =& \frac{13}{12} \left(f_{k; i-2} -2f_{k; i-1} + f_{k;i} \right)^2 
+ \frac{1}{4} \left( f_{k; i-2} - 4f_{k; i-1} + 3f_{k;i} \right)^2, 
\nonumber \\
\sigma_2 =& \frac{13}{12} \left(f_{k; i-1} -2f_{k; i} + f_{k; i+1} \right)^2 
+ \frac{1}{4} \left( f_{k; i-1} - f_{k; i+1} \right)^2,
\nonumber \\
\sigma_3 =& \frac{13}{12} \left(f_{k; i} -2f_{k; i+1} + f_{k; i+2} \right)^2 
+ \frac{1}{4} \left( 3f_{k; i} -4 f_{k; i+1} + f_{k; i+2} \right)^2.
\label{eq:weno_sigma}
\end{align}
It is customary to add a small quantity $\varepsilon$ (usually taken 
as $10^{-6}$) in the denominators of $\widetilde{\omega}_q$ to avoid divisions by $0$. 
However, as pointed out in Ref.~\onlinecite{henrick05}, the effect of 
this alteration of the indicators of smoothness is strongly dependent on the given problem, since 
$\varepsilon$ becomes a dimensional quantity. Furthermore, the accuracy of the resulting scheme depends on the value 
of $\varepsilon$. Since at higher orders, the distribution functions corresponding to large velocities can 
have values which are significantly smaller than those for smaller velocities, we cannot predict the effect 
of employing a unitary value for $\varepsilon$ for all distributions. Therefore, we prefer to follow 
Refs.~\onlinecite{blaga16jcp,busuioc17} and compute the limiting values of $\widetilde{\omega}_q$ 
when one, two or all three of the indicators of smoothness vanish, as indicated in Tab.~\ref{tab:weno}.

\subsubsection{Boundary conditions}

In this paper, we only consider the case of full diffuse reflection boundary conditions (i.e. corresponding to 
full accomodation at the walls). We implement these boundary conditions by ensuring that the 
flux of particles returning into the fluid domain through the cell interface between the fluid 
and the wall follows a Maxwellian distribution. For definiteness, let us consider the case of the left wall, 
in which case the above condition reads:
\begin{equation}
 \mathcal{F}_{k; 1/2} = \feq_+ \frac{p_k}{m}, \qquad
 (p_k / m > 0),
 \label{eq:weno_flux_eq}
\end{equation}
where $\feq_+$ represents the Maxwell-Boltzmann distribution of the emergent particles 
[this is defined later in Eq.~\eqref{eq:fwall}].
The flux in Eq.~\eqref{eq:weno_flux_eq} can be achieved analytically by populating the 
ghost nodes at $i = -2$, $-1$ and $0$ according to:
\begin{equation}
 f_{-2;k} = f_{-1;k} = f_{0; k} = \feq_+, \qquad (p_k / m > 0).\label{eq:weno_populate} 
\end{equation}
With the above definitions, Eq.~\eqref{eq:weno_sigma} shows that $\sigma_1 = 0$ for $i = 0$.
According to Tab.~\ref{tab:weno}, $\overline{\omega}_1 = 1$ and $\overline{\omega}_2 = \overline{\omega}_3 = 0$
when $\sigma_1 = 0$. Thus, Eq.~\eqref{eq:weno5_flux} implies that 
\begin{equation}
 \mathcal{F}_{k; 1/2} = \mathcal{F}_{k; 1/2}^1 = f_{k; 0} \frac{p_k}{m}.
\end{equation}
Thus, Eq.~\eqref{eq:weno_flux_eq} is established.

In order to calculate the fluxes at $i = 1/2$ and $i = 3/2$ for particles travelling towards the wall ($p_k <0$), 
a quadratic extrapolation is used:
\begin{align}
 f_{k;0} =& 3 f_{k;1} - 3 f_{k; 2} + f_{k; 3}, \nonumber\\
 f_{k; -1} =& 6 f_{k; 1} - 8 f_{k; 2} + 3f_{k; 3}, \\
 & (p_k < 0).\nonumber
\end{align}

Finally, mass conservation is ensured by requiring that:
\begin{equation}
 \sum_{k = 1}^{\mathcal{Q}} \mathcal{F}_{k; 1/2} = 0.
\end{equation}
This translates into the following equation for the particle number density $n_+$ of the emergent particles:
\begin{equation}
 n_+ = - \frac{\displaystyle \sum_{p_k < 0} \mathcal{F}_{k; 1/2}}
 {\displaystyle \sum_{p_k > 0} g_k^+},
 \label{eq:nplus_def}
\end{equation}
where $g_k^+ = \feq_+ / n_+$ is introduced in Eq.~\eqref{eq:H_gk} for the 
full-range Gauss-Hermite quadrature and in Eq.~\eqref{eq:hh_gk} for 
the case when half-range quadratures are employed.

\subsection{Notation}\label{sec:LB:notation}

The models considered in this paper have three free parameters: the type of quadrature 
(we will only consider the full-range Gauss-Hermite and the half-range Gauss-Hermite
quadratures in the remainder of this paper); the quadrature order $Q$; and the 
order $N$ of the expansion of the function $g$ in $\feq$. To denote an LB model 
based on the full-range Gauss-Hermite quadrature of order $Q$ (employing $Q$ discrete momentum vectors) 
and order $N$ for the expansion of $\feq$, we will use the notation ${\rm HLB}(N;Q)$. 
A similar model based on the half-range Gauss-Hermite quadrature will be denoted 
${\rm HHLB}(N; Q)$, but in this case, the size of the momentum set is $\mathcal{Q} = 2Q$, 
since a number of $Q$ quadrature points is required on each semiaxis.

The significance of the quadrature order $Q$ was discussed for the case of the 
full-range Gauss-Hermite quadrature in Subsec.~\ref{sec:LB:H:discrete}, 
while for the case of half-range quadratures (including the half-range Hermite quadrature),
more details can be found in Subsec.~\ref{sec:LB:hh:discrete}. 
The significance of the expansion order $N$ of the equilibrium distribution function 
was explained in Subsecs.~\ref{sec:LB:H:feq} and \ref{sec:LB:hh:feq} for the 
case of the full-range Gauss-Hermite quadrature and the half-range quadratures, respectively.

\section{Numerical results} \label{sec:num}

To illustrate the method introduced in section~\ref{sec:LB}, we consider in 
this paper the problem of a one-dimensional flow between two diffuse reflective boundaries 
at rest. The flow
is driven by a force derived from a linear potential $V(x) = mgx$, 
giving rise to a constant force pointing towards the right wall:
\begin{equation}
 F = -\partial_x V(x) = -mg,\label{eq:V}
\end{equation}
where $m$ is the mass of the particles and $g$ is the constant gravitational acceleration.

This section is divided as follows. 
In Subsec.~\ref{sec:num:macro}, we discuss the macroscopic equations governing the flow.
The particular case when the walls have equal temperatures admits an equilibrium solution 
which we discuss in Subsec.~\ref{sec:num:eq}. In Subsec.~\ref{sec:num:NS},
we validate our models in the Navier-Stokes (small ${\rm Kn}$) limit, where both the full-range 
and the half-range LB
models  adequately recover the analytic solution. In Subsec.~\ref{sec:num:bal}, we derive the 
analytic solution for the free molecular flow regime and demonstrate that in the ballistic 
regime, the models based on the full-range Gauss-Hermite quadrature are no longer adequate.
However, we find that the models based on the half-range Gauss-Hermite quadrature with the force 
term implemented as described in Sec.~\ref{sec:LB:hh:force} correctly recover the analytic solution.  
In Subsec.~\ref{sec:num:visc} we make an analysis of the capabilities of the full-range and half-range 
models to simulate flows in the transition regime.

\subsection{Macroscopic equations}\label{sec:num:macro} 

The Boltzmann equation \eqref{eq:boltz} in the presence of the force \eqref{eq:V} 
can be written as:
\begin{equation}
 \partial_t f + \frac{p}{m} \partial_x f - mg\frac{\partial f}{\partial p} = 
 -\frac{1}{\tau} (f - \feq).\label{eq:boltz_force}
\end{equation}
Fixing the origin of the coordinate system at $x = 0$, the channel walls are located at 
$x_+ = -L/2$ and $x_- = L/2$, where $f$ obeys diffuse reflection boundary conditions:
\begin{gather}
 f(x_+, p > 0) = \feq_+ = n_+ g_+,  \qquad
 f(x_-, p < 0) = \feq_- = n_- g_-, \nonumber\\
 g_\pm = \frac{1}{\sqrt{2\pi m T_\pm}} \exp\left(-\frac{p^2}{2mT_\pm}\right).\label{eq:fwall}
\end{gather}
In the above, $\feq_\pm$ are the Maxwell-Boltzmann distributions
\eqref{eq:feq} for the 
particles emerging from the wall, defined using the particle number densities $n_\pm$,
the vanishing velocity of the walls and the wall temperatures $T_\pm$. The densities $n_\pm$ at
the walls can be computed by requiring that the mass fluxes through the boundaries vanish, cf.~\eqref{eq:nplus_def}:
\begin{equation}
 n_+ = -\frac{\displaystyle \int_{-\infty}^0 f(x_+, p) \, p\, dp}
 {\displaystyle \int_0^\infty g(0, T_+; p)\,p\,dp}, \qquad 
 n_- = -\frac{\displaystyle \int_{0}^\infty f(x_-, p) \, p\, dp}
 {\displaystyle \int_0^\infty g(0, T_-; p)\,p\,dp},
\end{equation}
where $g(u, T; p)$ is defined in Eq.~\eqref{eq:feq}.

In the stationary state, Eq.~\eqref{eq:boltz_force} reduces to:
\begin{equation}
 \frac{p}{m} \partial_x f - mg\frac{\partial f}{\partial p} = 
 -\frac{1}{\tau} (f - \feq).\label{eq:boltz_force_stationary}
\end{equation}
Multiplying the above equation by $1$, $p$ and $p^2$ and integrating with respect to $p$ yields:
\begin{subequations}\label{eq:macro}
\begin{align}
 \partial_x(nu) =& 0, \label{eq:nu}\\
 \partial_x(nT + \rho u^2) =& -\rho g, \label{eq:nT}\\
 \partial_x\left(q_x + \frac{3}{2} nu T + \frac{1}{2}\rho u^3 \right) =& 
 -\rho u g, \label{eq:q}
\end{align}
\end{subequations}
where $\rho = nm$ is the mass density, while
the particle number density $n$, macroscopic velocity $u$, temperature $T$ and heat flux $q$ 
are obtained as moments of $f$, as follows:
\begin{gather}
 n = \int_{-\infty}^\infty dp\, f, \qquad
 u = \frac{1}{n} \int_{-\infty}^\infty dp\, p\, f, \nonumber\\
 T = \frac{1}{nm} \int_{-\infty}^\infty dp\, (p - mu)^2\, f,\qquad
 q = \int_{-\infty}^\infty dp\, \frac{(p - mu)^3}{2m^2}\, f. \label{eq:fmoms} 
\end{gather}
Furthermore, $n(x)$ can be used to compute the total number of particles 
in the channel, defined as:
\begin{equation}
 \mathcal{N} = \int_{-L/2}^{L/2} dx\, n(x).\label{eq:Ndef}
\end{equation}

Since $u =0$ on the boundaries, Eq.~\eqref{eq:nu} implies that
$u(x) = 0$ throughout the channel, 
such that the remaining two equations reduce as follows:
\begin{subequations}
\begin{align}
 \partial_x(nT) =& - nmg, \label{eq:partialnT}\\
 q =& {\rm const.}\label{eq:qconst}
\end{align}
\end{subequations}

\subsection{Equal wall temperatures}\label{sec:num:eq}

\begin{figure}
\begin{center}
\begin{tabular}{c}
 \includegraphics[angle=270,width=0.45\columnwidth]{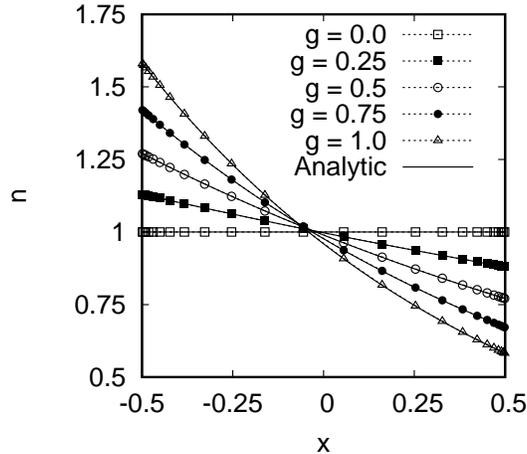}
\end{tabular}
\end{center}
\caption{Density profiles (dashed lines with points) at equilibrium ($T_- = T_+$) for the case 
when the potential has the linear form 
in Eq.~\eqref{eq:V}, obtained at ${\rm Kn} = 0.5$ using the ${\rm HLB}(4;5)$ model
(see Sec.~\ref{sec:LB:notation} for more details on this notation), 
compared with the analytic solution (lines) given in Eq.~\eqref{eq:iso_n}. 
Excellent agreement was found for all values of the gravitational constant $g$ considered 
here.
}
\label{fig:grav-iso}
\end{figure}

When the two walls are at the same temperature $T_+ = T_- = T_w$,
complete thermal equilibrium is achieved in the stationary state 
for all values of the relaxation time $\tau$. 
Indeed, setting $T = T_w$ in Eq.~\eqref{eq:partialnT} yields:
\begin{equation}
 f = \frac{n(x)}{\sqrt{2\pi m T_w}} \exp\left(-\frac{p^2}{2mT_w}\right),
\end{equation}
where
\begin{equation}
 n(x) = \frac{mg \mathcal{N}}{2T_w \sinh\frac{mgL}{2T_w}} \exp\left(-\frac{mg x}{T_w}\right).\label{eq:iso_n}
\end{equation}
The constant $\mathcal{N}$ represents the total number of particles between 
the walls and is defined in \eqref{eq:Ndef}. It can be checked that the above 
solution satisfies the Boltzmann equation \eqref{eq:boltz_force} for any value of $\tau$. 

Since the analytic result does not depend on the value of ${\rm Kn}$, we have 
chosen ${\rm Kn} = 0.5$ to validate our models. Our numerical results obtained 
using the ${\rm HLB}(4;5)$ model (see Sec.~\ref{sec:LB:notation} for more details on this notation)
are compared with the analytic expression \eqref{eq:iso_n} in Fig.~\ref{fig:grav-iso} and 
an excellent agreement can be observed. In our simulations,
we used a number of $N_x = 24$ nodes stretched according to Eq.~\eqref{eq:x_eta} with 
stretching parameter $A = 0.99$. The time step was set to $\delta t = 10^{-3}$.

\subsection{Different wall temperatures: Hydrodynamic regime}\label{sec:num:NS}

\begin{figure}
\begin{center}
\begin{tabular}{cc}
 \includegraphics[angle=270,width=0.45\columnwidth]{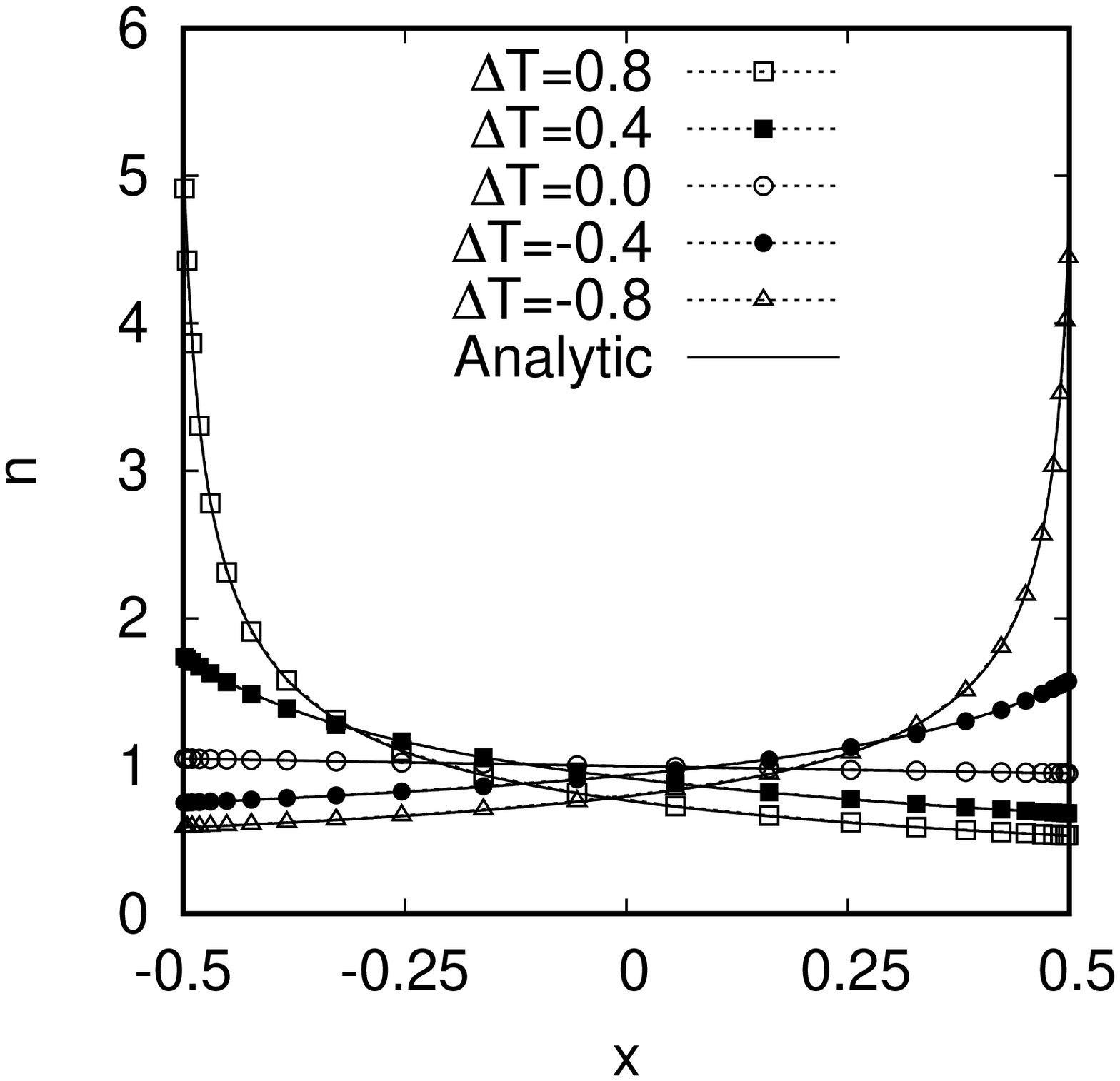} &
 \includegraphics[angle=270,width=0.45\columnwidth]{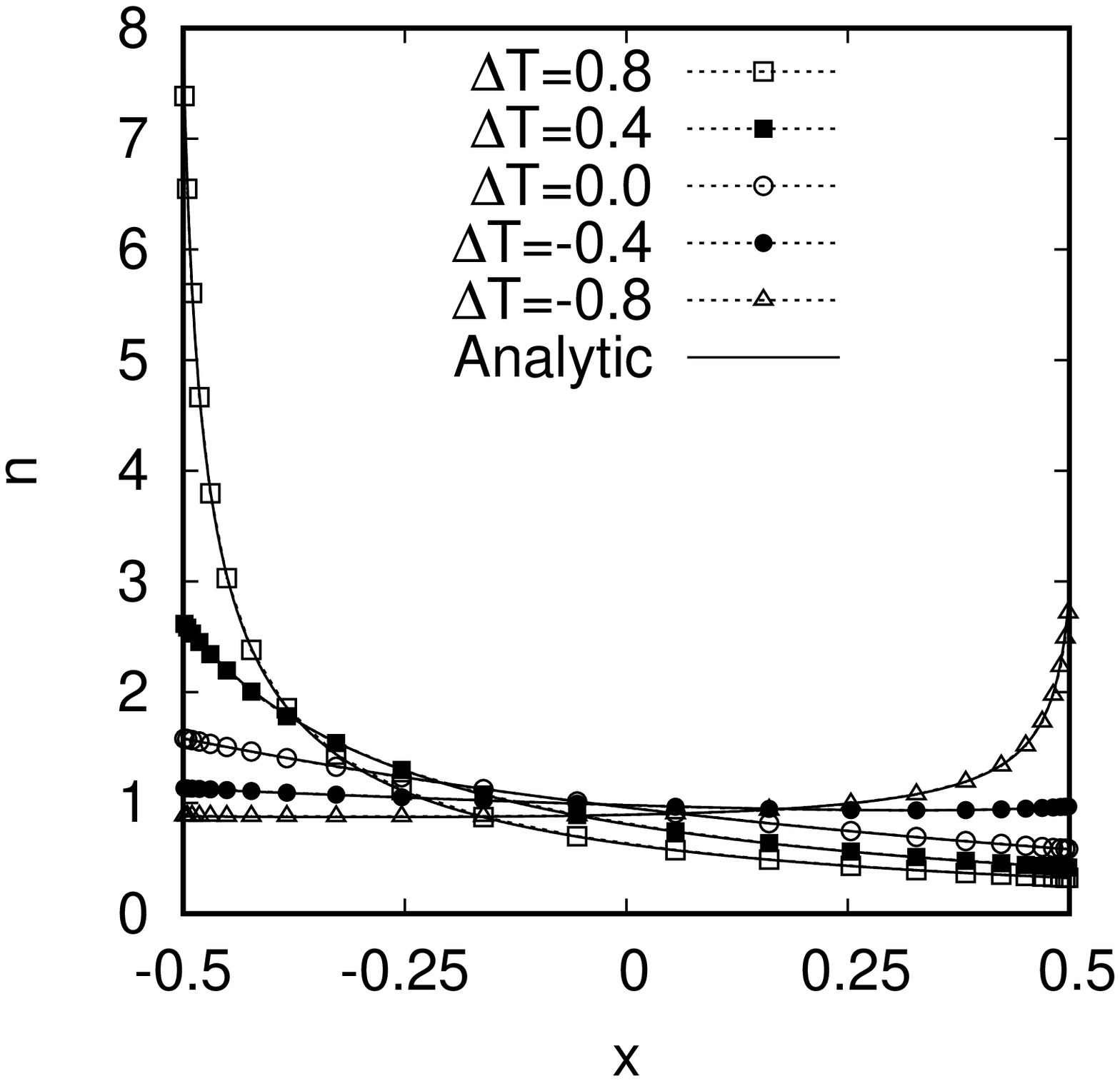} \\ 
 \quad\,\,\,\,(a) & \quad\,\,\,\,(b) 
\end{tabular}
\begin{tabular}{c}
 \includegraphics[angle=270,width=0.45\columnwidth]{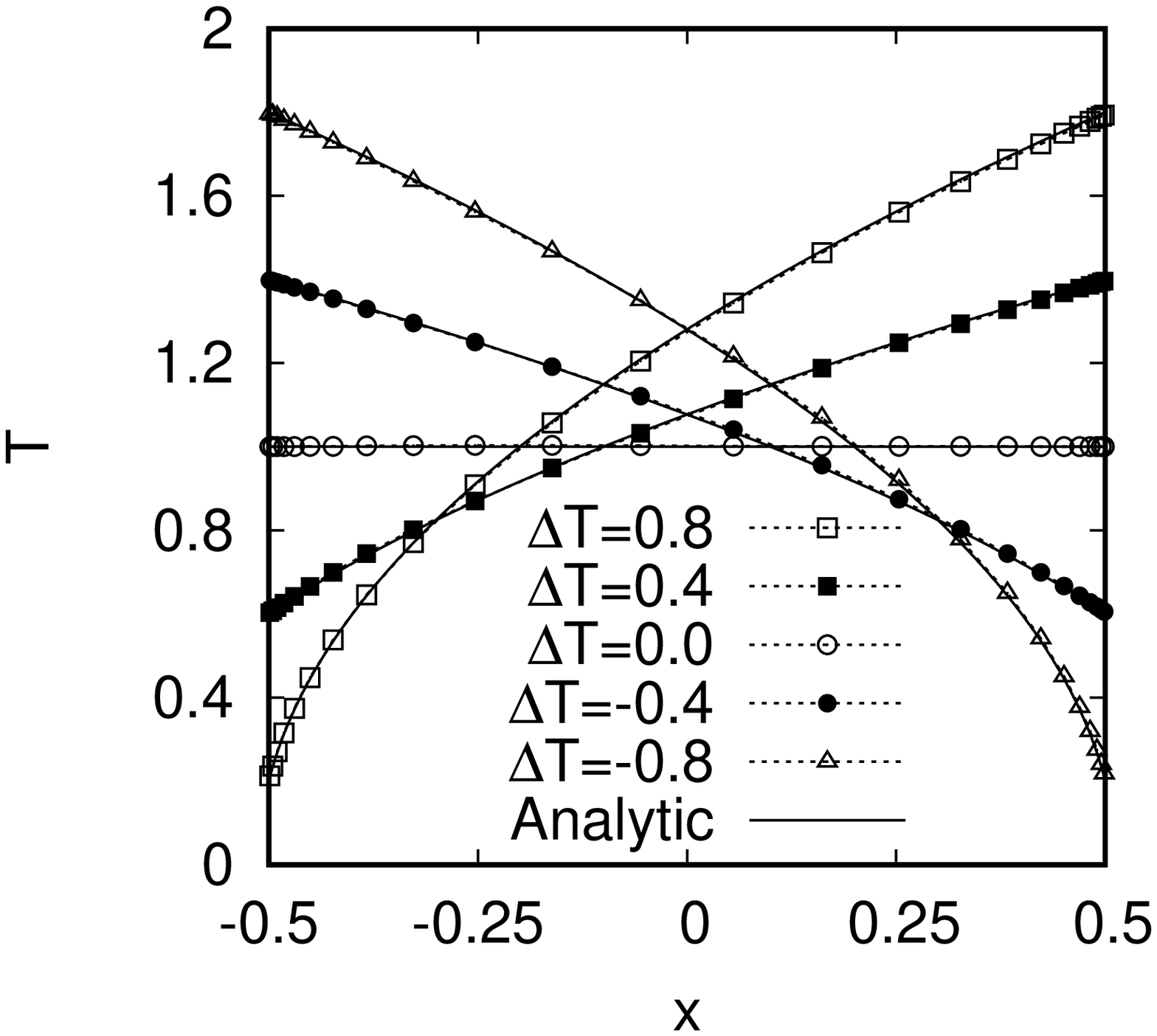} \\ 
 \qquad\,\,\,\,\,(c) 
\end{tabular}
\caption{Navier-Stokes regime (${\rm Kn} = 10^{-3}$): 
Comparison between the density $n$ (top) and temperature $T$ (bottom) profiles obtained 
using the ${\rm HLB}(4;5)$ model (dashed lines with points) and the 
analytic solutions \eqref{eq:iso_n} and \eqref{eq:NS_temp}, respectively (contiguous lines).
The plots for $n$ correspond to the accelerations $g = 0.1$ (left) and $g = 1.0$ (right).
Since the temperature profile in the Navier-Stokes regime does not depend on the 
potential (i.e.~on the value of $g$), (c) displays only the case when $g = 1.0$.
The curves correspond to various temperature differences 
$\Delta T = (T_- - T_+) / 2$ between the left and right walls.
}
\label{fig:grav-ns}
\end{center}
\end{figure}

At small values of ${\rm Kn}$, the flow is close to equilibrium and the Chapman-Enskog
expansion can be employed to obtain an approximate solution of the Boltzmann equation
\cite{huang63,harris71,cercignani88}. The heat flux can be computed using Fourier's law:
\begin{equation}
 q = -\kappa_T \partial_x T, \qquad 
 \kappa_T = \frac{3\tau n T}{2m}. \label{eq:NS_q_def}
\end{equation}
In this paper, we consider the following model for the relaxation time \cite{sone06}:
\begin{equation}
 \tau = \frac{\rm Kn}{n}, 
\end{equation}
such that Eq.~\eqref{eq:qconst} can be rearranged as follows:
\begin{equation}
 \partial_x T^2 = A,
\end{equation}
where $A$ is a constant. The solution of the above equation is:
\begin{equation}
 T(x) = \sqrt{A x + B},\label{eq:NS_temp}
\end{equation}
where the constants $A$ and $B$ are determined by the boundary conditions:
\begin{equation}
 A = \frac{T_-^2 - T_+^2}{L}, \qquad 
 B = \frac{T_-^2 + T_+^2}{2}.
\end{equation}
It is remarkable that, in the Navier-Stokes 
limit, the temperature profile does not depend on $g$. More generally, 
it can be shown that $T(x)$ does not depend on the form of the potential $V(x)$. 
Substituting Eq.~\eqref{eq:NS_temp} into Eq.~\eqref{eq:NS_q_def} yields:
\begin{equation}
 q(x) = \frac{3 {\rm Kn}}{4mL} (T_+^2 - T_-^2).\label{eq:NS_q}
\end{equation}

To find the expression for the particle number density
$n$, Eq.~\eqref{eq:partialnT} can be integrated, yielding:
\begin{equation}
 n(x) = \frac{mg \mathcal{N}}{2T(x) \sinh \frac{mgL}{T_+ + T_-}}
 \exp\left[-\frac{mgL}{T_-^2 - T_+^2}[2T(x) - T_- - T_+]\right],
 \label{eq:NS_n}
\end{equation}
where the total number of particles $\mathcal{N}$ is defined in Eq.~\eqref{eq:Ndef}.
It can be checked that Eq.~\eqref{eq:NS_n} reduces to Eq.~\eqref{eq:iso_n} when $T_- = T_+ = T_w$.

Figures~\ref{fig:grav-ns}(a) and (b) show a comparison between our numerical results and the 
analytic expression \eqref{eq:NS_n} obtained for $n$ in the 
Navier-Stokes regime ($\mathcal{N}$ was set to 1), for various values of the temperature 
difference $\Delta T = (T_- - T_+) / 2$
and for two values of the gravitational acceleration: $g = 0.1$ and $g = 1.0$.
In both cases, the agreement is excellent. 
Since in the Navier-Stokes limit, the temperature does not depend on $g$, 
Fig.~\ref{fig:grav-ns}(c) compares the temperature profile obtained using our models and 
the analytic expression in Eq.~\eqref{eq:NS_temp} only for the case when $g = 1.0$. Also in this case, 
it can be seen that the agreement is excellent.
The numerical results were obtained using the ${\rm HLB}(4;5)$ model (this notation is 
explained in Sec.~\ref{sec:LB:notation}. The time step 
was set to $\delta t= 10^{-4}$, the fluid domain was discretised using $N_x = 24$ nodes and 
we performed $5.000.000$ iterations to ensure that the stationary state was achieved.

\subsection{Ballistic regime: analytic analysis}\label{sec:num:bal}

Let us consider the stationary solutions of the Boltzmann equation
in the ballistic regime, when the right hand side of 
Eq.~\eqref{eq:boltz_force_stationary} vanishes:
\begin{equation}
 \frac{p}{m} \partial_x f - mg \partial_p f = 0,
 \label{eq:boltz_force_stationary_ballistic}
\end{equation}
where $f$ obeys the diffuse reflection boundary conditions \eqref{eq:fwall}.
During free-streaming, the total energy (Hamiltonian) of each particle, defined through:
\begin{equation}
 H = V(x) + \frac{p^2}{2m},\label{eq:Hamiltonian}
\end{equation}
is conserved. Indeed, it can be checked that $H$ satisfies Eq.~\eqref{eq:boltz_force_stationary_ballistic}, 
since $\partial_x H = mg$ and $\partial_p H = p / m$.

The distribution function can be written as:
\begin{subequations}\label{eq:bal_f_gen}
\begin{equation}
 f(x, p) = \theta_+ f^+ + \theta_- f^-,
\end{equation}
where $(\theta_+, \theta_-)$ are Heaviside step functions which reduce to $(1, 0)$ and $(0, 1)$ 
on the left and right walls, respectively. The functions $f^\pm \equiv f^\pm(x, p)$ 
are completely determined by the diffuse reflection boundary conditions 
\eqref{eq:fwall}, together with the requirement that the Hamiltonian \eqref{eq:Hamiltonian}
an invariant of the Boltzmann equation:
\begin{align}
 f^+ =& \frac{n_+}{\sqrt{2\pi m T_+}} \exp\left\{-\frac{1}{T_+} \left[
 \frac{p^2}{2m} + V(x) - V(x_+)\right]\right\},\\
 f^- =& \frac{n_-}{\sqrt{2\pi m T_-}} \exp\left\{-\frac{1}{T_-} \left[
 \frac{p^2}{2m} + V(x) - V(x_-)\right]\right\},
\end{align}
\end{subequations}
where $n_\pm$ and $T_\pm$ represent the densities and temperatures of the walls.

In order to find the form of the arguments of $\theta_\pm$, 
we note that the potential $V(x) =mgx$ has a maximum 
on the right wall. For a given position $x$, three situations can occur: a) the particle 
has a negative momentum, such that the particle will travel to the left boundary; 
(b) the particle is travelling rightwards and its kinetic 
energy is larger than the potential difference between the right wall and its current location:
in this case, the particle will hit the right wall; (c) the momentum of the particle is positive, 
however the value of its Hamiltonian $H$ is smaller than the potential at the right boundary: this implies 
that the particle will reach a minimum approach with respect to the right wall, after which its 
momentum will become negative and its behaviour will be similar to the one described for case (a). 
This suggests that $\theta_+$ and $\theta_-$ can be written as:
\begin{equation}
 \theta_+ \equiv \theta(p + m\sqrt{q(L - 2x)}), \qquad 
 \theta_- \equiv \theta(-p - m\sqrt{g(L - 2x)}).
 \label{eq:thetapm}
\end{equation}
Indeed, substituting $\theta_+\equiv \theta (p \pm \sqrt{2m[a_+ - V(x)]})$ instead of $f$ in 
Eq.~\eqref{eq:boltz_force_stationary_ballistic} gives:
\begin{multline}
 \left(\frac{p}{m} \frac{\partial}{\partial x} - mg \frac{\partial}{\partial p}\right)
 \theta(\pm p \pm m \sqrt{g(L - 2x)}) \\
 = \mp \sqrt{\frac{g}{L - 2x}} \delta[\pm p \pm m\sqrt{g(L - 2x)}]
 [p + m \sqrt{g(L - 2x)}].
\end{multline}
It can be seen that the right hand side of the above 
equation vanishes due to the presence of the delta function. 

Thus, the full solution of the stationary Boltzmann equation in the ballistic regime 
subject to a force derived from the potential $V(x) = mgx$ is:
\begin{multline}
 f(x, p) = \theta[p + m\sqrt{g(L - 2x)}] \frac{n_+ e^{-mg(L + 2x) / 2T_+}}{\sqrt{2\pi m T_+}} e^{-p^2/2mT_+}\\
 +\theta[-p - m\sqrt{g(L - 2x)}] \frac{n_- e^{mg(L-2x) / 2T_-}}{\sqrt{2\pi m T_-}} e^{-p^2/2mT_-}.
 \label{eq:f_bal}
\end{multline}
The mass flux inside the channel is given by:
\begin{align}
 \int_{-\infty}^\infty dp\, f\, p = \sqrt{\frac{m}{2\pi}}
 \left(n_+ \sqrt{T_+} e^{-2mgL/T_+} - n_- \sqrt{T_-}\right),
\end{align}
which is constant, as expected from the continuity equation. In order to prevent mass transfer through the
boundaries, the constants $n_\pm$ must satisfy:
\begin{equation}
 n_- \sqrt{T_-} = n_+ e^{-mgL / T_+} \sqrt{T_+} .
 \label{eq:bal_npnm}
\end{equation}
The particle number density can be found by substituting Eq.~\eqref{eq:f_bal} into Eq.~\eqref{eq:fmoms}:
\begin{multline}
 n(x) = \frac{n_+}{2} e^{-mg(L + 2x) / 2T_+} \erfc\left(-\sqrt{\frac{mg(L - 2x)}{2T_+}}\right) \\
 + \frac{n_-}{2} e^{mg(L - 2x) / 2T_-} \erfc\left(\sqrt{\frac{mg(L - 2x)}{2T_-}}\right).
 \label{eq:bal_n_aux}
\end{multline}
The constant $n_+$ can be found by considering the total number of particles $\mathcal{N}$ inside the channel,
which is defined in Eq.~\eqref{eq:Ndef}. Using the following formula:
\begin{equation}
 \int_0^a dz\, e^z \erf\sqrt{z} = e^a \erf\sqrt{a} - 2 \sqrt{\frac{a}{\pi}},
\end{equation}
the total number of particles $\mathcal{N}$ can be expressed in terms of $n_\pm$:
\begin{multline}
 \mathcal{N} = \frac{n_+ T_+}{2mg} e^{-mgL/T_+} \left[e^{mgL/T_+}\erfc\left(-\sqrt{\frac{mgL}{T_+}}\right) - 1\right] \\
 + \frac{n_- T_-}{2mg} \left[e^{mgL/T_-} \erfc\left(\sqrt{\frac{mgL}{T_-}}\right) - 1\right].
\end{multline}
Using Eq.~\eqref{eq:bal_npnm}, the following expressions for $n_-$ and $n_+$ can be obtained:
\begin{multline}
 n_- \sqrt{T_-} = n_+ 
 \sqrt{T_+} e^{-mgL/T_+} \\
 = \frac{2mg \mathcal{N}}{\sqrt{T_+} \left[e^{mgL/T_+}\erfc\left(-\sqrt{\frac{mgL}{T_+}}\right) 
 - 1\right] 
 + \sqrt{T_-} \left[e^{mgL/T_-}\erfc\left(\sqrt{\frac{mgL}{T_-}}\right) - 
 1\right]}.
\end{multline}
Thus, the paricle number density can be written as:
\begin{equation}
 n(x) = mg \mathcal{N} 
 \frac{\frac{1}{\sqrt{T_+}} e^{\frac{mg(L - 2x)}{2T_+}}  \erfc\left(-\sqrt{\frac{mg(L - 2x)}{2T_+}}\right)
 + \frac{1}{\sqrt{T_-}}  e^{\frac{mg(L - 2x)}{2T_-}} \erfc\left(\sqrt{\frac{mg(L - 2x)}{2T_-}}\right)}
 {\sqrt{T_+} \left[e^{mgL/T_+}\erfc\left(-\sqrt{\frac{mgL}{T_+}}\right) 
 - 1\right] 
 + \sqrt{T_-} \left[e^{mgL/T_-}\erfc\left(\sqrt{\frac{mgL}{T_-}}\right) - 
 1\right]}.\label{eq:bal_n}
\end{equation}

An expression for the temperature can be obtained using Eq.~\eqref{eq:fmoms}:
\begin{multline}
 \frac{nT}{2} = n_+ e^{-mgL/T_+} e^{mg(L-2x)/2T_+} 
 \int_{-m\sqrt{g(L-2x)}}^\infty \frac{dp}{\sqrt{2\pi m T_+}}\,\frac{p^2}{2m}\,e^{-p^2/2mT_+} \\
 + n_- e^{mg(L-2x)/2T_-} 
 \int_{-\infty}^{-m\sqrt{g(L-2x)}} \frac{dp}{\sqrt{2\pi m T_-}}\,\frac{p^2}{2m}\,e^{-p^2/2mT_-}.
\end{multline}
Using the following property:
\begin{equation}
 \frac{2}{\sqrt{\pi}}\int_a^\infty dz\, z^2\, e^{-z^2} = \frac{a}{\sqrt{\pi}} e^{-a^2} + \frac{1}{2} \erfc(a),
\end{equation}
the final expression for $T(x)$ can be obtained:
\begin{equation}
 T(x) = \frac{\sqrt{T_+} e^{\frac{mg(L - 2x)}{2T_+}} \erfc\left(-\sqrt{\frac{mg(L - 2x)}{2T_+}}\right) + 
 \sqrt{T_-} e^{\frac{mg(L  -2x)}{2T_-}} \erfc\sqrt{\frac{mg(L - 2x)}{2T_-}}}
 {\frac{1}{\sqrt{T_+}} e^{\frac{mg(L - 2x)}{2T_+}}\erfc\left(-\sqrt{\frac{mg(L - 2x)}{2T_+}}\right) + 
 \frac{1}{\sqrt{T_-}} e^{\frac{mg(L  -2x)}{2T_-}} \erfc\sqrt{\frac{mg(L - 2x)}{2T_-}}}.
 \label{eq:bal_temp}
\end{equation}

The heat flux can be obtained using Eq.~\eqref{eq:fmoms}:
\begin{align}
 q(x) =& \frac{n_- \sqrt{T_-}}{\sqrt{2\pi m}} (T_+ - T_-) \nonumber\\
 =& \frac{g \mathcal{N}(T_+ - T_-) \sqrt{2m/\pi}}
 {\sqrt{T_+} \left[e^{\frac{mgL}{T_+}}\erfc\left(-\sqrt{\frac{mgL}{T_+}}\right) 
 - 1\right] 
 + \sqrt{T_-} \left[e^{\frac{mgL}{T_-}}\erfc\left(\sqrt{\frac{mgL}{T_-}}\right) - 
 1\right]}\label{eq:bal_q}
\end{align}

It can be checked that $n$ \eqref{eq:bal_n}, $T$ \eqref{eq:bal_temp} and
$q$ \eqref{eq:bal_q} satisfy Eqs.~\eqref{eq:macro}.
In the limit $T_- = T_+ = T_w$, $n$ reduces to Eq.~\eqref{eq:iso_n}, 
while $T(x) \rightarrow T_w$ and $q\rightarrow 0$.

\subsection{Ballistic regime: ${\rm HLB}$ vs ${\rm HHLB}$ }\label{sec:num:balvs}

\begin{figure}[!ht]
\begin{center}
\begin{tabular}{cc}
 \includegraphics[angle=270,width=0.45\columnwidth]{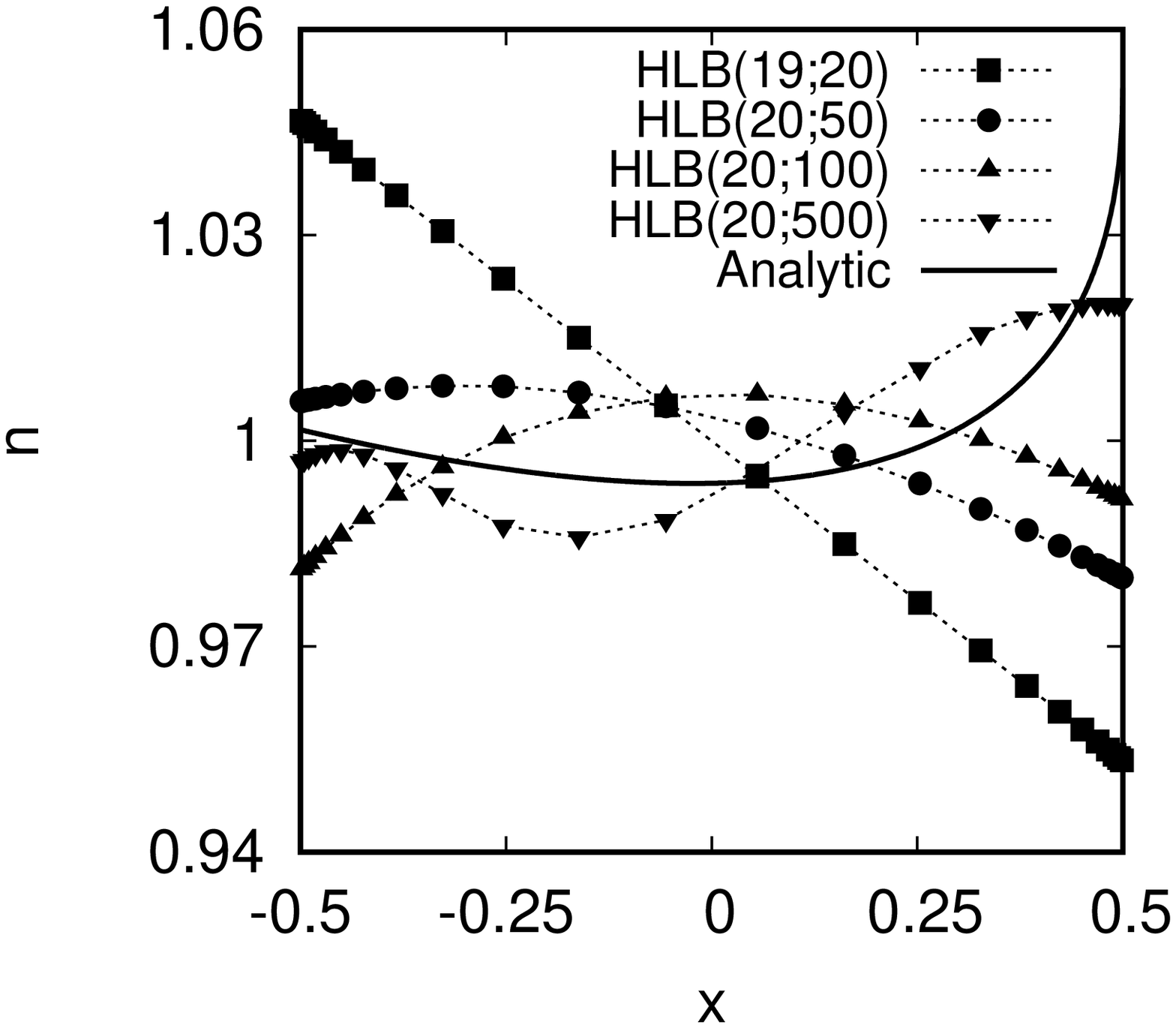} & 
 \includegraphics[angle=270,width=0.45\columnwidth]{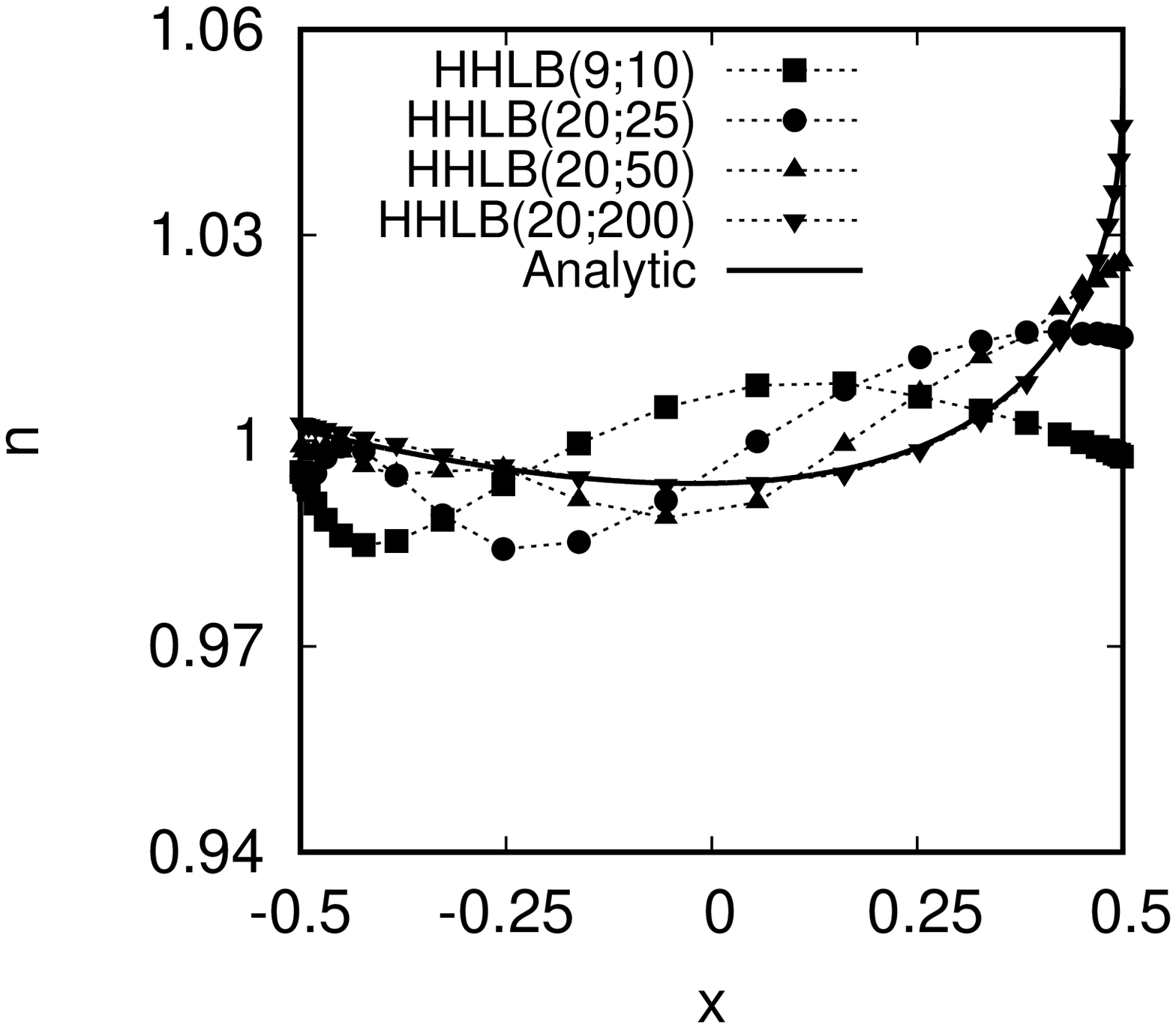} \\  
 (a) & (b) \\
 \includegraphics[angle=270,width=0.45\columnwidth]{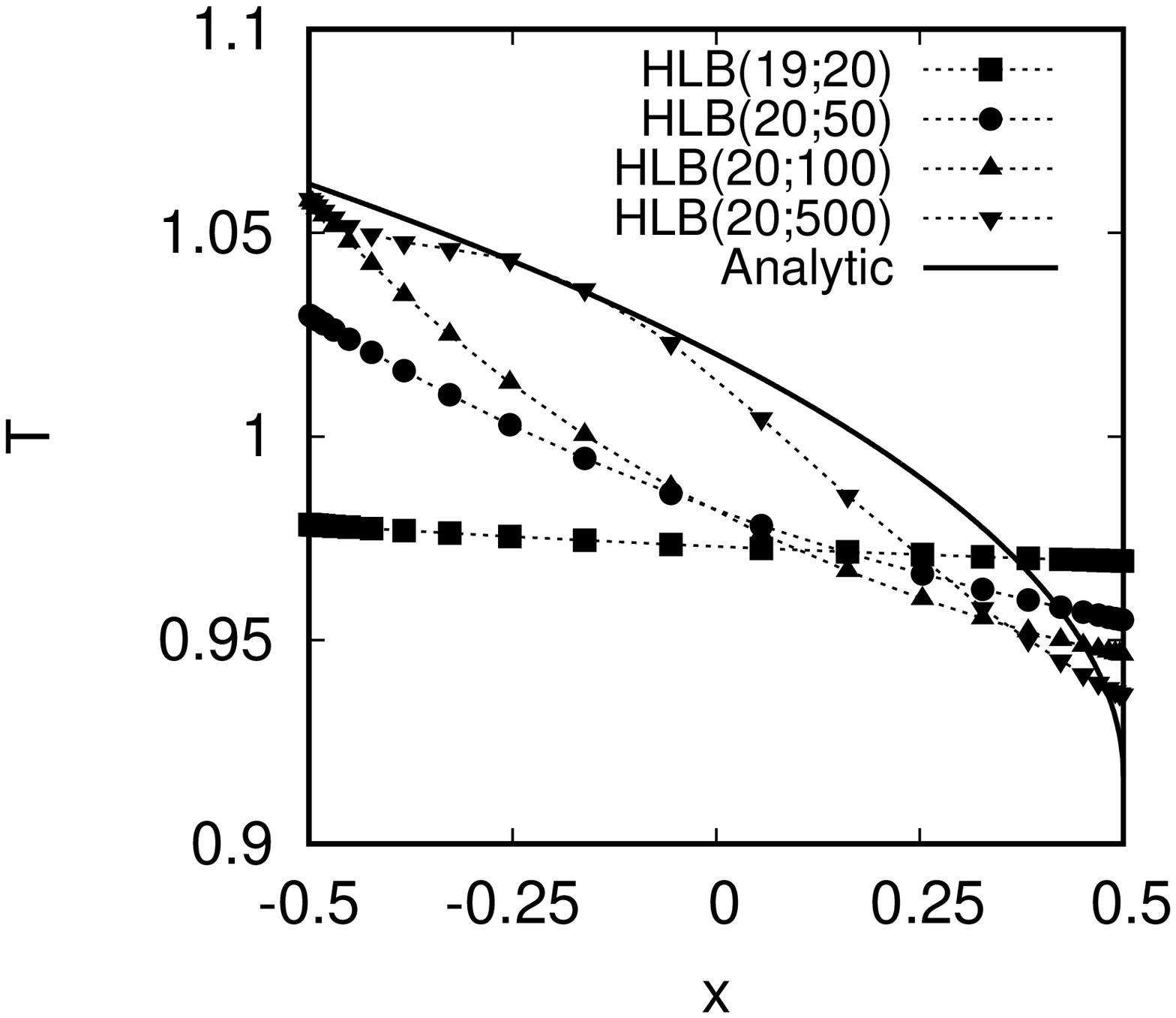} & 
 \includegraphics[angle=270,width=0.45\columnwidth]{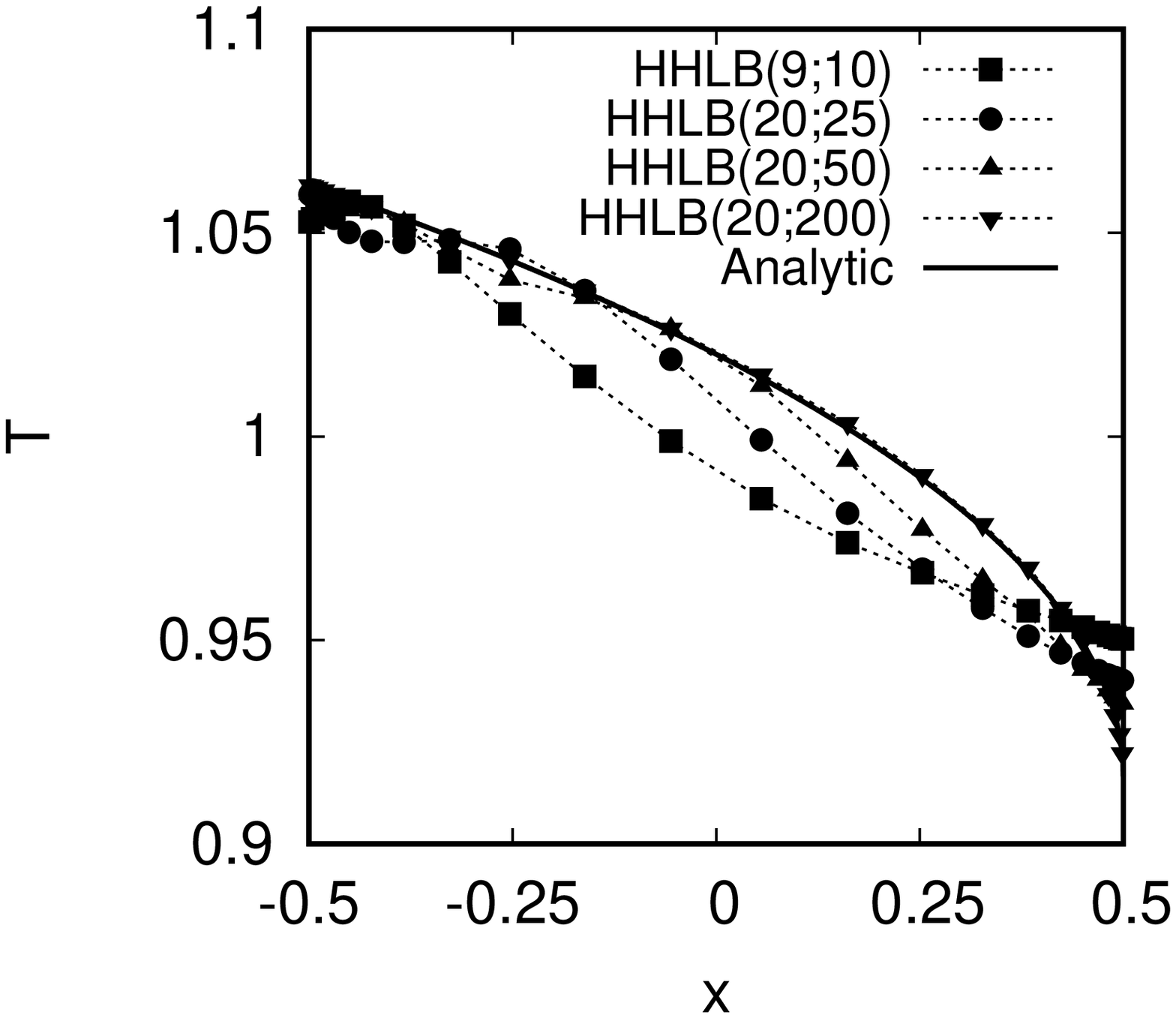} \\
 (c) & (d)
\end{tabular}
\end{center}
\caption{Density (top) and temperature (bottom) profiles in the ballistic regime
for $g = 1.0$ and $\Delta T = (T_- - T_+) / 2 = -0.4$, obtained using the 
${\rm HLB}$ (left) and ${\rm HHLB}$ (right) models.
The numerical results are shown using dashed lines and points, while 
the analytic solutions \eqref{eq:bal_n} and \eqref{eq:bal_temp}, 
are shown using continuous lines.
}
\label{fig:bal_vs}
\end{figure}

\begin{figure}[!ht]
\begin{center}
\begin{tabular}{cc}
 \includegraphics[angle=270,width=0.45\columnwidth]{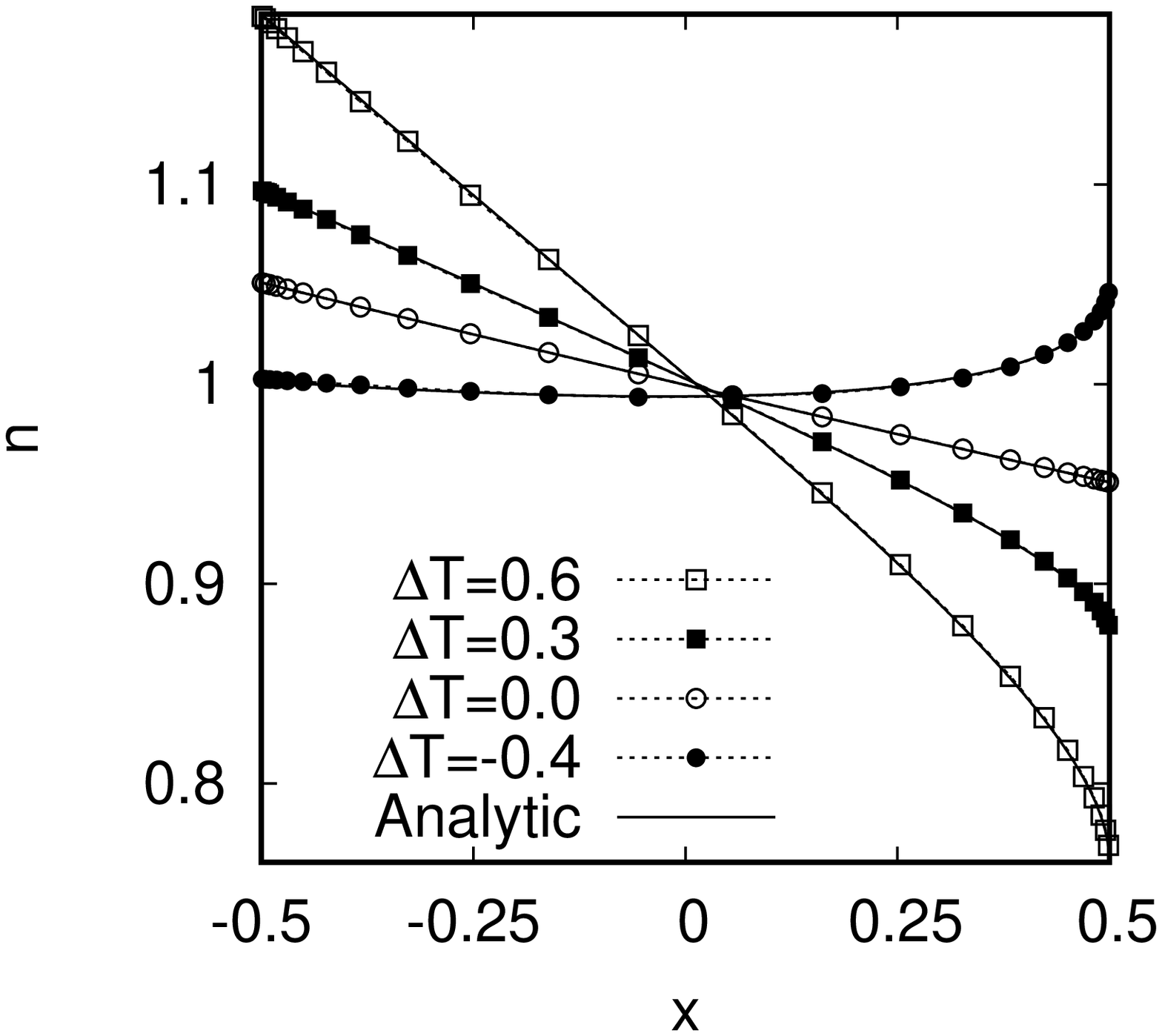} & 
 \includegraphics[angle=270,width=0.45\columnwidth]{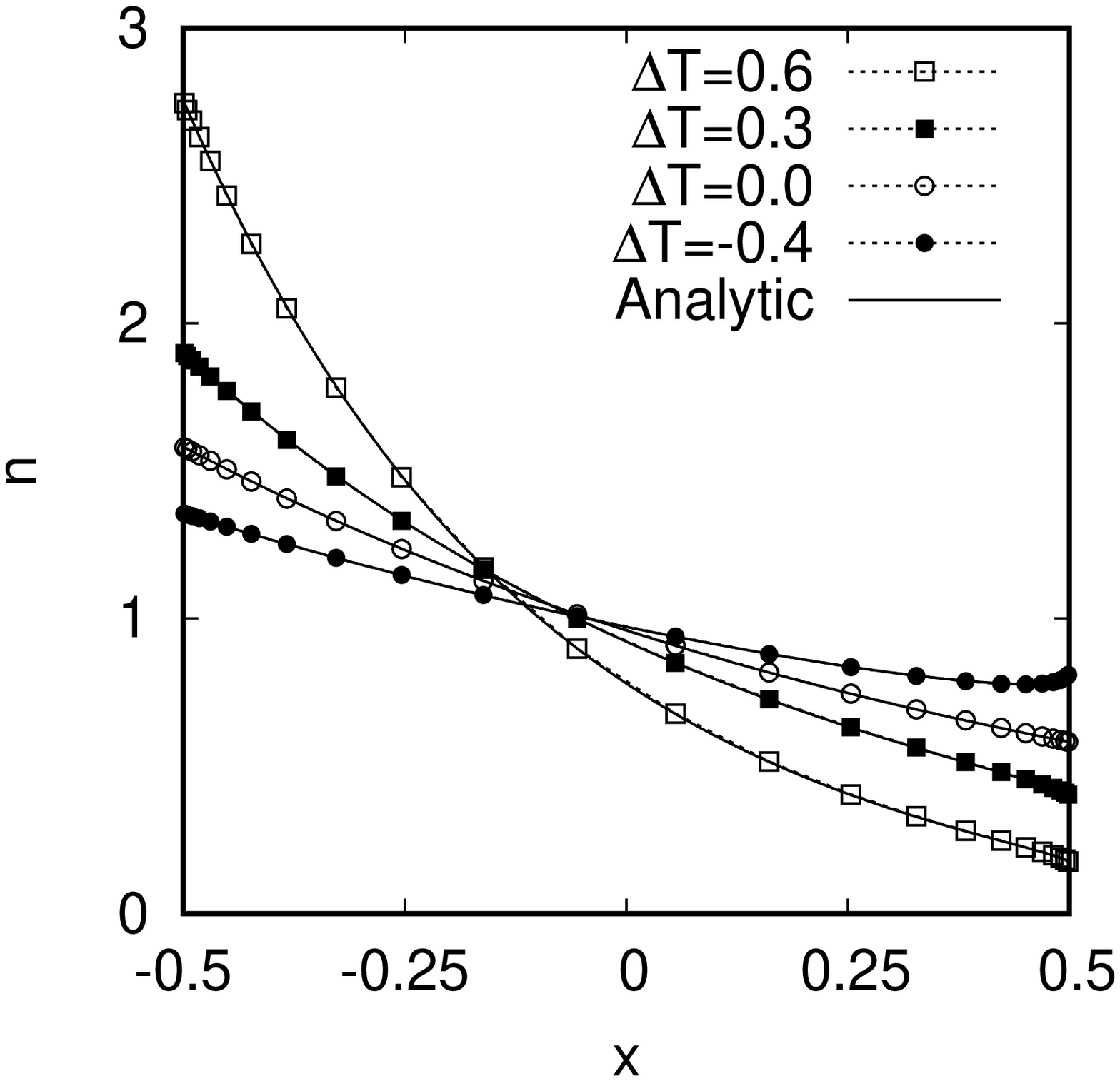}\\
 (a) & (b) \\
 \includegraphics[angle=270,width=0.45\columnwidth]{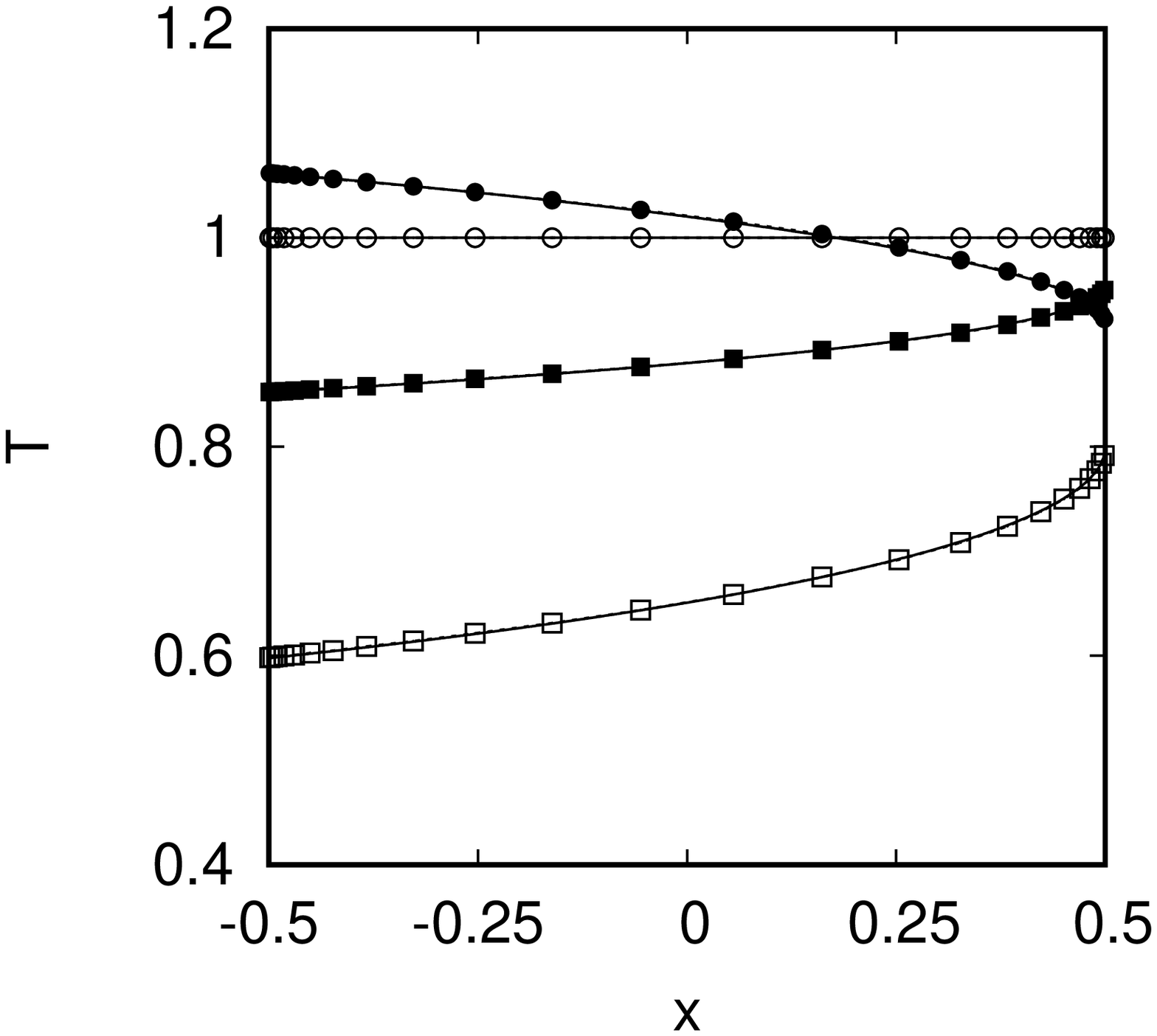} & 
 \includegraphics[angle=270,width=0.45\columnwidth]{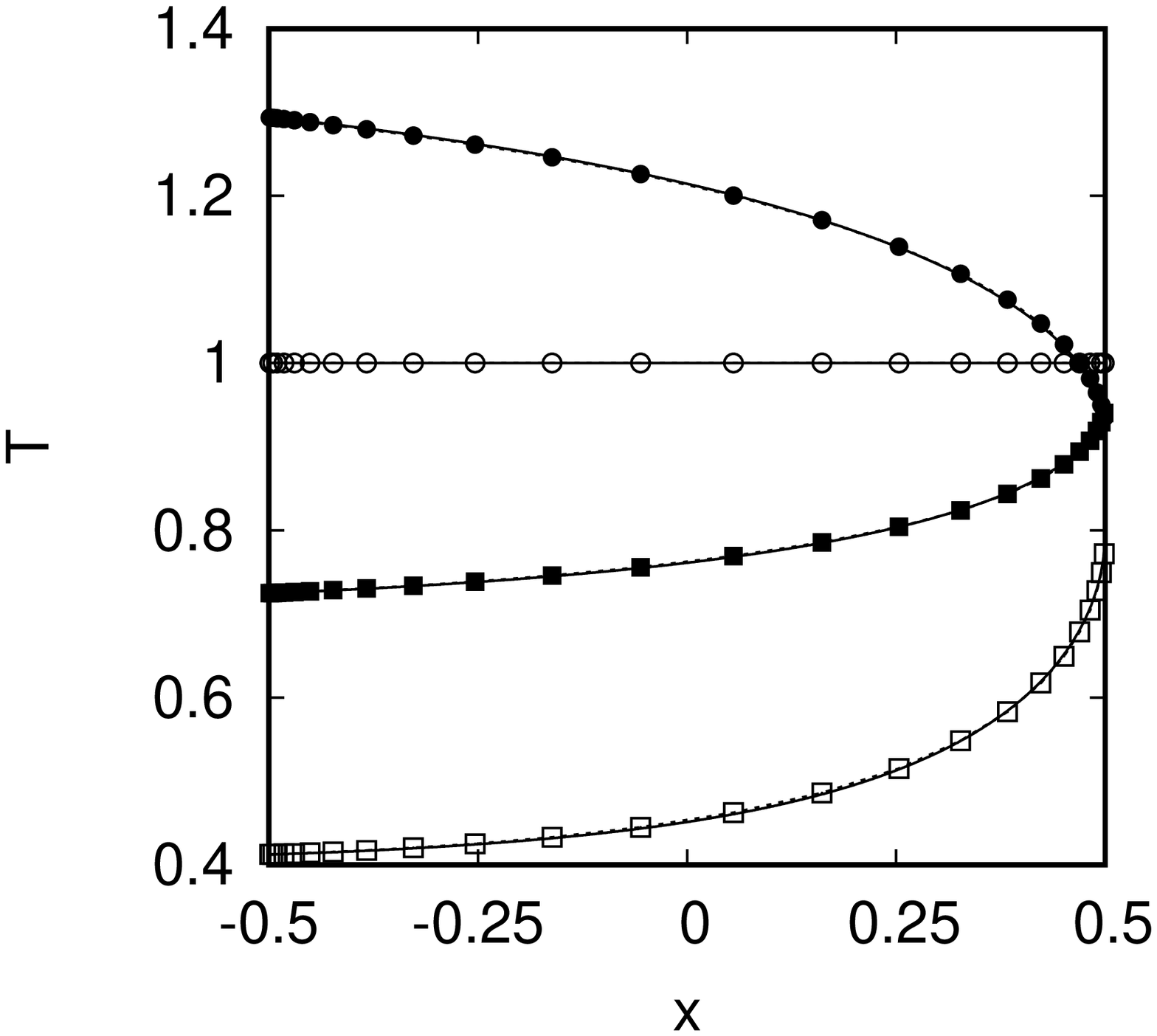}\\
 (c) & (d) \\
\end{tabular}
\end{center}
\caption{Density (top) and temperature (bottom) profiles in the ballistic regime
for $g = 0.1$ (left) and $g = 1.0$ (right). The curves correspond to various 
values of $\Delta T = (T_- - T_+) / 2$.
The numerical results are shown using dashed lines and points, being in excellent
agreement with the analytic solutions \eqref{eq:bal_n} and \eqref{eq:bal_temp}, 
which are shown using continuous lines. 
}
\label{fig:bal_hh}
\end{figure}

We now set $g = 0.1$ and $\Delta T = (T_- - T_+) / 2 = -0.4$. Because the left 
wall has a higher temperature than the right wall (i.e. $T_+ = 1.4$ compared to $T_- = 0.6$),
the gas particles tend to accumulate on the right wall. On the other hand, the gravitational force 
($g = 0.1$) acts from right to left, inducing an accumulation of particles on the 
left wall. The result is a non-trivial particle distribution having a minimum in the central region of 
the channel and local maxima in the vicinity of the walls. Our numerical results for the particle number 
density $n$ and temperature $T$ are compared in Fig.~\ref{fig:bal_vs}
with the corresponding analytic expressions \eqref{eq:bal_n} 
and \eqref{eq:bal_temp}, respectively. Throughout this section,
all simulation results presented were obtained using a time step of $\delta t = 10^{-4}$ 
on a grid of $N_x = 24$ points. The stretching parameter in Eq.~\eqref{eq:x_eta} is $A = 0.99$.

The plots on the left of Fig.~\ref{fig:bal_vs} show the numerical results obtained using the 
full-range Gauss-Hermite models ${\rm HLB}(N;Q)$ for 
$Q \in \{20, 50, 100, 500\}$ and $N = {\rm min}(20, Q - 1)$. At low quadrature orders, the density 
profile obtained using the ${\rm HLB}$ models decreases monotonically from the left to the right wall,
in qualitative contrast to the analytic prediction. At $Q = 500$, the density profile obtained using the 
${\rm HLB}$ model is closer to the analytic prediction, but it presents non-physical oscillations,
such that its shape is still qualitatively incorrect. 
The temperature profile decreases monotonically from the 
left wall to the right wall for all quadrature orders, as required analytically. 
The temperature on the walls slowly approaches the analytic value from below (above) 
on the left (right) wall, indicating that the temperature jump is overestimated at low quadrature orders. 
The temperature profile in the bulk still does not approach the form of the analytic profile even when $Q = 500$. 

The numerical results obtained in the same condition as described above using the half-range 
Gauss-Hermite models ${\rm HHLB}(N;Q)$ (employing a number of $\mathcal{Q} = 2Q$ velocities) with 
$Q \in \{10, 25, 50, 200\}$ and $N = {\rm min}(20, Q - 1)$ are presented in 
the plots on the right in Fig.~\ref{fig:bal_vs}. While even at low quadrature orders, the results obtained 
using the ${\rm HHLB}$ models are closer to the analytic prediction than those obtained using the ${\rm HLB}$ 
models, the resulting profiles exhibit unphysical oscillations. The quadrature order at 
which these oscillations disappear seems to depend on the grid size, as will be demonstrated later. 
It is important to note that when $Q = 200$, the agreement between the numerical and the analytic results is 
excellent.

The plots in Fig.~\ref{fig:bal_hh} show a comparison between our simulation results obtained 
using the ${\rm HHLB}(20; 200)$ model and the analytic profiles in Eqs.~\eqref{eq:bal_n} and 
\eqref{eq:bal_temp} for $g = 0.1$ (left) and $g = 1$ (right) at 
$\Delta T = (T_- - T_+) / 2 \in \{0.6, 0.3, 0.0, -0.4\}$. It can be seen that the 
agreement between our numerical solutions and the analytic profiles is excellent for all tested 
values of $g$ and $\Delta T$. A more quantitative analysis of the accuracy of our method is 
discussed in the following subsection.

\subsection{Ballistic regime: convergence tests}\label{sec:num:balconv}

\begin{figure}[!ht]
\begin{center}
\begin{tabular}{cc}
 \includegraphics[angle=270,width=0.45\columnwidth]{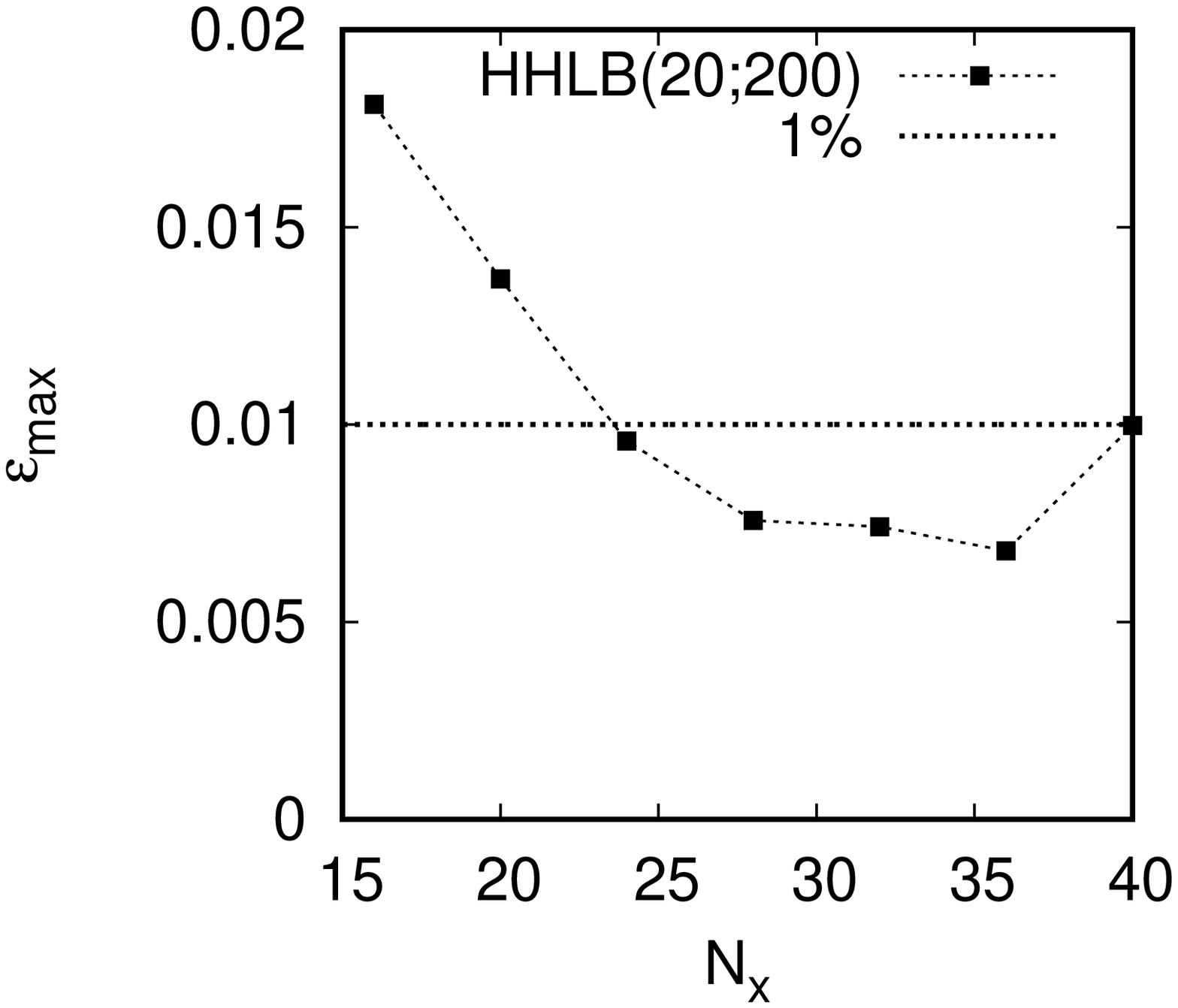}  &
 \includegraphics[angle=270,width=0.45\columnwidth]{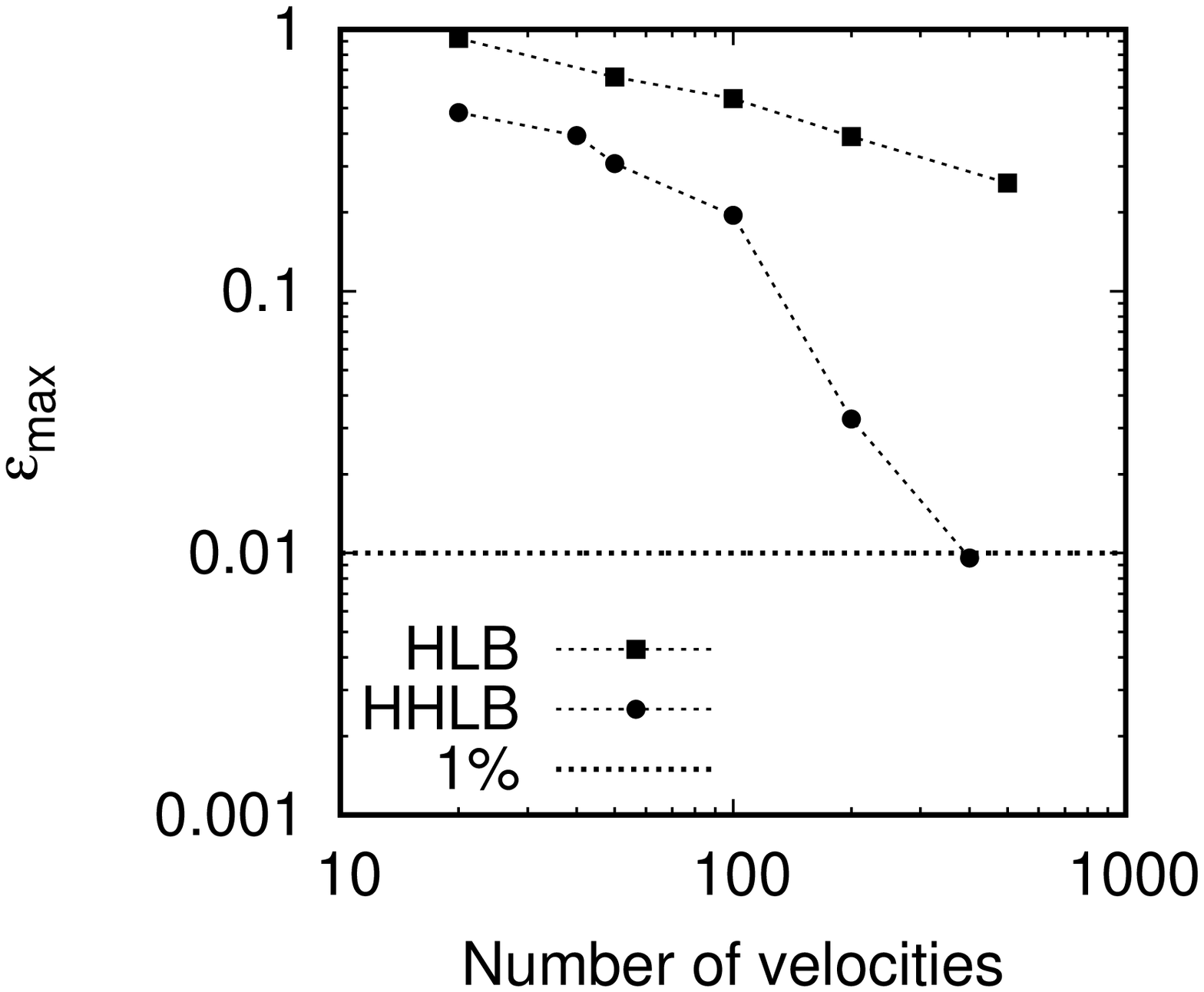} \\
 (a) & (b)
\end{tabular}
\end{center}
\caption{Relative error $\varepsilon_{\rm max}$ \eqref{eq:epsmax}
with respect to the analytic profiles in the free-streaming regime of: 
(a) the results obtained with the ${\rm HHLB}(20;200)$ model 
as a function of the number of grid points $N_x$;
(b) the results obtained using the models ${\rm HLB}(N; Q)$ and 
${\rm HHLB}(N; Q)$ at $N_x = 24$ as a function of the number of 
velocities ($Q$ for the ${\rm HLB}$ models and $2Q$ for the ${\rm HHLB}$ models).
}
\label{fig:bal_err}
\end{figure}

To gain some quantitative insight on the convergence properties of our LB models, we follow 
Refs.~\onlinecite{ambrus16jcp,ambrus16jocs} and perform the convergence test described below.
First, we fix as reference profiles $M_{\rm ref} \in \{n_{\rm ref}, T_{\rm ref}\}$
the profiles corresponding to the analytic solutions \eqref{eq:bal_n} and 
\eqref{eq:bal_temp}, evaluated at the points corresponding to our chosen grid.
The convergence of our models is tested by analysing the relative error 
between the profiles obtained using our models and the reference profiles.
We introduce for a given profile $M \in \{n, T\}$ the error $\varepsilon(M)$ through:
\begin{equation}
 \varepsilon(M) = \frac{\max_x [M(x) - M_{\rm ref}(x)]}{\max\{\max_x[M(x)] - \min_x[M(x)], 0.1\}}.
\end{equation}
where the denominator measures the spread of $M(x)$ (i.e. the difference between the maximum and the minimum 
values of $M(x)$), bounded from below by $0.1$. 
In this paper, we consider that convergence is achieved when the error $\varepsilon(M)$ is less than $1\%$ for 
all $M \in \{n, T\}$, i.e.:
\begin{equation}
 \varepsilon_{\rm max} \equiv \max_{M \in \{n, T\}} \varepsilon(M) < 0.01.
 \label{eq:epsmax}
\end{equation}

Figure~\ref{fig:bal_err}(a) shows that the error $\varepsilon_{\rm max}$ decreases with the number of 
grid points $N_x$ for $N_x \lesssim 36$. The increase in $\varepsilon_{\rm max}$ as $N_x$ is further 
increased is due to the development of spurious oscillations which seem to increase in amplitude as 
$N_x$ is increased at fixed values of $Q$. In Fig.~\ref{fig:bal_err}(b), 
we fix $N_x = 24$ and study the dependence of $\varepsilon_{\rm max}$ on 
the quadrature order $Q$ for the ${\rm HLB}(N; Q)$ and the ${\rm HHLB}(N; Q)$ models, where 
$N = \min(20, Q - 1)$. 
It can be seen that the error in the ${\rm HLB}$ models decreases at a much slower rate than the 
error corresponding to the ${\rm HHLB}$ models.

\begin{figure}[!ht]
\begin{center}
\begin{tabular}{cc}
 \includegraphics[angle=270,width=0.45\columnwidth]{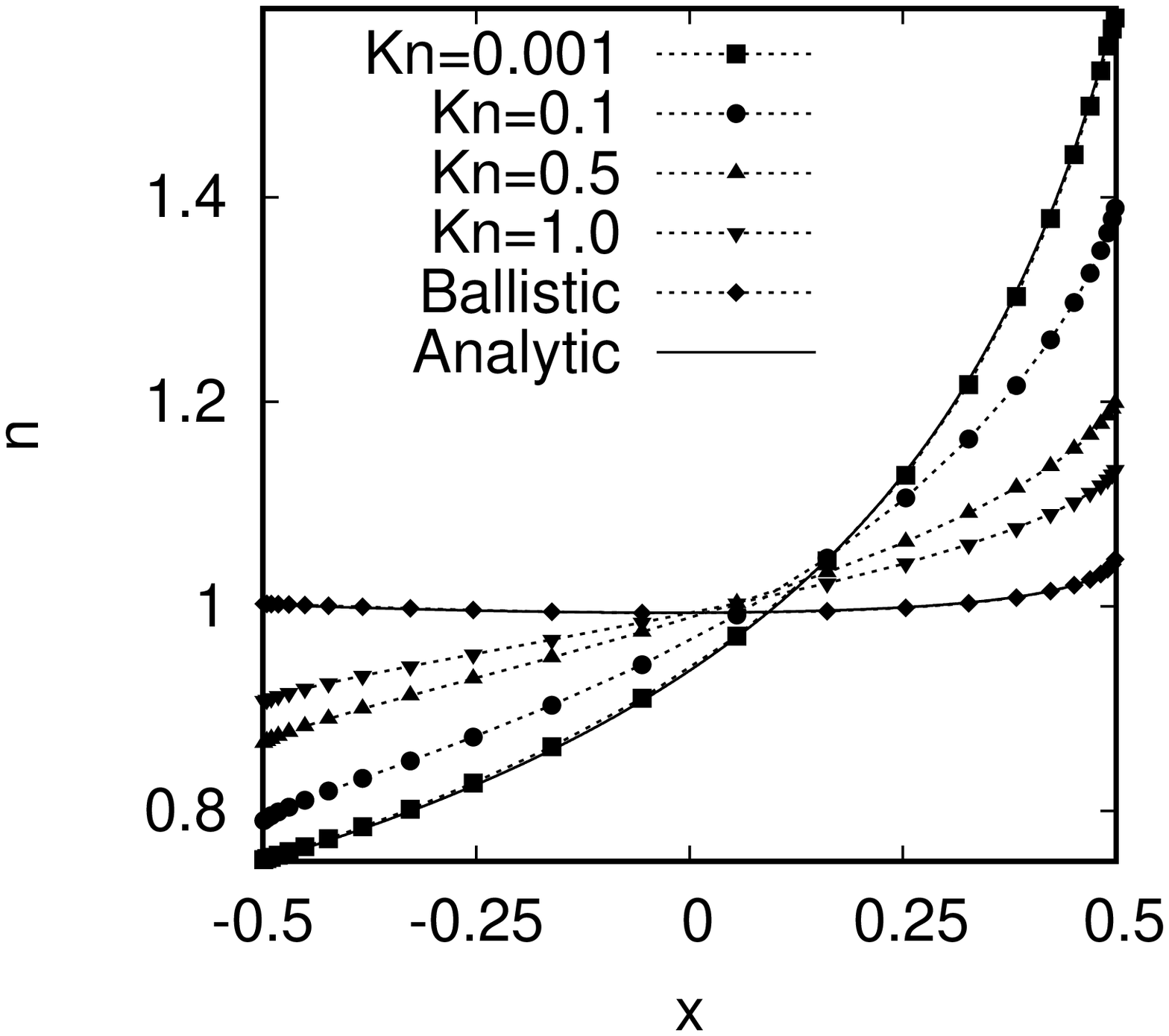} &
 \includegraphics[angle=270,width=0.45\columnwidth]{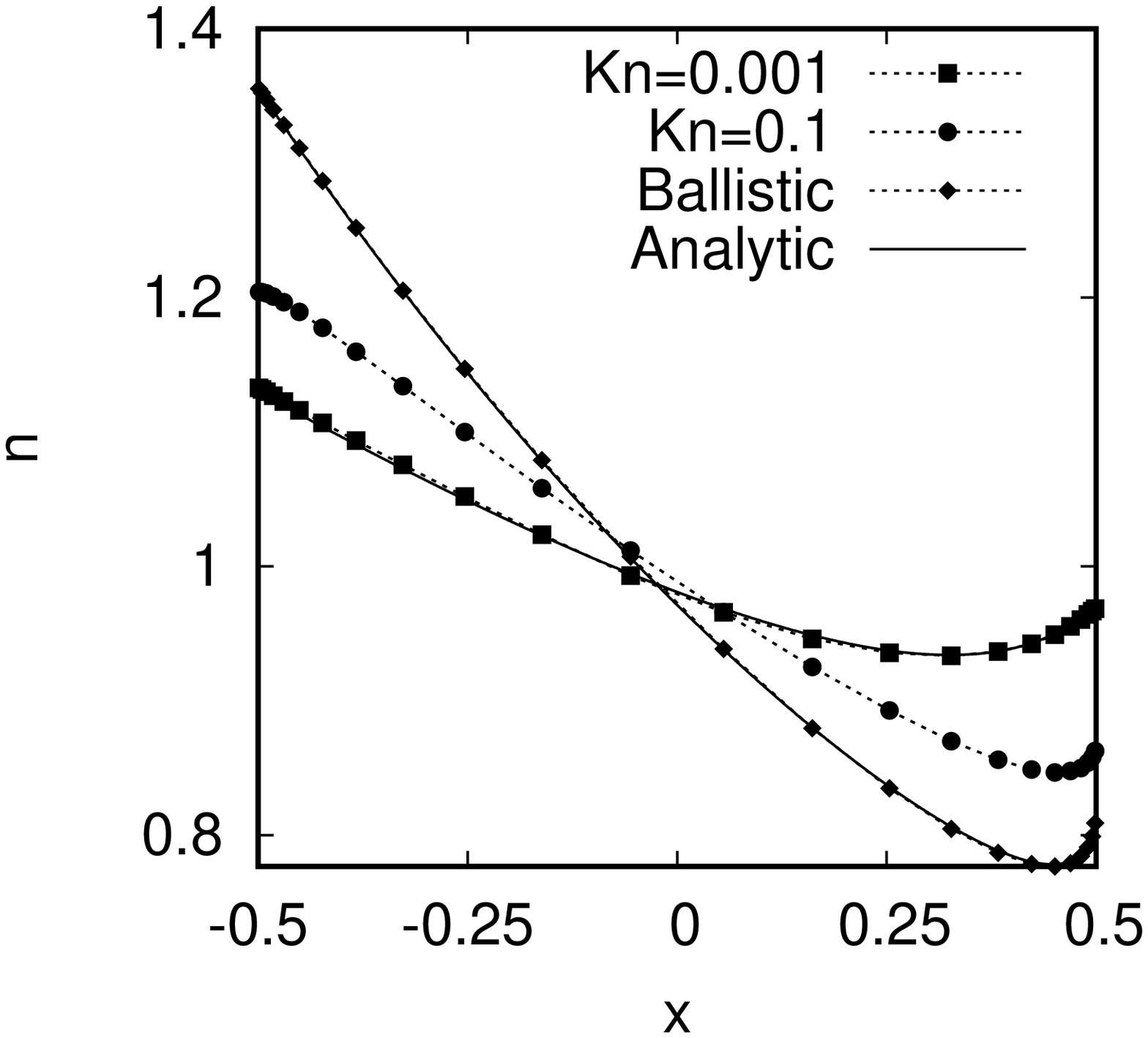} \\ 
 (a) & (b) \\
 \includegraphics[angle=270,width=0.45\columnwidth]{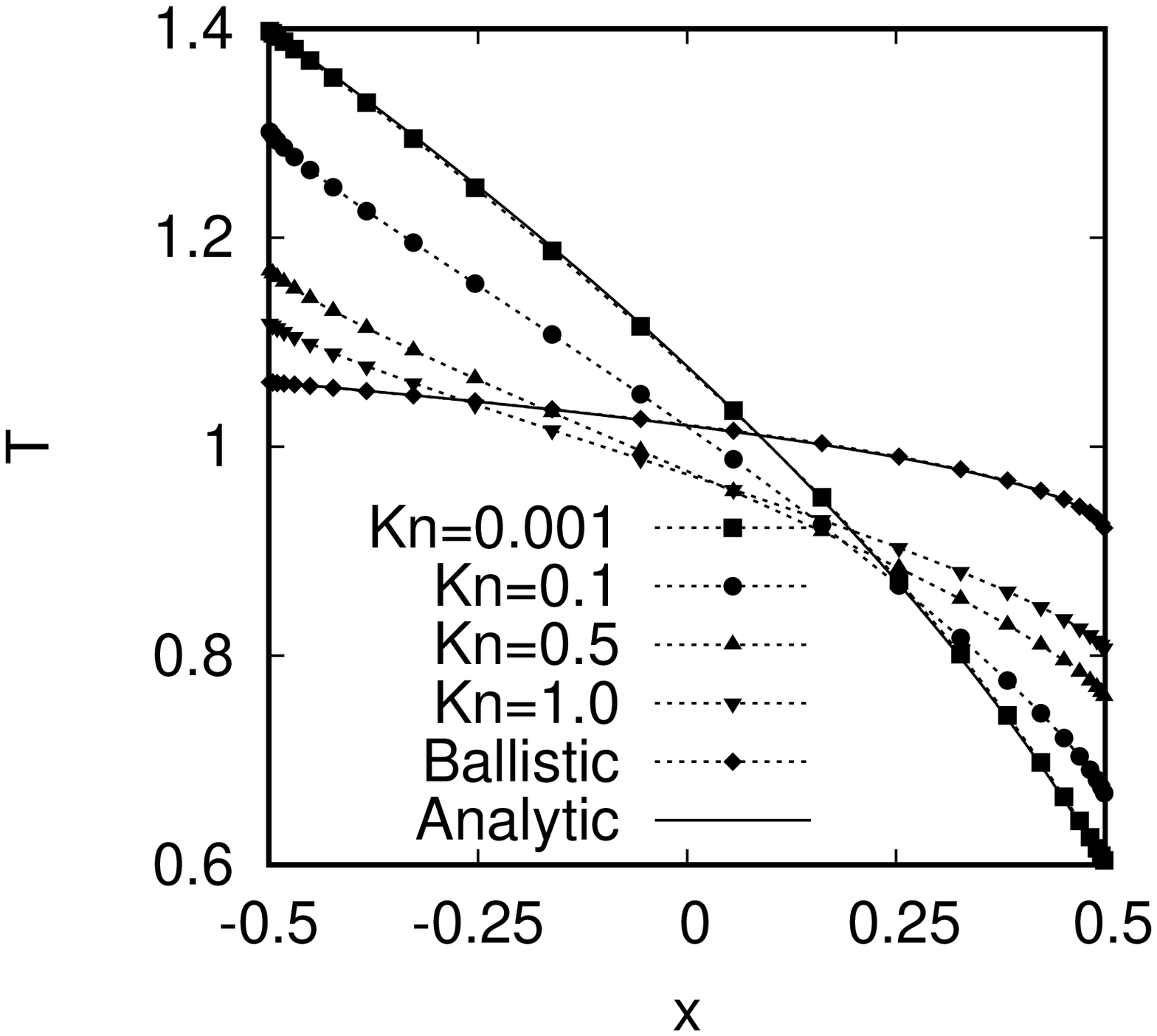} &
 \includegraphics[angle=270,width=0.45\columnwidth]{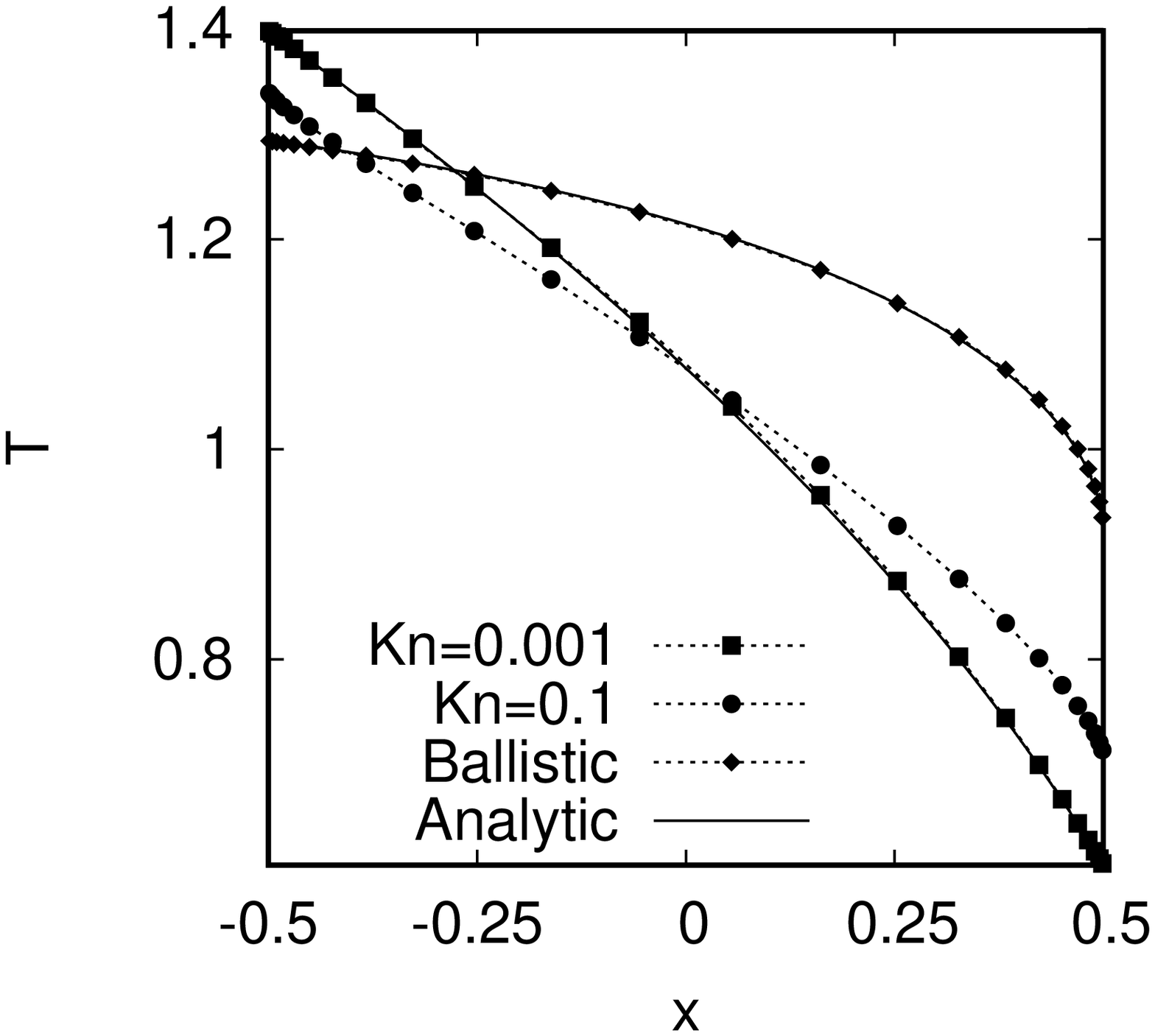} \\ 
 (c) & (d)
\end{tabular}
\caption{Density (top) and temperature (bottom) profiles obtained using our LB models (dotted lines and points) when 
$g = 0.1$ (left) and $g = 1.0$ (right). The analytic formulae 
represented using contiguous lines are given by 
Eqs.~\eqref{eq:NS_n} ($n$) and \eqref{eq:NS_temp} ($T$) in the hydrodynamic regime and 
by Eqs.~\eqref{eq:bal_n} ($n$) and \eqref{eq:bal_temp} ($T$) in the ballistic regime.
}
\label{fig:trans}
\end{center}
\end{figure}

\subsection{Transition regime}\label{sec:num:visc}

In Subsec.~\ref{sec:num:NS} we validated our implementation of the force term by comparing our numerical results 
for the particle number density $n$ and temperature $T$ in the Navier-Stokes regime with the analytic results 
\eqref{eq:NS_n} and \eqref{eq:NS_temp}, respectively. Next, we have shown in Subsec.~\ref{sec:num:balconv} 
that the ${\rm HHLB}(N; Q)$ models based on the half-range Gauss-Hermite quadrature
can reproduce, at large enough values of the quadrature $Q$, the analytic solutions \eqref{eq:bal_n} and 
\eqref{eq:bal_temp} for $n$ and $T$ in the ballistic regime. At the same time, we have shown that in 
the ballistic regime, the ${\rm HLB}(N; Q)$ based on the full-range Gauss-Hermite models exhibit a slow convergence 
towards the analytic solutions, with the error obtained using the test 
introduced in Refs.~\onlinecite{ambrus16jcp,ambrus16jocs} and described in Eq.~\eqref{eq:epsmax} being 
$\varepsilon_{\rm max} \simeq 25.9\%$ with respect to the analytic solution even at $Q = 500$. 

In this section, we attempt to analyse the ability of the ${\rm HLB}$ and ${\rm HHLB}$ models 
to simulate flows in the transition regime. In the absence of an analytic solution of 
the Boltzmann-BGK equation \eqref{eq:boltz} or of results obtained using other methods, 
we will limit the analysis to employing the convergence test introduced in Subsec.~\ref{sec:num:balconv},
but in this case with respect to the reference profiles obtained using the ${\rm HHLB}(20;200)$ model. 
According to Fig.~\ref{fig:bal_err}, this model produces profiles which are within the $1\%$ error limit 
of the analytic solution in the ballistic regime. 

Figure~\ref{fig:trans} shows the numerical results obtained using our models at various 
values of the Knudsen number ${\rm Kn}$. At ${\rm Kn} = 0.001$, the numerical results obtained using 
the ${\rm HLB}(4;5)$ model are compared with the analytic solutions derived in Subsec.~\ref{sec:num:NS}.
In the ballistic regime, the numerical results obtained using the ${\rm HHLB}(20; 200)$ model 
are compared with the analytic solutions derived in Subsec.~\ref{sec:num:bal}. In the transition regime
${\rm Kn} \in \{0.1, 0.5, 1\}$, the numerical 
results represented in Fig.~\ref{fig:trans} were obtained using the ${\rm HHLB}(20; 200)$ model. 

\begin{figure}
\begin{center}
\begin{tabular}{cc}
 \includegraphics[angle=270,width=0.45\columnwidth]{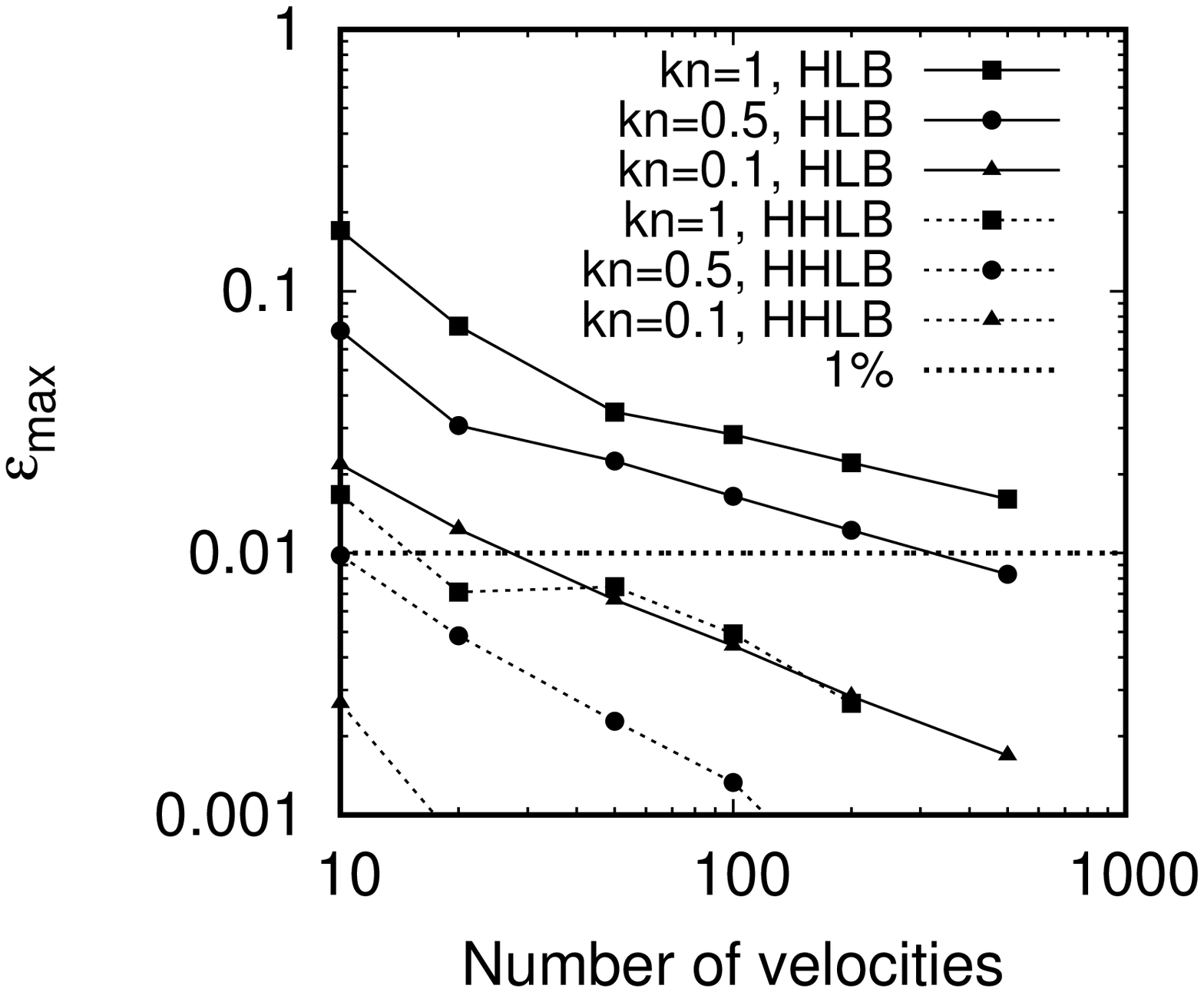} &
 \includegraphics[angle=270,width=0.45\columnwidth]{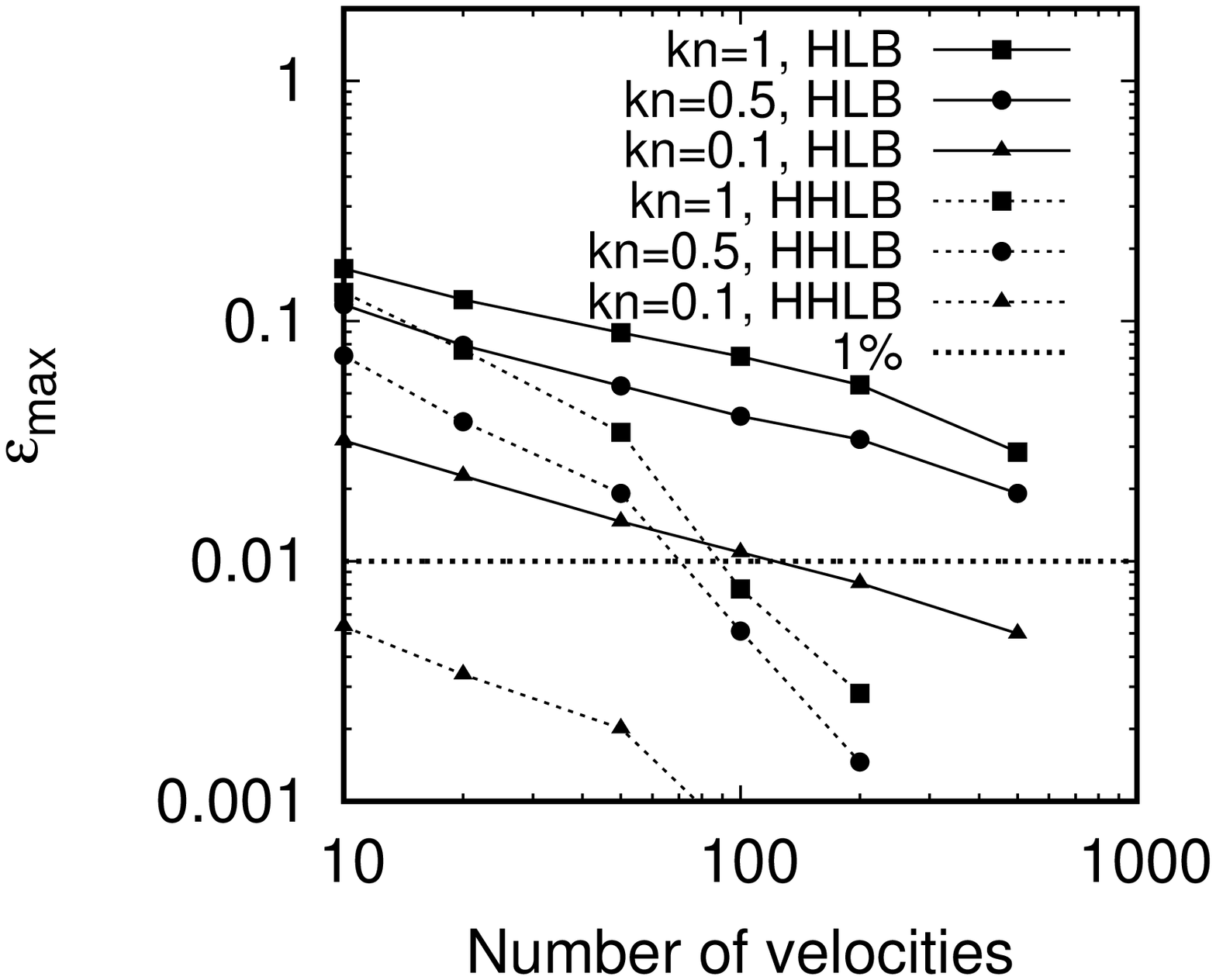} \\ 
 (a) & (b) 
\end{tabular}
\caption{Total error $\varepsilon_{\rm max}$ \eqref{eq:epsmax} computed with respect to the 
reference profiles obtained using the ${\rm HHLB}(20; 200)$ model for ${\rm Kn} = 0.5$ (dotted lines with points) 
and ${\rm Kn} = 1.0$ (continuous lines with points)
when $g = 0.1$ (left) and $g = 1.0)$ (right). The error $\varepsilon_{\rm max}$ is shown as a function of 
the number of velocities $\mathcal{Q} = Q$ and $2Q$ for the ${\rm HLB}(N; Q)$ and 
${\rm HHLB}(N; Q)$ models, respectively.
}
\label{fig:trans_err}
\end{center}
\end{figure}

Taking as reference the profiles obtained using the ${\rm HHLB}(20; 200)$ model at $N_x = 24$ nodes and 
a time step $\delta t = 10^{-4}$, we test the convergence of the ${\rm HLB}(N; Q)$ and ${\rm HHLB}(N; Q)$ 
models in the transition regime as $Q$ is increased, with $N = {\rm min}(Q - 1, 20)$. 
Figure~\ref{fig:trans_err} shows the 
error $\varepsilon_{\rm max}$ \eqref{eq:epsmax} computed with respect to the reference profiles.
It can be seen that the results obtained using the ${\rm HLB}(N; Q)$ models unambiguously converge towards 
the reference profiles, but the error at a fixed number of velocities is always larger than when the 
${\rm HHLB}(N; Q)$ models are considered. Out of the six convergence tests that we conducted, the 
${\rm HLB}(N; Q)$ models managed to satisfy the $1\%$ convergence test only in three cases, namely 
$(g, {\rm Kn}) \in \{(0.1, 0.1), (0.1, 0.5), (1, 0.1)\}$, failing when 
$(g, {\rm Kn}) \in \{(0.1, 1), (1, 0.5), (1, 1)\}$ for all tested $Q \le 500$.
It can be seen that at fixed ${\rm Kn}$, the convergence of the ${\rm HLB}$ and ${\rm HHLB}$ models 
is slower as $g$ is increased.

\section{Conclusion}\label{sec:conc}

In this paper, we have presented a systematic procedure for the construction of lattice Boltzmann models
based on half-range quadratures for force-driven flows in confined geometries. In order to ensure the 
correct recovery of the zeroth order half-range moment of the momentum derivative $\partial_p f$ of the 
distribution function, we have shown that the theory of distributions must be employed in order to 
cope with the potential discontinuity of the distribution function $f$ at $p = 0$. 

In order to validate our proposed scheme, we have considered the problem of a gas under the action of 
gravity (i.e.~$V(x) = mgx$, where $g$ is the constant gravitational acceleration) between two 
diffuse-reflecting walls. 
We validated our models using analytic solutions in three regimes: in the isothermal regime (i.e.~the 
walls have equal temperature); in the Navier-Stokes regime; and in the ballistic regime. 
In order for our models to accurately reproduce the analytic solutions, we employ a grid stretching 
algorithm which allows the mesh to be refined near the boundaries, where the gradients of the macroscopic 
fields are in general steeper. In all cases, our simulation results reproduce to high accuracy the analytic 
predictions. 

While in the isothermal and Navier-Stokes regimes, the models based on the full-range Gauss-Hermite
quadrature are sufficiently accurate to reproduce the analytic solution, as the value of ${\rm Kn}$ 
is increased, half-range capabilities are requried in order to deal with the discontinuity 
of the distribution function induced by the boundaries. In the ballistic regime, we were able to reduce 
the error with respect to the analytic solution to less than $1\%$ on a grid consisting of $N_x = 24$ nodes 
by using the 
${\rm HHLB}(20; 200)$ model, which employs an expansion of order $N = 20$ of the equilibrium distribution 
function $\feq$ and a quadrature order $Q = 200$. Our convergence test showed that the 
error of the ${\rm HLB}(N; Q)$ models based on the full-range Gauss-Hermite quadrature compared to the 
analytic result was over $25\%$ even when $Q = 500$.

We further performed an analysis of the properties of our models in the transition regime,
at ${\rm Kn} = 0.5$ and ${\rm Kn} = 1$. Since no analytic solution was available in this regime,
we performed a convergence study by taking the results obtained using the ${\rm HHLB}(20;200)$ model
as reference profiles. Our analysis showed that the error $\varepsilon_{\rm max}$ in the profiles obtained 
using the ${\rm HLB}(N; Q)$ models with respect to the reference profiles decreases monotonically as 
$Q$ is increased. The value of $\varepsilon_{\rm max}$ stays larger than $1\%$ for ${\rm Kn} \in \{0.5,1\}$ 
when $g = 1$ for all $Q \le 500$, while when $g = 0.1$, we were able to reduce $\varepsilon_{\rm max}$ under 
$1\%$ at $Q = 500$ only for ${\rm Kn} = 0.5$.

In this paper, we have only considered the academic example of a constant force acting between two
walls which have different temperatures. However, the purpose of this paper is to introduce the formalism 
necessary to implement the momentum derivative of the distribution function when half-range quadratures are 
employed. Furthermore, the example considered in this paper was taken for simplicity to be one-dimensional. 
While we highlight in Appendix~\ref{app:multi} that the methodology is easily extendible to higher dimensions,
we defer further validation tests for future publications. A direct application of the formalism introduced 
in this paper can be found in Ref.~\onlinecite{busuioc17}, where the circular Couette flow between concentric 
cylinders is considered.

\begin{acknowledgments} 
This work was supported by a grant of the Romanian National Authority
for Scientific Research, CNCS-UEFISCDI, project number
PN-II-ID-JRP-2011-2-0060 and
by the ANR DEPART project, Grant
ANR-12-IS04-0003-01 of the French Agence Nationale de la
Recherche.
\end{acknowledgments}

\appendix

\section{Extension to multiple dimensions}\label{app:multi}

In $d$ dimensions, the Boltzmann equation \eqref{eq:boltz} becomes:
\begin{equation}
 \partial_t f + \frac{\vp}{m} \cdot \nabla f + \bm{F} \cdot (\nabla_{\vp} f) = -\frac{1}{\tau} (f - \feq).
 \label{eq:boltz_multi}
\end{equation}
The numerical solution of Eq.~\eqref{eq:boltz_multi} can be obtained following the same steps described in 
Sec.~\ref{sec:LB}, which we summarise in what follows. We refer to the resulting models as mixed quadrature 
lattice Boltzmann models, following Refs.~\onlinecite{ambrus16jcp,ambrus16jocs}. For definiteness, we will always
refer to the case $d = 3$. Since there is no direct generalisation of one-dimensional 
advection schemes, we will restrict the discussion in this appendix to the discretisation of the 
momentum space (Sec.~\ref{app:multi:discrete}), the construction of $\feq$ (Sec.~\ref{app:multi:feq})
and the construction of the force term (Sec.~\ref{app:multi:force}). 

\subsection{Discretisation of the momentum space}\label{app:multi:discrete}

In the mixed quadrature lattice Boltzmann models, the momentum space is constructed using a direct product rule.
This allows the quadrature on each axis to be constructed independently by taking into account the 
characteristics of the flow, as employed, e.g., in 
for the 2D Couette \cite{ambrus16jcp}, 2D Poiseuille \cite{ambrus16jocs} and 
3D Couette \cite{ambrus17couette} flows.

Since in this paper we propose a Cartesian split of the momentum space, the elements of the 
discrete set of momentum vectors can be written as $\vp_{ijk} = (p_{x,i}, p_{y,j}, p_{z,k})$. 
The indices $i$, $j$ and $k$ run from 
$1$ to $\mathcal{Q}_{\alpha}$ ($\alpha \in \{x, y, z\}$), where $\mathcal{Q}_\alpha = Q_\alpha$ 
or $\mathcal{Q}_{\alpha} = 2Q_\alpha$ when a full-range or half-range quadrature of order $Q_\alpha$ 
is employed on the $\alpha$ axis. The distribution function $f_{ijk}$ corresponding to the 
momentum vector $\vp_{ijk}$ is linked to the Boltzmann distribution function through the direct extension 
of Eq.~\eqref{eq:H_fk}:
\begin{equation}
 f_{ijk} = \left(\frac{w_i^x p_{0,x}}{\omega(p_{x, i})}\right)
 \left(\frac{w_j^y p_{0,y}}{\omega(p_{y, j})}\right)
 \left(\frac{w_k^z p_{0,z}}{\omega(p_{z, k})}\right)
 f^{Q_x, Q_y, Q_z}(p_{x,i}, p_{y,j}, p_{z,k}). 
\end{equation}
The weights $w_s^\alpha$ are given by Eq.~\eqref{eq:H_wk} for the case of the full-range Gauss-Hermite 
quadrature and by Eq.~\eqref{eq:wk_half} when the half-range Gauss-Laguerre or Gauss-Hermite 
quadratures are employed, respectively.

After the discretisation of the momentum space, Eq.~\eqref{eq:boltz_multi} becomes:
\begin{equation}
 \partial_t f_{ijk} + \frac{\vp_{ijk}}{m} \cdot \nabla f_{ijk} + \bm{F} \cdot (\nabla_{\vp} f)_{ijk} = 
 -\frac{1}{\tau} (f_{ijk} - \feq_{ijk}).
 \label{eq:boltz_multi_k}
\end{equation}

\subsection{Expansion of the equilibrium distribution function} \label{app:multi:feq}

The Maxwell-Boltzmann equilibrium distribution can be factorised with respect to the 
Cartesian axes as follows \cite{ambrus16jcp,ambrus16jocs,ambrus17couette}:
\begin{equation}
 \feq = n g_x g_y g_z, 
\end{equation}
where $g_\alpha \equiv g(u_\alpha, T; p_\alpha)$ is defined in Eq.~\eqref{eq:feq}.
After choosing the type of quadrature, the quadrature order $Q_\alpha$ and the expansion order 
$N_\alpha$ on each axis $\alpha$, $\feq_{ijk}$ is obtained by multiplying the corresponding 
truncations of $g_{x,i}$, $g_{y,j}$ and $g_{z,k}$:
\begin{equation}
 \feq_{ijk} = n g_{x,i} g_{y,j} g_{z,k},
\end{equation}
where $g_{\alpha,s}$ is given by Eqs.~\eqref{eq:H_gk} and \eqref{eq:hh_gk} 
when the full-range Gauss-Hermite quadrature and half-range quadratures are employed.

\subsubsection{Force term}\label{app:multi:force}

In order to implement the force term in multiple dimensions, the inner product 
between the force $\bm{F}$ and the momentum derivative $\nabla_{\vp} f$ can be 
expanded as follows:
\begin{equation}
 \bm{F} \cdot \nabla_{\vp} f = F_x \partial_{p_x} f + 
 F_y \partial_{p_y} f + F_z \partial_{p_z} f.
 \label{eq:Fnp_exp}
\end{equation}
Each momentum derivative can be computed using the corresponding kernel function, following the 
prescription in Eqs.~\eqref{eq:H_K_def} and \eqref{eq:hh_K_def}:
\begin{align}
 \partial_{p_x} f =& \int_{-\infty}^\infty dp_x \, \mathcal{K}(p_x, p_x') 
 f(p_x', p_y, p_z), \nonumber\\
 \partial_{p_y} f =& \int_{-\infty}^\infty dp_y \, \mathcal{K}(p_y, p_y') 
 f(p_x, p_y', p_z), \nonumber\\
 \partial_{p_z} f =& \int_{-\infty}^\infty dp_z \, \mathcal{K}(p_z, p_z') 
 f(p_x, p_y, p_z').
\end{align}
The kernel function $\mathcal{K}(p,p')$ is given in Eqs.~\eqref{eq:H_K} 
when the full-range Gauss-Hermite is employed. For a generic half-range quadrature,
the kernel function is given in Eq.~\eqref{eq:hh_K}.

After the discretisation of the momentum space, the momentum derivatives of 
$f$ can be written as:
\begin{align}
 \left(\frac{ \partial f}{\partial p_x}\right)_{ijk} =& 
 \sum_{i' = 1}^{\mathcal{Q}_x} \mathcal{K}_{i,i'} f_{i'jk}, \nonumber\\
 \left(\frac{ \partial f}{\partial p_y}\right)_{ijk} =& 
 \sum_{j' = 1}^{\mathcal{Q}_y} \mathcal{K}_{j,j'} f_{ij'k}, \nonumber\\
 \left(\frac{ \partial f}{\partial p_z}\right)_{ijk} =& 
 \sum_{k' = 1}^{\mathcal{Q}_z} \mathcal{K}_{k,k'} f_{ijk'},
\end{align}
where the kernel function $\mathcal{K}_{s,s'}$ for the full-range Gauss-Hermite 
quadrature is given in Eq.~\eqref{eq:H_Kk}. For a general half-range quadrature, 
the kernel is given in Eq.~\eqref{eq:hh_Kk}, while for the specific cases of the 
Gauss-Laguerre and half-range Gauss-Hermite quadratures, the kernel is given in 
Eqs.~\eqref{eq:hh_Kk_L} and \eqref{eq:hh_Kk_hh}.

Finally, the analogue of Eq.~\eqref{eq:Fnp_exp} after the discretisation of the momentum 
space reads:
\begin{equation}
 \bm{F} \cdot (\nabla_{\vp} f)_{ijk} = F_x \left(\frac{\partial f}{\partial p_x}\right)_{ijk} +
 F_y \left(\frac{\partial f}{\partial p_y}\right)_{ijk} +
 F_z \left(\frac{\partial f}{\partial p_z}\right)_{ijk}.
\end{equation}

\section{Half-range orthogonal polynomials}\label{app:hrange}

Let $\{\phi_\ell(z)\}$ be the set of polynomials employed in Sec.~\ref{sec:hh_force},
which satisfy the orthogonality relation \eqref{eq:phi_ortho} on the domain $[0, \infty)$, forming
a complete set in the sense of Eq.~\eqref{eq:phi_compl}
The polynomial $\phi_\ell(z)$ is of order $\ell$ in $z$ and admits the following expansion:
\begin{equation}
 \phi_\ell(z) = \sum_{s = 0}^\ell \phi_{\ell, s} z^s, 
 \label{eq:phi_expl}
\end{equation}
where $\phi_{\ell, s}$ are constant coefficients. In particular, it can be seen that:
\begin{equation}
 \phi_\ell(0) = \phi_{\ell, 0}.
 \label{eq:phil0}
\end{equation}

The derivative of $\phi_\ell(z)$ with respect to $z$ is a polynomial of order $\ell - 1$, 
which can be expressed with respect to the (complete) set $\{\phi_\ell(z)\}$ as follows:
\begin{equation}
 \phi_\ell'(z) = - \sum_{s = 0}^{\ell - 1} \varphi_{\ell, s} \phi_s(z),
 \label{eq:phip}
\end{equation}
where $\varphi_{\ell, s}$ are constant coefficients. The minus sign was inserted for future convenience.

In the construction of the momentum derivative of the distribution function $f$, 
it will be useful to compute the expansion of the derivative of $\omega(z) \phi_\ell(z)$
in the following form:
\begin{equation}
 \frac{\partial [\omega(z) \phi_\ell(z)]}{\partial z} = \omega(z) 
 \sum_{s = 0}^\infty \frac{1}{\gamma_s} \psi_{\ell, s} \phi_s(z),
 \label{eq:ophip_def}
\end{equation}
where $\psi_{\ell, s}$ can be calculated as in Eq.~\eqref{eq:hh_F}, using an integration by parts:
\begin{equation}
 \psi_{\ell, s} = - \omega(0) \phi_\ell(0) \phi_s(0) + 
 \sum_{j = 0}^{s - 1} \varphi_{s, j} \braket{\phi_\ell, \phi_j}.
\end{equation}
Equation~\eqref{eq:phip} was used to eliminate $\phi'_s(x)$. The inner product above vanishes for all values of $j$ which 
are smaller than $\ell$. Hence, it only contributes to $\psi_{\ell, s}$ when $s > \ell$. 
Substituting the above result back into Eq.~\eqref{eq:ophip_def}, the following result can be obtained:
\begin{equation}
 \frac{\partial [\omega(z) \phi_\ell(z)]}{\partial z} = - \omega(z) \omega(0) \phi_{\ell,0} 
 \sum_{s = 0}^\infty \frac{1}{\gamma_s} \phi_{s,0} \phi_s(z) + \omega(z) \gamma_\ell 
 \sum_{s = \ell + 1}^\infty \frac{1}{\gamma_s} \varphi_{s, \ell} \phi_s(z). \label{eq:ophip}
\end{equation}

Let us now specialise the above to the case of the half-range Hermite polynomials 
$\{\hh_\ell(z)\}$, for which the weight function $\omega(z)$ and norm $\gamma_\ell$ 
have the following expressions:
\begin{equation}
 \omega(z) = \frac{1}{\sqrt{2\pi}} e^{-z^2/2}, \qquad \gamma_{\ell} = 1.
 \label{eq:hh_omega}
\end{equation}
The coefficients $\varphi_{\ell,s}$ introduced in Eq.~\eqref{eq:phip} can be obtained 
by multiplying Eq.~\eqref{eq:phip} by $\omega(z) \phi_s(z)$ and integrating with respect to $z$:
\begin{equation}
 \varphi_{\ell, s} = \omega(0) \hh_{\ell,0} \hh_{s,0}
 - \braket{\hh_\ell, z \hh_s} + \braket{\hh_\ell, \hh_s'},\label{eq:hh_varphi_aux}
\end{equation}
where an integration by parts was used. Since $\hh_s'$ is a polynomial of order $s - 1 < \ell$,
the last term in the above equality vanishes. To evaluate the second term, the following 
recursion relation can be employed \cite{ambrus16jcp}:
\begin{equation}
 \hh_{s + 1}(z) = (a_s z + b_s) \hh_s(z) + c_s \hh_{s - 1}(z).
\end{equation}
Since $s < \ell$, $\hh_\ell$ is orthogonal on $\hh_s$ and $\hh_{s -1}$, allowing 
Eq.~\eqref{eq:hh_varphi_aux} to be written as:
\begin{equation}
 \varphi_{\ell, s} = \omega(0) \hh_{\ell,0} \hh_{s,0} -
 \frac{1}{a_s} \delta_{\ell, s + 1},\label{eq:hh_varphi}
\end{equation}
where $a_s$ is given by \cite{ambrus16jcp}:
\begin{equation}
 a_s = \frac{\hh_{s + 1, s + 1}}{\hh_{s, s}}.\label{eq:a_def}
\end{equation}

\section{Note on supplemental material}\label{app:supp}

The supplemental material available at the publisher's website consists of a collection of data files organised as 
follows.

The files \texttt{roots-herm.txt}, \texttt{weights-herm.txt}, \texttt{roots-hh.txt} and \texttt{weights-hh.txt} 
contain the roots of the full-range and half-range Hermite polynomials employed in this paper, as well as their 
corresponding weights. Each of these files comprise $200$ lines corresponding to $1 \le Q \le 200$. Line $Q$ consists of 
$Q$ entries corresponding to the roots or weights for the quadrature order $Q$, as follows:
\begin{equation*}
 \begin{matrix}
 x_{1,1} & & & \\
 x_{2,1} &  x_{2,2} & \\
 x_{3,1} &  x_{3,2} &  x_{3,3} &\\
 \hdotsfor{4}
 \end{matrix}
\end{equation*}
In the above, $x_{Q,k}$ represents the $k$'th root or weight for quadrature order $Q$.

The files \texttt{kherm-Q*.dat} contain, for a given value of $Q$, the $Q \times Q$ matrix elements of the 
kernel $\mathcal{K}^{\rm H}_{k,k'}$ \eqref{eq:H_Kk}. The files consist of $Q$ rows corresponding to $1 \le k \le Q$ with 
$Q$ entries each, corresponding to $1 \le k' \le Q$:
\begin{equation*}
 \begin{matrix}
 \mathcal{K}^{\rm H}_{1,1} & \mathcal{K}^{\rm H}_{1,2} & \dots & \mathcal{K}^{\rm H}_{1,Q}\\
 \mathcal{K}^{\rm H}_{2,1} & \mathcal{K}^{\rm H}_{2,2} & \dots & \mathcal{K}^{\rm H}_{2,Q}\\
 \hdotsfor{4}\\
 \mathcal{K}^{\rm H}_{Q,1} & \mathcal{K}^{\rm H}_{Q,2} & \dots & \mathcal{K}^{\rm H}_{Q,Q}
 \end{matrix}
\end{equation*}

Finaly, the files \texttt{khh-Q*.dat} contain, for a given value of $Q$, the $2Q \times 2Q$ matrix elements of the 
kernel $\mathcal{K}^{\hh}_{k,k'}$ \eqref{eq:hh_Kk}, with $1 \le k \le 2Q$ and $1 \le k' \le 2Q$. 
The $(k')$'th element on line $k$ corresponds to $\mathcal{K}^{\hh}_{k,k'}$.

\bibliographystyle{elsarticle-num} 

\bibliography{lb}

\begin{thebibliography}{10}
\expandafter\ifx\csname url\endcsname\relax
  \def\url#1{\texttt{#1}}\fi
\expandafter\ifx\csname urlprefix\endcsname\relax\def\urlprefix{URL }\fi
\expandafter\ifx\csname href\endcsname\relax
  \def\href#1#2{#2} \def\path#1{#1}\fi

\bibitem{gadelhaq06app}
M.~G. el~Haq, {MEMS} Applications, CRC Press, Boca Raton, 2006.

\bibitem{tabeling11}
P.~Tabeling, Introduction to Microfluidics (English translation), Oxford
  University Press, Oxford, 2011.

\bibitem{stone01}
H.~A. Stone, S.~Kim, Microfluidics: Basic issues, applications, and challenges,
  AIChE Journal 47 (2001) 1250--1254.
\newblock \href {http://dx.doi.org/10.1002/aic.690470602}
  {\path{doi:10.1002/aic.690470602}}.

\bibitem{delgado01}
A.~V. Delgado, Interfacial Electrokinetics and Electrophoresis, CRC Press, New
  York, 2001.

\bibitem{voldman06}
Voldman, Electrical forces for microscale cell manipulation, Annual Review of
  Biomedical Engineering 8 (2006) 425--454.
\newblock \href {http://dx.doi.org/10.1146/annurev.bioeng.8.061505.095739}
  {\path{doi:10.1146/annurev.bioeng.8.061505.095739}}.

\bibitem{fede15}
P.~Fede, V.~Sofonea, R.~Fournier, S.~Blanco, O.~Simonin, G.~Lepout\'{e}re,
  V.~E. Ambru\cb{s}, Lattice {B}oltzmann model for predicting the deposition of
  inertial particles transported by a turbulent flow, Int. J. Multiph. Flow 76
  (2015) 187--197.
\newblock \href {http://dx.doi.org/10.1016/j.ijmultiphaseflow.2015.07.004}
  {\path{doi:10.1016/j.ijmultiphaseflow.2015.07.004}}.

\bibitem{balescu05}
R.~Balescu, Aspects of anomalous transport in plasmas, IOP Publishing Ltd,
  Cornwall (UK), 2005.

\bibitem{bangalore}
B.~Piaud, S.~Blanco, R.~Fournier, V.~E. Ambru\cb{s}, V.~Sofonea, {G}auss
  quadratures - the keystone of lattice {B}oltzmann models, Int. J. Mod. Phys.
  C 25 (2014) 1340016.
\newblock \href {http://dx.doi.org/10.1142/S0129183113400160}
  {\path{doi:10.1142/S0129183113400160}}.

\bibitem{shan06}
X.~W. Shan, X.~F. Yuan, H.~D. Chen, Kinetic theory representation of
  hydrodynamics: a way beyond the {N}avier-{S}tokes equation, J. Fluid. Mech.
  550 (2006) 413--441.
\newblock \href {http://dx.doi.org/10.1017/S0022112005008153}
  {\path{doi:10.1017/S0022112005008153}}.

\bibitem{deville}
M.~O. Deville, T.~B. Gatski, Mathematical Modeling for Complex Fluids and
  Flows, Springer, Berlin, 2012.

\bibitem{grad}
H.~Grad, Principles of the Kinetic Theory of Gases, Vol. XII of Encyclopedia of
  Physics (editor: S. Fl\"{u}gge), Springer, Berlin, 1958.

\bibitem{sone02}
Y.~Sone, Kinetic Theory and Fluid Dynamics, Birkh\"{a}user, Boston, 2002.

\bibitem{karniadakis05}
G.~Karniadakis, A.~Beskok, N.~Aluru, Microflows and Nanoflows: Fundamentals and
  Simulation, Springer, Berlin, 2005.

\bibitem{struchtrup05}
H.~Struchtrup, Macroscopic Transport Equations for Rarefied Gas Flows,
  Springer, Berlin, 2005.

\bibitem{shen05}
C.~Shen, Rarefied Gas Dynamics: Fundamentals, Simulations and Micro Flows,
  Springer, Berlin, 2005.

\bibitem{sone06}
Y.~Sone, Molecular Gas Dynamics: Theory, Techniques and Applications,
  Birkh\"{a}user, Boston, 2007.

\bibitem{maxwell1879}
J.~C. Maxwell, On stresses in rarified gases arising from inequalities of
  temperature, Philos. Trans. R. Soc. London 170 (1879) 231--256.

\bibitem{meng11jcp}
J.~P. Meng, Y.~H. Zhang, Accuracy analysis of high-order lattice {B}oltzmann
  models for rarefied gas flows, J. Comput. Phys. 230 (2011) 835--849.
\newblock \href {http://dx.doi.org/10.1016/j.jcp.2010.10.023}
  {\path{doi:10.1016/j.jcp.2010.10.023}}.

\bibitem{adif}
S.~Ansumali, I.~V. Karlin, Kinetic boundary conditions in the lattice
  {B}oltzmann method, Phys. Rev. E 66 (2002) 026311.
\newblock \href {http://dx.doi.org/10.1103/PhysRevE.66.026311}
  {\path{doi:10.1103/PhysRevE.66.026311}}.

\bibitem{meng11pre}
J.~P. Meng, Y.~H. Zhang, {G}auss-{H}ermite quadratures and accuracy of lattice
  {B}oltzmann models for nonequilibrium gas flows, Phys. Rev. E 83 (2011)
  036704.
\newblock \href {http://dx.doi.org/10.1103/PhysRevE.83.036704}
  {\path{doi:10.1103/PhysRevE.83.036704}}.

\bibitem{meng11pre2}
J.~P. Meng, Y.~H. Zhang, X.~W. Shan, Multiscale lattice {B}oltzmann approach to
  modeling gas flows, Phys. Rev. E 83 (2011) 046701.
\newblock \href {http://dx.doi.org/10.1103/PhysRevE.83.046701}
  {\path{doi:10.1103/PhysRevE.83.046701}}.

\bibitem{meng13jfm}
J.~P. Meng, Y.~H. Zhang, N.~G. Hadjiconstantinou, G.~A. Radtke, X.~W. Shan,
  Lattice ellipsoidal statistical {BGK} model for thermal non-equilibrium
  flows, J. Fluid. Mech. 718 (2013) 347--370.
\newblock \href {http://dx.doi.org/10.1017/jfm.2012.616}
  {\path{doi:10.1017/jfm.2012.616}}.

\bibitem{ambrus12}
V.~E. Ambru\cb{s}, V.~Sofonea, High-order thermal lattice {B}oltzmann models
  derived by means of {G}auss quadrature in the spherical coordinate system,
  Phys. Rev. E 86 (2012) 016708.
\newblock \href {http://dx.doi.org/10.1103/PhysRevE.86.016708}
  {\path{doi:10.1103/PhysRevE.86.016708}}.

\bibitem{yang95}
J.~Y. Yang, J.~C. Huang, L.~Tsuei, Numerical solutions of the nonlinear model
  {B}oltzmann equations, Proc. R. Soc. Lond. A 448 (1995) 55--80.
\newblock \href {http://dx.doi.org/10.1098/rspa.1995.0003}
  {\path{doi:10.1098/rspa.1995.0003}}.

\bibitem{li03}
Z.-H. Li, H.-X. Zhang, Numerical investigation from rarefied flow to continuum
  by solving the boltzmann model equation, Int. J. Numer. Meth. Fluids 42
  (2003) 361--382.
\newblock \href {http://dx.doi.org/10.1002/fld.517}
  {\path{doi:10.1002/fld.517}}.

\bibitem{li04}
Z.-H. Li, H.-X. Zhang, Study on gas kinetic unified algorithm for flows from
  rarefied transition to continuum, J. Comput. Phys 193 (2004) 708--738.
\newblock \href {http://dx.doi.org/10.1016/j.jcp.2003.08.022}
  {\path{doi:10.1016/j.jcp.2003.08.022}}.

\bibitem{lorenzani07}
S.~Lorenzani, L.~Gibelli, A.~Frezzotti, A.~Frangi, C.~Cercignani, Kinetic
  approach to gas flows in microchannels, Nanoscale and Microscale
  Thermophysical Engineering 11 (2007) 211--226.
\newblock \href {http://dx.doi.org/10.1080/15567260701333489}
  {\path{doi:10.1080/15567260701333489}}.

\bibitem{li09}
Z.-H. Li, H.-X. Zhang, Gas-kinetic numerical studies of three-dimensional
  complex flows on spacecraft re-entry, J. Comput. Phys 228 (2009) 1116--1138.
\newblock \href {http://dx.doi.org/10.1016/j.jcp.2008.10.013}
  {\path{doi:10.1016/j.jcp.2008.10.013}}.

\bibitem{frezzotti09}
A.~Frezzotti, L.~Gibelli, B.~Franzelli, A moment method for low speed
  microflows, Continuum~Mech. Thermodyn. 21 (2009) 495--509.
\newblock \href {http://dx.doi.org/10.1007/s00161-009-0128-y}
  {\path{doi:10.1007/s00161-009-0128-y}}.

\bibitem{frezzotti11}
A.~Frezzotti, G.~P. Ghiroldi, L.~Gibelli, Solving the {B}oltzmann equation on
  {GPU}s, Comput. Phys. Comm. 182 (2011) 2445--2453.
\newblock \href {http://dx.doi.org/10.1016/j.cpc.2011.07.002}
  {\path{doi:10.1016/j.cpc.2011.07.002}}.

\bibitem{gibelli12}
L.~Gibelli, Velocity slip coefficients based on the hard-sphere {B}oltzmann
  equation, Phys. Fluids 24 (2012) 022001.
\newblock \href {http://dx.doi.org/10.1063/1.3680873}
  {\path{doi:10.1063/1.3680873}}.

\bibitem{guo13pre}
Z.~Guo, K.~Xu, R.~Wang, Discrete unified gas kinetic scheme for all {K}nudsen
  number flows: Low-speed isothermal case, Phys. Rev. E 88 (2013) 033305.
\newblock \href {http://dx.doi.org/10.1103/PhysRevE.88.033305}
  {\path{doi:10.1103/PhysRevE.88.033305}}.

\bibitem{ghiroldi14}
G.~P. Ghiroldi, L.~Gibelli, A direct method for the {B}oltzmann equation based
  on a pseudo-spectral velocity space discretization, J. Comput. Phys. 258
  (2014) 568--584.
\newblock \href {http://dx.doi.org/10.1016/j.jcp.2013.10.055}
  {\path{doi:10.1016/j.jcp.2013.10.055}}.

\bibitem{guo15pre}
Z.~Guo, R.~Wang, K.~Xu, Discrete unified gas kinetic scheme for all knudsen
  number flows. ii. thermal compressible case, Phys. Rev. E 91 (2015) 033313.
\newblock \href {http://dx.doi.org/10.1103/PhysRevE.91.033313}
  {\path{doi:10.1103/PhysRevE.91.033313}}.

\bibitem{gibelli15}
G.~P. Ghiroldi, L.~Gibelli, A finite-difference lattice {B}oltzmann approach
  for gas microflows, Commun. Comput. Phys. 17 (2015) 1007--1018.
\newblock \href {http://dx.doi.org/10.4208/cicp.2014.m424}
  {\path{doi:10.4208/cicp.2014.m424}}.

\bibitem{sader15}
Y.~Shi, Y.~W. Yap, J.~E. Sader, Linearized lattice {B}oltzmann method for
  micro- and nanoscale flow and heat transfer, Phys. Rev. E 92 (2015) 013307.
\newblock \href {http://dx.doi.org/10.1103/PhysRevE.92.013307}
  {\path{doi:10.1103/PhysRevE.92.013307}}.

\bibitem{ambrus14ipht}
V.~E. Ambru\cb{s}, V.~Sofonea, Application of lattice {B}oltzmann models based
  on {L}aguerre quadratures in complex flows, Interfac. Phenom. Heat~Transfer 2
  (2014) 235--251.
\newblock \href
  {http://dx.doi.org/10.1615/InterfacPhenomHeatTransfer.2015011655}
  {\path{doi:10.1615/InterfacPhenomHeatTransfer.2015011655}}.

\bibitem{ambrus14pre}
V.~E. Ambru\cb{s}, V.~Sofonea, Implementation of diffuse-reflection boundary
  conditions using lattice {B}oltzmann models based on half-space
  {G}auss-{L}aguerre quadratures, Phys. Rev. E 89 (2014) 041301(R).
\newblock \href {http://dx.doi.org/10.1103/PhysRevE.89.041301}
  {\path{doi:10.1103/PhysRevE.89.041301}}.

\bibitem{ambrus14ijmpc}
V.~E. Ambru\cb{s}, V.~Sofonea, Lattice {B}oltzmann models based on {G}auss
  quadratures, Int. J. Mod. Phys. C 25 (2014) 1441011.
\newblock \href {http://dx.doi.org/10.1142/S0129183114410113}
  {\path{doi:10.1142/S0129183114410113}}.

\bibitem{ambrus16jcp}
V.~E. Ambru\cb{s}, V.~Sofonea, Lattice {B}oltzmann models based on half-range
  {G}auss-{H}ermite quadratures, J. Comput. Phys. 316 (2016) 760--788.
\newblock \href {http://dx.doi.org/10.1016/j.jcp.2016.04.010}
  {\path{doi:10.1016/j.jcp.2016.04.010}}.

\bibitem{ambrus16jocs}
V.~E. Ambru\cb{s}, V.~Sofonea, Application of mixed quadrature lattice
  {B}oltzmann models for the simulation of {P}oiseuille flow at non-negligible
  values of the {K}nudsen number, J. Comput. Sci.\href
  {http://dx.doi.org/10.1016/j.jocs.2016.03.016}
  {\path{doi:10.1016/j.jocs.2016.03.016}}.

\bibitem{ambrus17couette}
V.~E. Ambru\cb{s}, V.~Sofonea, Half-range lattice {B}oltzmann models for the
  simulation of {C}ouette flow using the {S}hakhov collision term,
  arXiv:1702.01335 [physics.flu-dyn].

\bibitem{martys}
N.~S. Martys, X.~Shan, H.~Chen, Evaluation of the external force term in the
  discrete {B}oltzmann equation, Phys. Rev. E 58 (1998) 6855--6857.
\newblock \href {http://dx.doi.org/10.1103/PhysRevE.58.6855}
  {\path{doi:10.1103/PhysRevE.58.6855}}.

\bibitem{nist}
F.~W.~J. Olver, D.~W. Lozier, R.~F. Boisvert, C.~W. Clark, {NIST} Handbook of
  Mathematical Functions, Cambridge University Press, New York, 2010.

\bibitem{gross}
E.~P. Gross, E.~A. Jackson, S.~Ziering, Boundary value problems in kinetic
  theory of gases, Ann. Phys. 1 (1957) 141--167.
\newblock \href {http://dx.doi.org/10.1016/0003-4916(57)90056-8}
  {\path{doi:10.1016/0003-4916(57)90056-8}}.

\bibitem{graur09}
I.~A. Graur, A.~P. Polikarpov, Comparison of different kinetic models for the
  heat transfer problem, Heat Mass Transf. 46 (2009) 237--244.
\newblock \href {http://dx.doi.org/10.1007/s00231-009-0558-x}
  {\path{doi:10.1007/s00231-009-0558-x}}.

\bibitem{kim08}
S.~H. Kim, H.~Pitsch, I.~D. Boyd, Accuracy of higher-order lattice {B}oltzmann
  methods for microscale flows with finite {K}nudsen numbers, J. Comput. Phys.
  227 (2008) 8655--8671.
\newblock \href {http://dx.doi.org/10.1016/j.jcp.2008.06.012}
  {\path{doi:10.1016/j.jcp.2008.06.012}}.

\bibitem{watari07}
M.~Watari, Finite difference lattice {B}oltzmann method with arbitrary specific
  heat ratio applicable to supersonic flow simulations, Physica A 382 (2007)
  502--522.
\newblock \href {http://dx.doi.org/10.1016/j.physa.2007.03.037}
  {\path{doi:10.1016/j.physa.2007.03.037}}.

\bibitem{watari09}
M.~Watari, Velocity slip and temperature jump simulations by the
  three-dimensional thermal finite-difference lattice {B}oltzmann method, Phys.
  Rev. E 79 (2009) 066706.
\newblock \href {http://dx.doi.org/10.1103/PhysRevE.79.066706}
  {\path{doi:10.1103/PhysRevE.79.066706}}.

\bibitem{watari10}
M.~Watari, Relationship between accuracy and number of velocity particles of
  the finite-difference lattice {B}oltzmann method in velocity slip
  simulations, J. Fluid Eng. 132 (2010) 101401.
\newblock \href {http://dx.doi.org/10.1115/1.4002359}
  {\path{doi:10.1115/1.4002359}}.

\bibitem{shi11}
Y.~Shi, P.~L. Brookes, Y.~W. Wap, J.~E. Sader, Accuracy of the lattice
  {B}oltzmann method for low-speed noncontinuum flows, Phys. Rev. E 83 (2011)
  045701(R).
\newblock \href {http://dx.doi.org/10.1103/PhysRevE.83.045701}
  {\path{doi:10.1103/PhysRevE.83.045701}}.

\bibitem{izarra11}
L.~de~Izarra, J.~L. Rouet, B.~Izrar, High-order lattice {B}oltzmann models for
  gas flow for a wide range of {K}nudsen numbers, Phys. Rev. E 84 (2011)
  066705.
\newblock \href {http://dx.doi.org/10.1103/PhysRevE.84.066705}
  {\path{doi:10.1103/PhysRevE.84.066705}}.

\bibitem{toro09}
E.~F. Toro, Riemann Solvers and Numerical Methods for Fluid Dynamics: A
  Practical Introduction, 3rd Edition, Springer, Berlin, 2009.

\bibitem{rezzolla13}
L.~Rezzolla, O.~Zanotti, Relativistic hydrodynamics, Oxford University Press,
  Oxford, UK, 2013.

\bibitem{hildebrand87}
F.~B. Hildebrand, Introduction to Numerical Analysis, second edition Edition,
  Dover Publications, 1987.

\bibitem{shizgal15}
B.~Shizgal, Spectral Methods in Chemistry and Physics: Applications to Kinetic
  Theory and Quantum Mechanics (Scientific Computation), Springer, 2015.

\bibitem{shu88}
C.-W. Shu, S.~Osher, Efficient implementation of essentially non-oscillatory
  shock-capturing schemes, J. Comput. Phys. 77 (1988) 439--471.
\newblock \href {http://dx.doi.org/10.1016/0021-9991(88)90177-5}
  {\path{doi:10.1016/0021-9991(88)90177-5}}.

\bibitem{gottlieb98}
S.~Gottlieb, C.-W. Shu, Total variation diminishing {R}unge-{K}utta schemes,
  Math. Comp. 67 (1998) 73--85.
\newblock \href {http://dx.doi.org/10.1090/S0025-5718-98-00913-2}
  {\path{doi:10.1090/S0025-5718-98-00913-2}}.

\bibitem{henrick05}
A.~K. Henrick, T.~D. Aslam, J.~M. Powers, Mapped weighted essentially
  non-oscillatory schemes: {A}chieving optimal order near critical points, J.
  Comput. Phys. 207 (2005) 542--567.
\newblock \href {http://dx.doi.org/10.1016/j.jcp.2005.01.023}
  {\path{doi:10.1016/j.jcp.2005.01.023}}.

\bibitem{trangenstein07}
J.~A. Trangenstein, Numerical solution of hy\-per\-bo\-lic par\-tial
  differential equations, Cam\-bridge Uni\-ver\-si\-ty Press, New York, USA,
  2007.

\bibitem{blaga16jcp}
R.~Blaga, V.~E. Ambru\cb{s}, High-order quadrature-based lattice {B}oltzmann
  models for the flow of ultrarelativistic rarefied gases, arXiv:1612.01287
  [physics.flu-dyn].

\bibitem{jiang96}
G.~S. Jiang, C.~W. Shu, Efficient implementation of {W}eighted {ENO} schemes,
  J. Comput. Phys. 126 (1996) 202.
\newblock \href {http://dx.doi.org/10.1006/jcph.1996.0130}
  {\path{doi:10.1006/jcph.1996.0130}}.

\bibitem{shu99}
C.-W. Shu, High order {ENO} and {WENO} schemes for computational fluid
  dynamics, in: T.~J. Barth, H.~Deconinck (Eds.), High-order methods for
  computational physics, Springer-Verlag, Berlin, 1999, pp. 439--582.

\bibitem{gan11}
Y.~Gan, A.~Xu, G.~Zhang, Y.~Li, Lattice {B}oltzmann study on kelvin-helmholtz
  instability: Roles of velocity and density gradients, Phys. Rev. E 83 (2011)
  056704.
\newblock \href {http://dx.doi.org/10.1103/PhysRevE.83.056704}
  {\path{doi:10.1103/PhysRevE.83.056704}}.

\bibitem{hejranfar17pre}
K.~Hejranfar, M.~H. Saadat, S.~Taheri, High-order weighted essentially
  nonoscillatory finite-difference formulation of the lattice {B}oltzmann
  method in generalized curvilinear coordinates, Phys. Rev. E 95 (2017) 023314.
\newblock \href {http://dx.doi.org/10.1103/PhysRevE.95.023314}
  {\path{doi:10.1103/PhysRevE.95.023314}}.

\bibitem{busuioc17}
S.~Busuioc, V.~E. Ambru\cb{s}, Lattice {B}oltzmann models based on the vielbein
  formalism for the simulation of the circular {C}ouette flow, arXiv:1708.05944
  [physics.flu-dyn].

\bibitem{mei98}
R.~Mei, W.~Shyy, On the finite difference-based lattice {B}oltzmann method in
  curvilinear coordinates, J. Comput. Phys. 143 (1998) 426--448.
\newblock \href {http://dx.doi.org/10.1006/jcph.1998.5984}
  {\path{doi:10.1006/jcph.1998.5984}}.

\bibitem{guo03}
Z.~Guo, T.~S. Zhao, Explicit finite-difference lattice {B}oltzmann method for
  curvilinear coordinates, Phys. Rev. E 67 (2003) 066709.
\newblock \href {http://dx.doi.org/10.1103/PhysRevE.67.066709}
  {\path{doi:10.1103/PhysRevE.67.066709}}.

\bibitem{huang63}
K.~Huang, Statistical Mechanics, Wiley, New York, 1963.

\bibitem{harris71}
S.~Harris, An Introduction to the Theory of the Boltzmann Equation, Holt,
  Rinehart and Winston, New York, 1971.

\bibitem{cercignani88}
C.~Cercignani, The {B}oltzmann Equation and Its Applications, Springer-Verlag,
  New York, 1988.

\end{thebibliography}

\end{document}